\documentclass[journal]{IEEEtran}
\usepackage{amsmath,amsthm}
\usepackage{graphicx}
\usepackage{epstopdf}
\usepackage{amsfonts,amsmath,amssymb}
\usepackage{cite}
\usepackage{graphicx}
\usepackage{url}
\usepackage{bm}
\usepackage{bbm}
\usepackage{amssymb}
\usepackage{array}
\usepackage{booktabs}
\usepackage{multirow}
\begin{document}
%\newpage
%
%\pagenumbering{arabic}
%\setcounter{page}{1}    % set page to 1 again to start arabic count

\title{{Satellite-Based Continuous-Variable Quantum Communications:\\State-of-the-Art and a Predictive Outlook}}
\author{
\IEEEauthorblockN{Nedasadat Hosseinidehaj$^1$, Robert Malaney$^1$}\\
\IEEEauthorblockA{$^1$School of Electrical Engineering  \& Telecommunications,\\
The University of New South Wales, Sydney, NSW 2052, Australia.}\\
%\\neda.hosseini@unsw.edu.au, r.malaney@unsw.edu.au}
\and
\IEEEauthorblockN{Soon Xin Ng$^2$, Lajos Hanzo$^{2}$}\\
\IEEEauthorblockA{$^2$School of Electronics and Computer Science,\\
University of Southampton, Southampton SO17 1BJ, U.K.\\
}}

%\author{
%\IEEEauthorblockN{Shihao Yan$^1$, Robert Malaney$^1$}
%\IEEEauthorblockA{$^1$School of Electrical Engineering  \& Telecommunications,\\
%The University of New South Wales,\\
%Sydney, NSW 2052, Australia}
%%Email: shihao.yan@student.unsw.edu.au;  r.malaney@unsw.edu.au}
%\and
%\IEEEauthorblockN{Ido Nevat$^2$, Gareth W. Peters$^{3}$}
%\IEEEauthorblockA{$^2$Institute for Infocom Research, A$^{\star}$STAR, Singapore.\\
%%$^3$School of Mathematics and Statistics, University of NSW\\
%$^3$Department of Statistical Science,\\
%University College London, London, UK
%}}

\vspace{-1cm}
\maketitle
\begin{abstract}
The recent launch of the Micius quantum-enabled satellite heralds a major step forward for long-range quantum communication. Using single-photon discrete-variable quantum states, this exciting new development proves beyond any doubt that all of the quantum protocols previously deployed over limited ranges in terrestrial experiments can in fact be translated to global distances via the use of low-orbit satellites. In this work we survey the imminent extension of space-based quantum communication to the continuous-variable regime -  the quantum regime perhaps most closely related to classical wireless communications. The CV regime offers the potential for increased communication performance, and represents the next major step forward for quantum communications and the development of the global quantum internet.
\end{abstract}

\section{Motivation and Introduction}

{\em  Moore's Law has remained valid for half-a-century! As a result,
  contemporary semi-conductor technology is approaching nano-scale
  integration.  Hence nano-technology is about to enter the realms of
  quantum physics, where many of the physical phenomena are rather
  different from those of classical physics. Hence this treatise
  contributes towards completing the `quantum jig-saw puzzle' by
  paving the way from classical wireless systems to their perfectly
  secure quantum-communications counterparts, as
  heralded in~\cite{6191306,6515077}. }

\begin{figure}[tbh]
\begin{center}
\includegraphics[width=\linewidth]{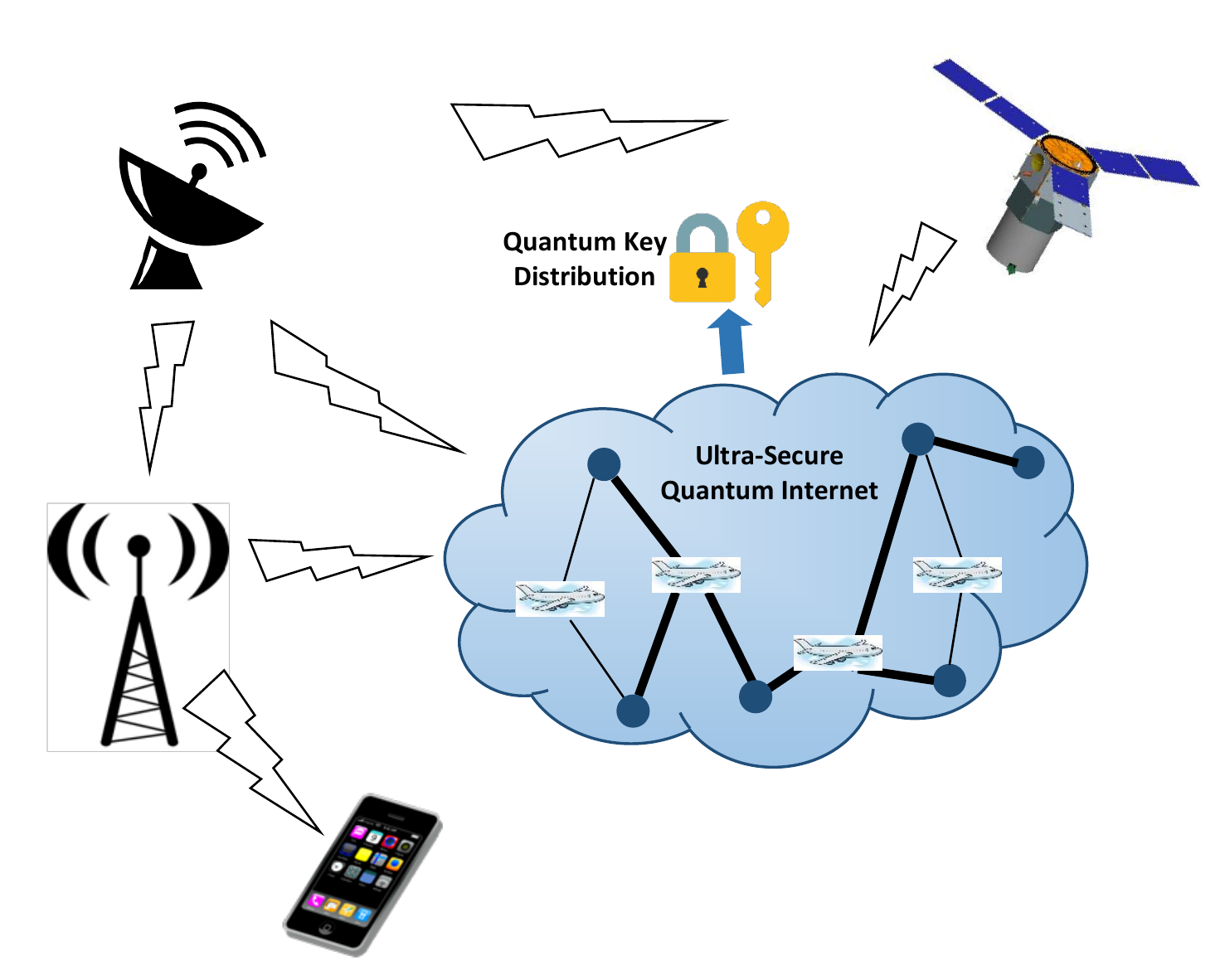}
\end{center}
\caption{Stylized vision of future global quantum communications}
\label{moores}
\end{figure}

\begin{itemize}
\item {\bf {\em The Inspiration:}} In order to circumvent the specific
  limitations of the classical wireless systems detailed
  in~\cite{6191306}, we set out to bridge the separate
  classical and quantum worlds into a joint universe, with the
  objective of contributing to perfectly secure quantum-aided
  communications for anyone, anywhere, anytime across the globe, as
  indicated by the stylized vision of the near-future quantum
  communications scenario seen in Fig.~\ref{moores}.

\item {\bf {\em The Reality:}} However, quantum processing is far from
  being flawless - it has substantial challenges, as detailed in this
  contribution.  Nonetheless, at the time of writing long-range
  quantum communications via satellites has become a reality.
\end{itemize}

%%%%%%%%%%%%%%%%%%%%%%%%%%%%%%%%%%%%%%%%%%%%%%%%%%%%%%%%%%%%%%%%
\begin{figure*}
    \begin{center}
   {\includegraphics[width=150mm]{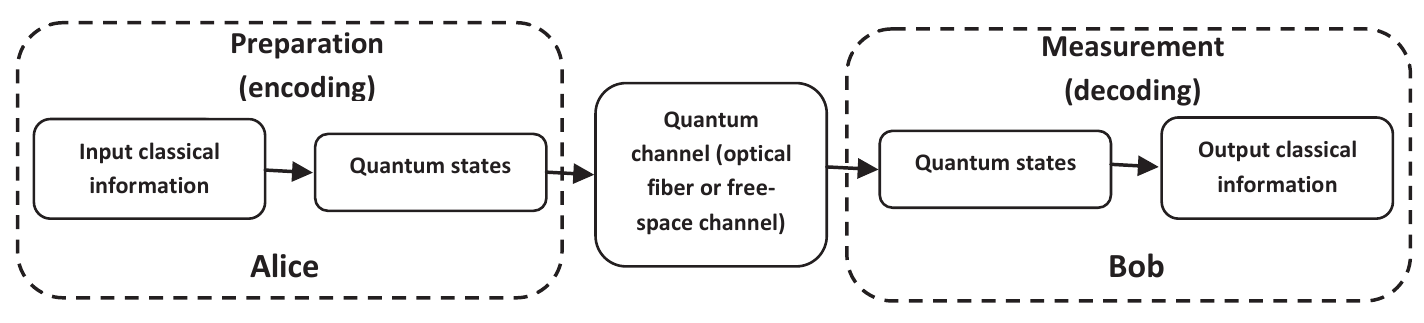}}
    \caption{Basic quantum communications schematic.}\label{fig:quantum communication}
    \end{center}
\end{figure*}

Amongst its numerous intriguing attributes, quantum communication
has the potential to achieve secure communications at confidence
levels simply unattainable in classical communications settings. This
is due to the fact that quantum physics introduces a range of
phenomena which have no counterpart in the classical domain, such as
quantum entanglement and the superposition of quantum
states\footnote{The superposition of a logical one and zero may be
  viewed as a coin spinning in a box, where we cannot claim to show
  its state being `head' or `tail'. When we stop spinning the coin,
  and lift the lid of the box, the superposition-based quantum state
  collapses back into the classical domain as a consequence of us
  observing it.}  The exploitation of such effects, both before and
after the transmission of information in the quantum domain, can in
effect lead to communications possessing `unconditional' security.

Quantum communication entails the transfer of quantum states from one place to another via a quantum channel. In a generic form, quantum communication consists of three steps: (i) the preparation of quantum states  - where the original classical information is encoded into quantum states; (ii) the transmission of the prepared quantum states over a quantum channel such as optical fiber or a free-space optical (FSO) channel - where the states are transmitted from a transmitter, held by Alice, to a receiver, named Bob; and (iii) detection - where the received states are decoded using quantum measurement resulting in some output classical information. A schematic including these three steps is shown in Fig.~\ref{fig:quantum communication}.

A key motivation for quantum communication is that the quantum information, mapped for example to the polarization of a photon, can be shared more securely than classical information. %allows tasks to be performed that could only be achieved far less efficiently using classical information.
The well-known example of this is quantum key distribution (QKD) \cite{BB84}, whose unconditional security has been theoretically proved (classical cryptography schemes are not proved to be secure). We also note the close connection between quantum communication and quantum entanglement. A pair of quantum states are said to be entangled if, for example, changing the polarization of a photon results in an instantaneous polarization change for its entangled pair. Einstein referred to this as a `spooky action at a distance.' Important quantum communication protocols utilizing entangled states include QKD, quantum teleportation \cite{teleB, vaid, OP-Telep}, and entanglement swapping (teleportation of entanglement) \cite{1st-tele}. %, quantum super dense coding, and quantum error correction.

%There are two main technologies to implement quantum communication, discrete-variable (DV) where the information is encoded on the properties of single photons such as the phase or polarization e.g., \cite{BB84, 144km, f2}, and continuous-variable (CV) where the information is encoded on the quadrature variables of the optical field e.g., \cite{1st-gaussQKD, RR2002, RR, 1st-EB, Scarani}. In the former technology detection is realized by single-photon detectors, while in the latter technology detection is realized by homodyne (or heterodyne) detectors which are faster and more efficient relative to single-photon detectors.

In terms of representing the quantum states in quantum communications, discrete-variable (DV) and continuous-variable (CV) descriptions have been used \cite{book-DV,CV-review}. In the former, information is mapped to discrete features  such as the polarization of single photons \cite{BB84}. The detection of such features would then be realized by single-photon detectors. In DV technology information is mapped to two (or to a finite number) of basis states. The standard unit of DV quantum information in the two basis form is the quantum bit, also known as the `qubit.' In a qubit, information is carried as a superposition of two orthogonal  quantum states which can be represented mathematically as $\left| \psi  \right\rangle  = a_1\left| 0 \right\rangle  + a_2\left| 1 \right\rangle $ with ${\left| a_1 \right|^2} + {\left| a_2 \right|^2} = 1$, where the complex numbers  $a_1$ and $a_2$ can be considered as probability amplitudes.
% ${\left| a_1 \right|^2}$ (${\left| a_2 \right|^2}$) defined as the probability of finding the quantum state $\left| \psi  \right\rangle$ in $\left| 0 \right\rangle$ state ($\left| 1 \right\rangle$ state),
The notation $\left| . \right\rangle $ is used to indicate that the object is a vector\footnote{Note we have utilised the standard quantum mechanical notation for a vector in a vector space, i.e. $\left| \psi  \right\rangle $, where $\psi$ is a label for the vector (any label is valid). The entire object $\left| \psi  \right\rangle $ is sometimes called a `ket'. Note also that $\left\langle \psi \right|$ is called a `bra' which is the Hermitian conjugate or adjoint of the ket $\left| \psi  \right\rangle $. In quantum mechanics, bra-ket notation is a standard notation for describing quantum states.}.

As an alternative approach, CV encoding has also been introduced  \cite{1st-CVQKD,2nd-CVQKD}, and it is this form of encoding that forms the focus of this work. Such encoding is more appropriate for quantum information carriers such as laser light. In  CV technology, information is usually encoded onto the quadrature variables of the optical field \cite{1st-CVQKD,2nd-CVQKD,1st-gaussQKD,RR2002,RR,1st-EB}, which constitute an infinite-dimensional Hilbert space. Detection of these variables is normally realized by high-efficiency homodyne (or heterodyne) detectors, which can be capable of operating at a faster transmission rate than single-photon detectors \cite{Elser,sem,Rf2016}. The field's quadrature components (representing the quantum state) can be considered as related to the amplitude and phase of the laser light. In quantum mechanics, the quadrature components can also be considered as corresponding to the position and momentum of a harmonic oscillator.

There are generally three quantum communication scenarios, namely, the use of optical fibers,  the use of terrestrial FSO channels, or the use of FSO channels to satellites. These scenarios are complementary and all may be expected to play a role in the emerging global quantum communication infrastructure. Fiber technology has the key advantage that once in place, an unperturbed channel from A to B exists. In fact, in fiber links the photon transfer is hardly affected by external conditions such as background light, the weather or other environmental obstructions. However, fiber suffers both from optical attenuation and polarization-preservation problems, which therefore limit its attainable distance to a few hundred kilometers \cite{LDPC, fiber-QKD-2006, exp-CVQKD2007-1, fiber-QKD-2009, exp-CVQKD2009-2, exp-PM, exp-CVQKD-2016-fiber100, Takesue, Stucki, fiber2014, fiber2015, MDI2016-404}. These distance limitations may be overcome by the development of suitable quantum repeaters \cite{repeater}. Losses in fiber are due to  inherent random scattering processes, which increase exponentially with the fiber length. Explicitly, the transmissivity determining the fraction of energy received at the output of a fiber link of length $L$ is given by $\tau=10^{-\alpha L/10}$, where the value of $\alpha$ is highly dependent on the wavelength. Losses are minimised at the wavelength of 1550 nm, where for silicon fiber $\alpha  \simeq 0.2$ dB/km.

Replacing the fiber channel with a FSO channel has the immediate advantage of lower losses \cite{Peng-2005, freespace-entanglement-2005, freespace-QKD-2006, f2}, largely because the atmosphere provides for low absorption. The atmosphere also provides  for almost unperturbed propagation of the polarization states. Additionally, FSO channels offer convenient flexibility in terms of infrastructure establishment, with links to moving objects  also feasible \cite{aircraft, moving-patforms,truck}. However, terrestrial FSO quantum communications remain ultimately distance-limited,  due to (amongst other issues) the curvature of the Earth,  potential ground-dwelling line-of-sight (LoS) blockages, as well as atmospheric attenuation and turbulence.

FSO quantum communication via satellites \cite{pp1,pp2,R2,r6,r7,r8,r9,r10,r11,r12,r13,r14,r15,r16,r17,r18,nano-satellite,micro-satellite,China1,China2,China3,proposal-2014, proposal-2008,Boone,IEEE1,IEEE2,IEEE3,IEEE4,satellite-survey,IEEE-s,Bacsardi2} has the additional advantage that communications can still take place, even when there is no direct free-space LoS from A to B. That is, assuming that  LoS paths from a satellite to  two ground stations exist, satellite-based FSO communication can still proceed. The range of satellite-based communication is also potentially much larger than that allowed for by direct terrestrial FSO connections, since the former circumvents the terrestrial horizon limit and there are lower photonic losses at high altitudes. In satellite-based FSO communications, only a small fraction of the propagation path (less than 10 km) is through the atmosphere - meaning most of the propagation path  experiences no absorption and no turbulence-induced losses. The utilisation of satellites also allows for fundamental studies on the impact of relativity on quantum communications \cite{R2}. %Geostationary satellites are too distant to implement a quantum communication link between Earth and Space, therefore fast-moving low-Earth-orbit (LEO) satellites (from 500 km to 2000 km above Earth surface) must be employed.
The key disadvantage of satellite-based quantum communications is, however, atmospheric turbulence-induced loss.

QKD constitutes the most studied quantum communication protocol, and has been deployed over both fiber and FSO channels. Indeed, the implementation of QKD over optical fibers has already been commercialised \cite{SECOQC-Vienna, Tokyo, quantum-access-network}. Terrestrial FSO quantum communications have been successfully deployed over very long distances \cite{Peng-2005, freespace-entanglement-2005, freespace-QKD-2006, f2}. In 2007 entanglement-based QKD and decoy-state QKD was realized over a 144 km FSO link between the Canary Islands of La Palma and Tenerife \cite{144km, decoy, freespace-QKD-2009}. In addition to QKD, long-distance terrestrial FSO experiments have also been carried out to implement both entanglement distribution \cite{freespace-QKD-2009, freespace-entanglement-2009} and quantum teleportation \cite{freespace-teleportation100-2012, freespace-teleportation143-2012}. The above long-distance FSO quantum communication experiments have been implemented at night. However, in a recent experiment (by choosing an appropriate wavelength, spectrum filtering and spatial filtering) FSO terrestrial QKD over 53 km has also been demonstrated during the day \cite{daylight-QKD}. Nonetheless, in both fiber and FSO QKD implementations, the increasing levels of channel attenuation and noise tend to limit the maximum distance of successful key distribution to a few hundred kilometers.

%All these free-space experiments have used a direct LoS between fixed locations (fixed transmitter and receivers) to implement quantum communication.

%QKD as the most developed protocol of quantum communications has matured to commercial applications and a number of QKD schemes have been implemented both over optical fibers e.g., \cite{LDPC, exp-CVQKD2007-1, exp-CVQKD2009-2, exp-PM, Takesue, Stucki, fiber2014, fiber2015} and terrestrial free-space channels e.g., \cite{144km, f2, decoy}, however, QKD transmission distances are still limited.

A promising way of extending the deployment range of QKD is through the use of satellites. Indeed, it is widely anticipated that the reliance on satellites will assist in the expansion of quantum communication to global scales \cite{pp1,pp2,R2,r6,r7,r8,r9,r10,r11,r12,r13,r14,r15,r16,r17,r18,nano-satellite,micro-satellite,China1,China2,China3,proposal-2014, proposal-2008,Boone,IEEE1,IEEE2,IEEE3,IEEE4,satellite-survey,IEEE-s,Bacsardi2}. %The first significant step towards satellite-based quantum communication was \cite{Peng-2005}, since it has demonstrated that entanglement can survive after propagating through a 13 km noisy terrestrial atmospheric channel (a distance beyond the effective thickness of the atmosphere).
Full-scale verifications of satellite-based QKD have been reported in \cite{aircraft} (by demonstration of QKD between an aeroplane and a ground station), in \cite{moving-patforms} (by demonstration of QKD using a moving platform on a turntable, and a floating platform on a hot-air balloon), and in \cite{truck} (by demonstration of QKD from a stationary transmitter to a moving receiver platform traveling at an angular speed equivalent to a 600 km altitude satellite). Furthermore, several satellite-based quantum communication projects have been reported in \cite{r11,r12,r13,r14,proposal-2014,proposal-2008}. In \cite{r15,r16,r17}, a satellite-to-ground single-photon downlink was simulated by reflecting weak laser (coherent) pulses (emitted by the ground-based station) off a low-Earth-orbit (LEO) satellite. In addition to experimental demonstrations, quantum communications with orbiting satellites have also been investigated by a growing number of feasibility studies \cite{pp1,pp2, R2,r6,r7,r8,r9,r10,r18,IEEE1,IEEE2,IEEE3,IEEE4}. Recently, the in-orbit operation of a photon-pair source aboard a nano-satellite has been reported, which demonstrates photon-pair generation and polarization correlation under space conditions \cite{nano-satellite}.

 Quantum communication via satellites has very recently been given an enormous boost with the launch of the world's first quantum satellite, Micius, by China \cite{China1}. Building on the previously mentioned experiments, this new LEO satellite creates entangled photon pairs, sending them down to Earth for subsequent processing  in a diverse range of communication scenarios. For example, using Micius, satellite-based distribution of entangled photon pairs in the downlink to two terrestrial locations separated by 1203 km  has  been demonstrated \cite{China2}.  Quantum teleportation of single-photon qubits from a ground station to Micius through an uplink channel has also been demonstrated \cite{China3}. Extensions of this technology to significantly smaller satellites has just been reported for a Japanese micro-satellite  and an optical ground station \cite{micro-satellite}.

%Thus, it is important to analyse the effectiveness of quantum communication protocols over free-space channels towards (and from) satellites.

%The feasibility of QKD between an aeroplane and a ground station was demonstrated for the first time in \cite{aircraft}, where the results are representative of typical communication links to satellites or to high-altitude platforms. In \cite{moving-platform} three independent experiments with a decoy-state QKD system was carried out. The system was operated on a moving platform (using a turntable), on a floating platform (using a hot-air balloon), and with a high-loss channel to demonstrate performances under conditions of rapid motion, attitude change, vibration, random movement of satellites, and a high-loss regime.

All of the previous FSO quantum communication systems referred to above have been focussed on DV technologies \cite{aircraft,moving-patforms,truck,144km,decoy,f2,freespace-entanglement-2005,Peng-2005,freespace-QKD-2006,freespace-QKD-2009, freespace-entanglement-2009,freespace-teleportation100-2012,freespace-teleportation143-2012,daylight-QKD,pp1,pp2,R2,r6,r7,r8,r9,r10,r11, r12,r13,r14,r15,r16,r17,r18, nano-satellite,micro-satellite,China1,China2,China3,proposal-2014,proposal-2008,Boone, IEEE1,IEEE2,IEEE3,IEEE4,satellite-survey,IEEE-s,Bacsardi2}. They are based on single-photon  technology and use single-photon detectors. Such detectors are impaired by  background light, and involve spatial, spectral and/or temporal filtering in order to reduce this noise \cite{daylight-QKD}. By contrast, in CV quantum communication, homodyne detection (in which the signal field is mixed with a strong coherent laser pulse, called the ``local oscillator'') is used for determining the field quadratures of light. Homodyne detectors offer better immunity to stray light \cite{Elser}, since the local oscillator is also capable of assisting in both spatial and spectral filtering. Also, such homodyne detectors are more efficient than single-photon detectors, since the PIN photodiodes used in them offer higher quantum efficiencies than the avalanche photodiodes of single-photon detectors. Hence,  CV-QKD can  generally be considered to be more robust against background noise than DV-QKD.

\begin{table*}
\centering
\caption{Comparison of this study with available surveys}
\begin{tabular}[t]{|m{1.5cm}| m{1cm}| m{1.3cm}| m{1cm}| m{1cm}| m{1cm}| m{1cm}| m{1cm}| m{1cm}| m{1cm}| m{1cm}| m{1cm}|}
\hline
Approach & Satellite-based quantum communication & Atmospheric fading quantum channels & CV quantum systems & Quantum communication protocols & \multicolumn{3}{c|}{QKD} & Gaussian CV quantum communication & Non-Gaussian CV quantum communication & CV quantum teleportation & CV entanglement swapping\\
\cline{6-8}
%\hline
&&&&& DV-QKD & CV-QKD & Security analysis &&&&\\
\hline
\hline
Braunstein and van Loock \cite{CV-review}&&&\checkmark&\checkmark&&\checkmark&&\checkmark&&\checkmark&\checkmark \\
\hline
Pirandola and Mancini \cite{Pirandola-survey1}, and Pirandola \emph{et al} \cite{Pirandola-survey2} &&&\checkmark&\checkmark&&&&\checkmark&&\checkmark&\checkmark \\
\hline
Adesso and Illuminati \cite{rr1}&&&\checkmark&&&&&\checkmark&\checkmark&& \\
\hline
Gisin and Thew \cite{Gisin-survey}&&&&\checkmark&\checkmark&&&&&& \\
\hline
Scarani \emph{et al} \cite{Scarani}&&&\checkmark&\checkmark&\checkmark&\checkmark&\checkmark&\checkmark&&& \\
\hline
Andersen \emph{et al} \cite{review-CV-2010}&&&\checkmark&\checkmark&&\checkmark&&\checkmark&&\checkmark& \\
\hline
Wang \emph{et al} \cite{rr2}&&&\checkmark&\checkmark&&\checkmark&\checkmark&\checkmark&&\checkmark&\checkmark \\
\hline
Weedbrook \emph{et al} \cite{Weedbrook2012}&&&\checkmark&\checkmark&&\checkmark&\checkmark&\checkmark&&\checkmark& \\
\hline
Lo \emph{et al} \cite{QKD-survey}, and Diamanti \emph{et al} \cite{QKD-survey2}&&&\checkmark&\checkmark&\checkmark&\checkmark&&&&& \\
\hline
Diamanti and Leverrier \cite{QKD-survey3}&&&\checkmark&\checkmark&&\checkmark&\checkmark&\checkmark&&& \\
\hline
Marshall and Weedbrook \cite{swapping-survey}&&&\checkmark&\checkmark&&&&\checkmark&&&\checkmark \\
\hline
Bedington \emph{et al} \cite{satellite-survey}&\checkmark&&&\checkmark&\checkmark&&&&&& \\
\hline
Li \emph{et al} \cite{China-CVQKD-survey}&&&\checkmark&\checkmark&&\checkmark&&\checkmark&&& \\
\hline
Shenoy-Hejamadi \emph{et al} \cite{India-CVQKD-survey}&&&\checkmark&\checkmark&\checkmark&\checkmark&&&&& \\
\hline
This survey&\checkmark&\checkmark&\checkmark&\checkmark&&\checkmark&\checkmark&\checkmark&\checkmark&&\checkmark \\
\hline
\end{tabular}
\end{table*}

In \cite{Elser, Heim-2009} the feasibility of a point-to-point CV-QKD (with coherent polarization states of light) has been demonstrated over a 100 m FSO link. In \cite{Semenov2009,Wander2012,2016} the nonclassical properties of CV quantum states propagating through the turbulent atmosphere have been analysed. Gaussian\footnote{Gaussian quantum states are CV states with field quadratures exhibiting a Gaussian probability distribution.} entanglement distribution through a single point-to-point atmospheric channel and its applicability to CV-QKD have been studied in \cite{Usenko}. The entanglement properties of quantum states in the turbulent atmosphere have also been studied in  \cite{Bohmann1,Bohmann2}. Satellite-based CV quantum communication in the context of Gaussian and non-Gaussian entanglement distribution, and its application to CV-QKD, have been investigated in detail in \cite{Neda1,Neda2,Neda3,Neda4,Neda5}. The results presented in \cite{Neda1,Neda2,Neda3,Neda4,Neda5} apply for both a single point-to-point atmospheric channel, and in combined satellite-based atmospheric channels where the satellite acts as a relay. Recently, a point-to-point CV quantum communication experiment relying on the coherent polarization states of light has been established over a 1.6 km FSO link in an urban environment \cite{Heim}. The distribution of polarization squeezed states\footnote{In quantum optics, there is an uncertainty relationship for the quadrature components of the light field, stating that the product of the uncertainties in both quadrature components is at least some quantity times Planck's constant. Hence, the uncertainty relationship dictates some lowest possible noise (i.e., uncertainty) amplitudes for the quadrature components of the light. In squeezed light, a further reduction in the noise amplitude of one quadrature component is carried out by squeezing the uncertainty region of that quadrature component, which is at the expense of an increased noise level  in the other quadrature component.} of light through an urban  1.6 km  FSO link  has also been demonstrated \cite{R4}. Recently, an experiment has been carried out   relying on homodyne detection at a ground station of optical signals transmitted from a geostationary satellite \cite{Geo-2017}. This experiment is important in that it clearly  demonstrates  the feasibility and potential of satellite-based CV-QKD implementations.

%\begin{figure*}[t]
%\begin{small}
%\begin{timeline}{2005}{2017}{2cm}{2cm}{15cm}{0.6\textheight}
%\entry{2005}{Peng et al \cite{Peng-2005} reported free-space distribution of entangled photon pairs over a noisy ground atmosphere of 13 km, a distance beyond the effective thickness of the atmosphere}
%\entry{2013}{Nauerth et al \cite{aircraft} have demonstrated the possibility of BB84 key exchange (at the single-photon level) between a fast-moving airborne and a ground station. Wang et al \cite{moving-patforms} have carried out three independent experiments with a decoy-state QKD system. The system is operated on a moving platform (using a turntable), on a floating platform (using a hot-air balloon), and with a high-loss channel to demonstrate performances under conditions of rapid motion, attitude change, vibration, random movement of satellites, and a high-loss regime.
%
%\entry{2007} a free-space demonstration of secure key
%distribution was performed between two ground stations10,
%over a distance of 144 km. This scenario is comparable to links
%between satellites in low Earth orbit and ground stations with
%respect to both attenuation and fluctuations.
%
%\entry{2017}{Alanis, Botsinis, Ng and Hanzo \cite{alanis2014ndqo} propose the \emph{Non-Dominated Quantum Optimization} (NDQO) algorithm, which is an extension to the DHA, for solving mutli-objective Pareto optimal problems.}
%
%
%\end{timeline}
%\end{small}
%\caption{Timeline of satellite-based quantum communication milestones.}\label{fig:timeline}
%\end{figure*}

The current work aims to survey and  characterise the capabilities of CV quantum technology in satellite-based quantum communications. Since CV entanglement has been widely known as a basic resource for CV-QKD \cite{exp-EB}, our survey is focussed on satellite-based CV quantum communication in the context of CV entanglement distribution and its application to CV-QKD. A brief comparison of this survey to the other published surveys on topics related to CV quantum communication is presented in Table I.

\begin{figure}
    \begin{center}
      {\includegraphics[width=3 in]{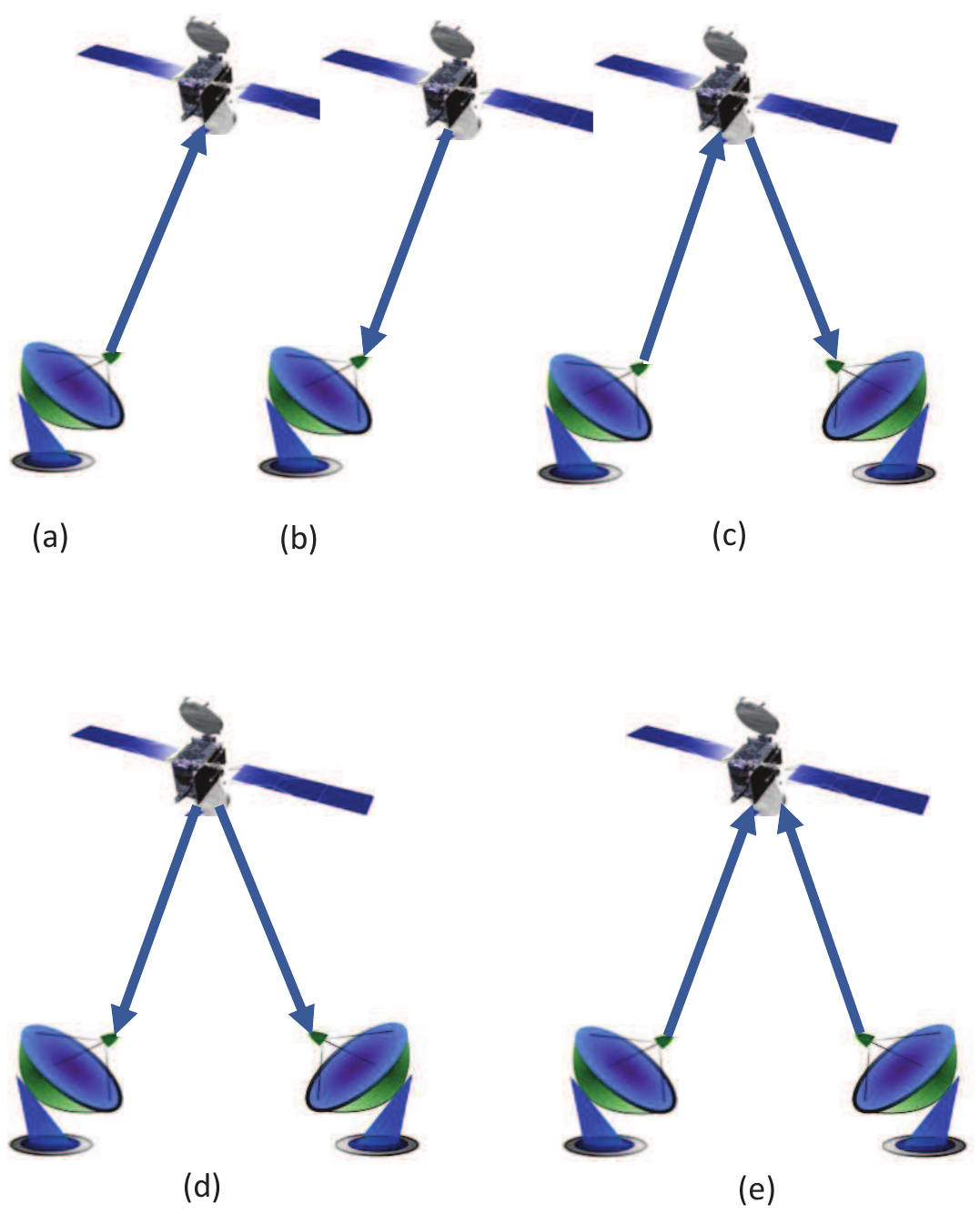}}
    \caption{Illustration of various architectures for implementing satellite-based quantum communication.}\label{fig:satellite-based-confs}
    \end{center}
\end{figure}

In the context of satellite-based quantum communication we are faced with two different channels, namely, the uplink (ground-to-satellite) channels and the downlink (satellite-to-ground) channels. In the uplink, the ground station transmits signals to the satellite receiver, and in the downlink, the satellite transmits signals to the ground station receiver. Correspondingly, there are several possible architectures for implementing satellite-based quantum communication depending on the types of links utilized. Some of these configurations are illustrated in Fig.~\ref{fig:satellite-based-confs}, and will be studied in this treatise in terms of entanglement distribution and CV-QKD implementation.

%The Performance of the architectures considered will be discussed later in terms of entanglement distribution and CV quantum key rates.
In Fig.~\ref{fig:satellite-based-confs}, the schemes (a) and (b) illustrate the uplink and downlink channels, respectively (both links have been demonstrated in the DV domain \cite{micro-satellite, China1, China3}). In scheme (c) of Fig.~\ref{fig:satellite-based-confs}, the deployment of quantum technology at the satellite is minimized, since the satellite is utilized only in a reflector mode (i.e. a simple relay). As a proof of concept for the reflecting paradigm, we note the recent experimental tests of \cite{r15,r16,r17}, where single photons (weak laser coherent pulses) emitted by the ground station were reflected (and subsequently detected on the ground) by a LEO satellite via the satellite's cube retro-reflectors. In  scheme (c) the complex quantum engineering components are  limited to the ground stations, since the source of quantum states is located in one of the ground stations and the receiver of quantum states is located in the other ground station. Although satellite reflection towards another station constitutes a sophisticated engineering task in its own right, it does not require onboard generation of quantum communication information. There are many practical advantages in deploying quantum communication technology at the ground stations, such as lower-cost maintenance, and the ability to rapidly upgrade as new quantum technology matures.

The other schemes, (d) and (e), in Fig.~\ref{fig:satellite-based-confs} can be considered as space-based high-complexity schemes, since they involve the deployment of quantum technology at the satellite. In scheme (d)  (again already demonstrated for DV states \cite{China2}) the source of quantum states is located on board  the satellite, with both ground stations acting as receivers. In  scheme (e)  the two ground stations transmit quantum states to the satellite. In the satellite, quantum measurements are performed on the received states and the classical measurement results are communicated back to the ground stations. Scheme (e) can be utilized in support of entanglement swapping and measurement-device-independent protocols so as to implement QKD between the two ground stations. %It has also been suggested that sources of entangled photons in space could be employed as quantum repeater stations, enabling entanglement swapping among more distant locations on the ground.

For the readers' convenience, the outline of this paper is
listed below.

\tableofcontents

\section{Free-space channels to and from satellites}

\subsection{Sources of loss in FSO channels}

The main sources of loss in FSO communication are diffraction, absorption, scattering and atmospheric turbulence \cite{Shaik, Scintillation2001, fso, Scintillation&wander}. %Note, in this work atmospheric channels and free-space channels are used interchangeably. In fact, free-space channels include terrestrial atmospheric channels and channels between ground stations and satellites.

\emph{Diffraction}: Diffraction is a ubiquitous form of natural wave propagation phenomenon  experienced by light beams, and leads to beam-spreading (beam-broadening). %of the wave as it propagates.

\emph{Absorption and scattering}: Absorption and scattering are imposed by the constituent gases and particles of the atmosphere. Both effects are strongly wavelength-dependent, and both impose attenuation on an optical wave. However, in this treatise we will assume that both scattering and absorption can be neglected, since they can be largely mitigated by an appropriate choice of the communication wavelength. Explicitly, there is a negligible absorption at the visible wavelengths spanning from 0.4 to 0.7~mm. For these reasons, scattering and absorption was also neglected in \cite{pp1, Rf2012, Wander2012, Usenko, Heim, Rf2015, pinting-error, Rf2016, 2016}.

\emph{Atmospheric turbulence}: Atmospheric turbulence arises due to random fluctuations in the refractive index caused by stochastic variations of temperature. The atmosphere contains turbulent random inhomogeneities of various scales - also referred to as turbulent eddies \cite{fso}. They range from a large-scale  (the outer scale of turbulence) to a small-scale (the inner scale of turbulence). These eddies affect optical wave-propagation through the atmosphere in different ways, depending on their size. In general, large scales produce refractive effects and hence predominately distort the phase of the propagating wave, while small scales are mostly diffractive in nature and therefore distort the amplitude of the wave \cite{Scintillation2001, fso}. The most important effects resulting from the atmospheric eddies are beam-wandering, beam-spreading and beam-scintillation \cite{Shaik, Scintillation2001, fso, Scintillation&wander}. We describe each of these three effects in more detail: (i) Random deviation of the beam from its original path is referred to as beam-wandering, which is caused by large-scale turbulent eddies, whose size is large compared to the beam-width. Beam-wandering causes time-varying power fades \cite{Shaik, fso, Scintillation&wander, pp1}. (ii) Atmospheric turbulence results in a randomly fluctuating beam-width in the receiver aperture plane. The broadening of the beam-width (when averaged over time) beyond that due to diffraction is termed as turbulence-induced beam-spreading \cite{Shaik, fso, pp1, 2016, eddies, r10}. (iii) We define scintillation by fluctuations in the received irradiance (intensity) within the beam cross section. Scintillation includes the temporal variation in the received irradiance and spatial variation within the receiver aperture. Scintillation is mainly caused by small-scale turbulent eddies \cite{Shaik, Scintillation2001, fso, Scintillation&wander}.

\subsection{Sources of loss in FSO channels to and from satellites}\label{Sources of loss to and from satellites}

In satellite-based quantum communications, the uplink and downlink channels are very different, since the atmospheric turbulence layer only occurs near the transmitter on an uplink, and only near the terrestrial receiver on a downlink. In the following, we briefly highlight how these two channels are affected by the above-mentioned turbulence-induced effects.

\begin{figure}[t!]
    \begin{center}
      {\includegraphics[width=3 in]{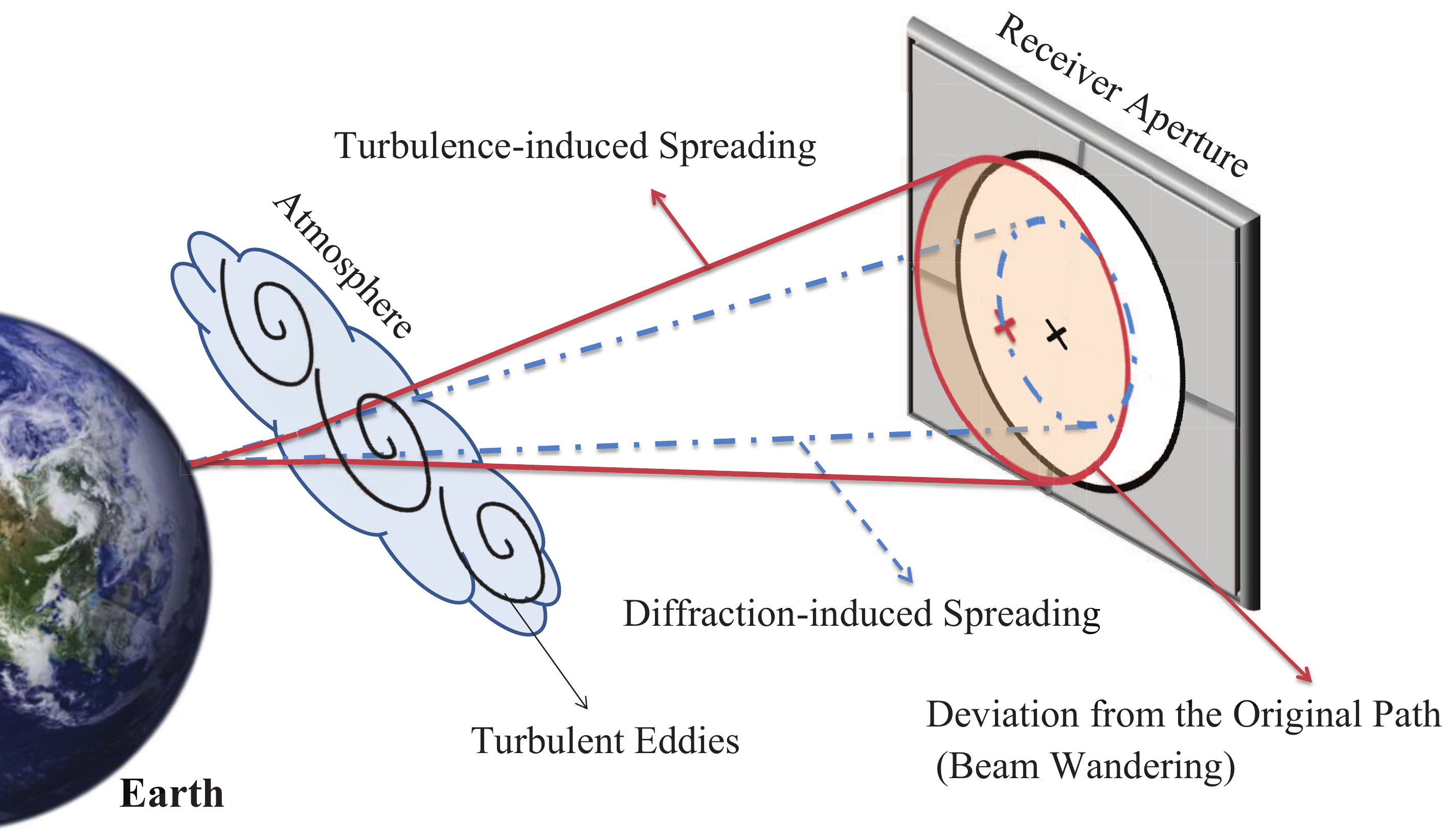}}
    \caption{Illustration of beam-wandering and beam-spreading in uplink channels.}\label{fig:eddies}
    \end{center}
\end{figure}

\emph{Uplink channels}: For typical dimensions of the aperture size embedded in the ground station, the uplink optical beam first propagates through the turbulent atmosphere and its beam-width is much narrower than the large-scale turbulent eddies \cite{Shaik, fso, Scintillation&wander, pp1}. This makes beam-wandering the dominant effect in the uplink \cite{Shaik, fso, Scintillation&wander, pp1}. Turbulence-induced beam-spreading also occurs to some extent in the uplink \cite{fso, pp1}. As a result, the beam received by the satellite (when averaged over time) is wider than that associated with diffraction \cite{fso, pp1}. Fig.~\ref{fig:eddies} illustrates these two atmospheric effects, namely beam-wandering and beam-spreading in the uplink. Scintillation is not dominant in the uplink \cite{Shaik, fso}.

\emph{Downlink channels}: In contrast to the uplink case, the downlink optical beam propagates through the turbulent atmosphere only in the final part of its path. Considering the typical aperture size of the optical system embedded in the satellite, the beam-width at its entry into the atmosphere is likely to be larger than the scale of the turbulent eddies. As such, beam-wandering in the downlink tends to be less important relative to uplink channels \cite{Shaik, fso, Scintillation&wander, pp1}. The photonic losses in the downlink are likely to be dominated by diffraction effects \cite{pp1, r10}. Scintillation can occur to some extent in the downlink \cite{Shaik, fso}. However, as a consequence of aperture averaging, the downlink scintillation effects imposed on the detector can be assumed negligible when the receiver includes a large-diameter ($>0.5$~m) telescope   \cite{Shaik, fso, Scintillation2001}.

\subsection{Atmospheric fading channels}
In atmospheric channels the transmissivity, $\eta_t$, fluctuates due to  turbulence-induced effects. These fading channels can be characterized by the probability distribution of the transmission coefficients, $\eta$ (where $\eta=\sqrt{\eta_t}$), which is denoted by $p(\eta)$. For a fading channel associated with the probability distribution $p(\eta)$ the mean fading loss in dB is given by $ - 10{\log _{10}}\left( {\int_0^{{\eta _0}} {{\eta ^2}p(\eta )d\eta } } \right) $, where $\eta _0$ is the maximum value of $\eta$.

As discussed in Sec.~\ref{Sources of loss to and from satellites}, beam-wandering is the dominant turbulence-induced effect in the uplink.  As an aside, we  note beam-wandering is expected to dominate the fading contributions in many terrestrial atmospheric communication scenarios \cite{Wander2012, Usenko, Heim, Rf2015, R4}. %Note the inclusion effect of beam-width fluctuations and pointing error in beam wandering will be discussed later in Chapter~\ref{C2.1:chapter2.1}.

\subsection{Beam-wandering model}
Here, we describe the probability distribution of the channel coefficients when the channel effects are dominated by beam-wandering. In the first instance we will assume that the beam-width at the receiver aperture is fixed. That is, initially we will ignore any fluctuations in the beam-width caused by atmospheric turbulence.

\begin{figure}[t!]
    \begin{center}
      {\includegraphics[width=3 in]{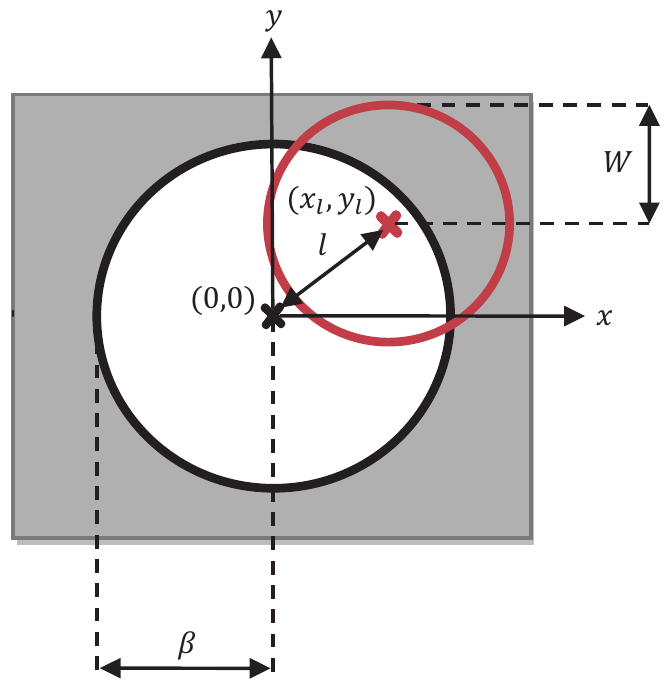}}
    \caption{A schematic illustration of beam-wandering in the receiver aperture plane.}\label{fig:beam-wandering}
    \end{center}
\end{figure}

\begin{figure}[t!]
    \begin{center}
      {\includegraphics[width=3 in]{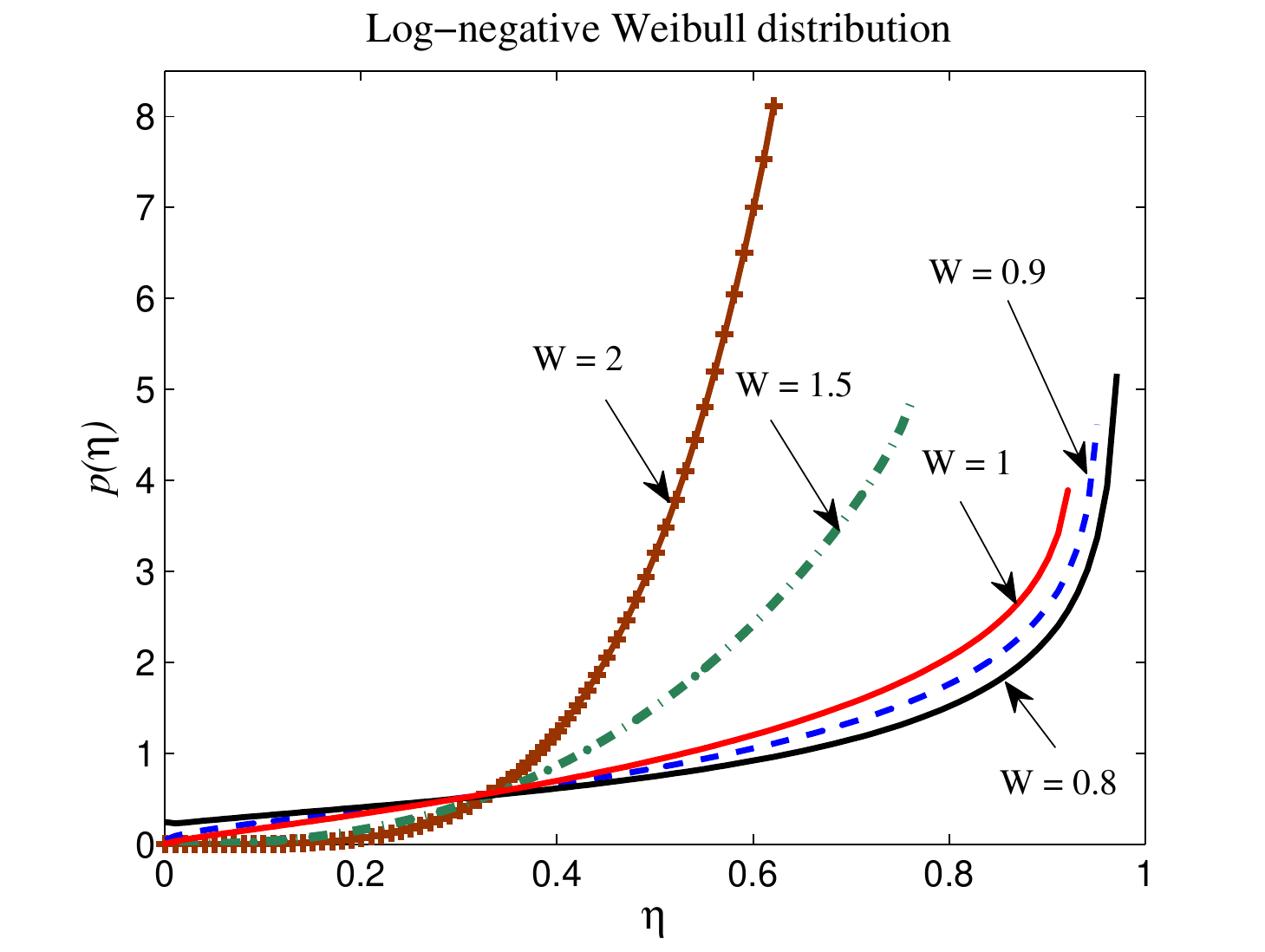}}
    \caption{The log-negative Weibull distribution for $\sigma_b=0.7$, $\beta=1$, and $d=0$ with different values of $W$. For these parameters, $W = 0.8$ leads to a mean fading loss of 2.7 dB and $W = 2$ leads to a mean fading loss of 5.5 dB.}\label{fig:fixed-sigma_b}
    \end{center}
\end{figure}

\begin{figure}[t!]
    \begin{center}
      {\includegraphics[width=3 in]{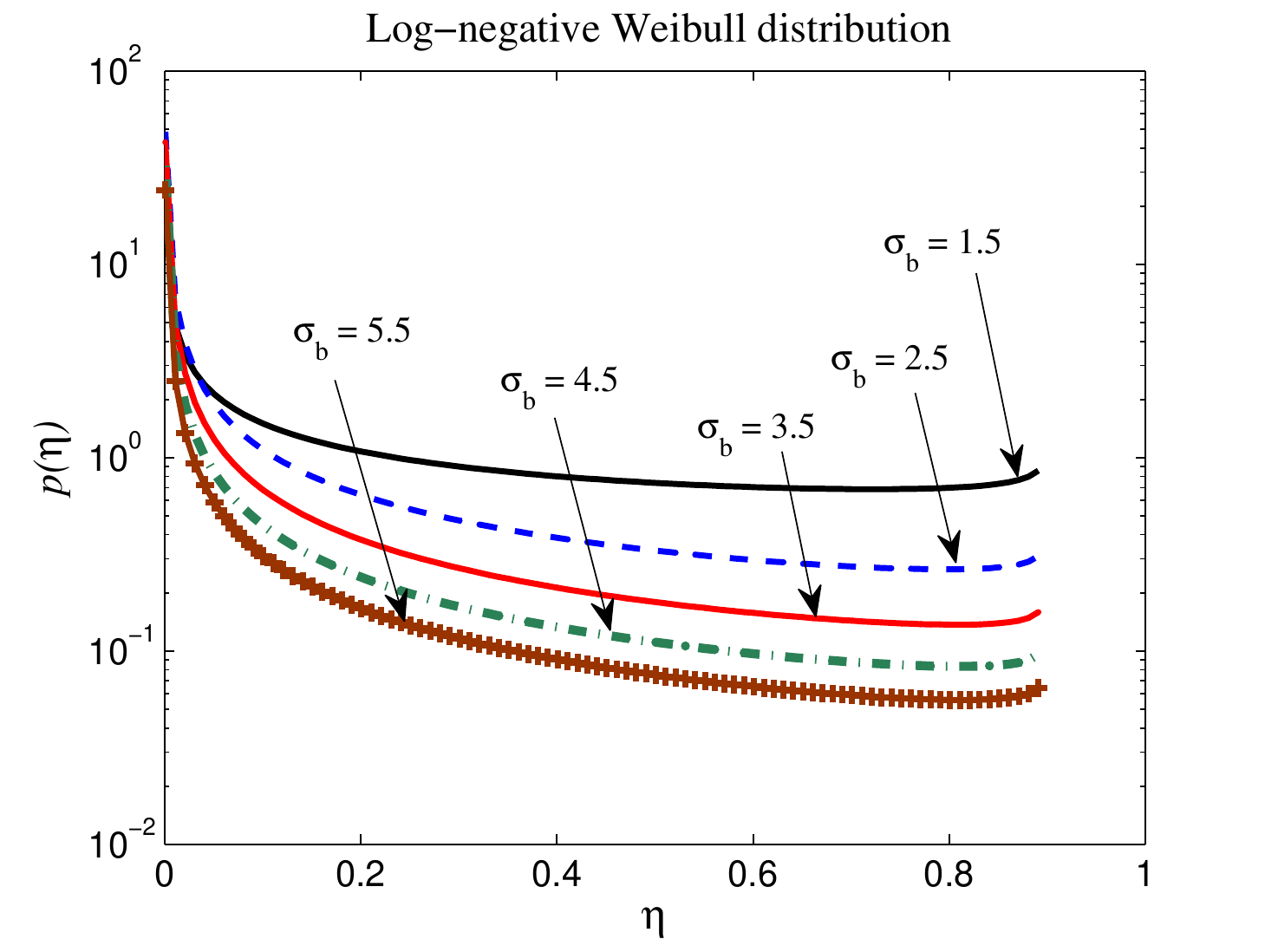}}
    \caption{The log-negative Weibull distribution for $W=1.1$ and $\beta=1$, and $d=0$ with different values of $\sigma_b$. For these parameters, $\sigma_b = 1.5$ leads to a mean fading loss of 7.4 dB and $\sigma_b = 5.5$ leads to a mean fading loss of 17.8 dB.}\label{fig:fixed-W}
    \end{center}
\end{figure}

In practice, beam-wandering causes the beam-center to be randomly displaced (along the $x$ and $y$ coordinates) from the center of the receiver aperture plane.
% the beam-center position fluctuates around a point at a distance of $d$ from the aperture centre \cite{beamwander}.
More explicitly,
%Assuming zero pointing error,
the beam-center position, $\left( {x_l,y_l} \right)$ randomly fluctuates around a fixed point, $({x_d},{y_d})$, in the receiver aperture plane according to a two-dimensional Gaussian distribution \cite{Wander2012}
\begin{equation}\label{2D-Gaussian_d}
p({x_l},{y_l}) = \frac{1}{{2\pi \sigma _b^2}}\exp \left( { - \frac{{{(x_l-x_d)}^2 + {(y_l-y_d)}^2}}{{2\sigma _b^2}}} \right),
\end{equation}
%the aperture center at $\left( {0,0} \right)$ according to a two-dimensional Gaussian distribution \cite{Wander2012},
%\begin{equation}\label{2D-Gaussian}
%p({x_l},{y_l}) = \frac{1}{{2\pi \sigma _b^2}}\exp \left( { - \frac{{{x_l}^2 + {y_l}^2}}{{2\sigma _b^2}}} \right),
%\end{equation}
where $\sigma _b$ is the beam-wandering standard deviation. Thus, the beam-deflection distance, $l = \sqrt {{x_l^2} + {y_l^2}} $, i.e. the distance between the beam-center and the aperture-center at $\left( {0,0} \right)$ fluctuates according to the Ricean distribution \cite{Wander2012}
\begin{equation}\label{Rice-d}
p(l) = \frac{l}{{\sigma _b^2}}{I_0}\left[ {\frac{{ld}}{{\sigma _b^2}}} \right]\exp \left( { - \frac{{{l^2} + {d^2}}}{{2\sigma _b^2}}} \right),
\end{equation}
where $d = \sqrt {x_d^2 + y_d^2} $ is the distance between the aperture-center and the fluctuation-center $({x_d},{y_d})$, and ${I_0}\left[ . \right]$ is the modified Bessel function. Note that $d=0$ means that the beam-center fluctuates around the aperture-center.
%aperture-center fluctuates according to the Ricean distribution \cite{Wander2012},
%\begin{equation}\label{Rice}
%p(l) = \frac{l}{{\sigma _b^2}}\exp \left( { - \frac{{{l^2}}}{{2\sigma _b^2}}} \right).
%\end{equation}
%In our calculations, $d$ is assumed to be zero, which means the beam-center position fluctuates around the aperture center.
In beam-wandering the channel transmission coefficient, $\eta$, is a function of the beam-deflection distance, $l$, and is given by \cite{Wander2012}
\begin{equation}\label{relation}
{\eta ^2} = \eta _0^2\exp \left( { - {{( {\frac{l}{S}})}^\gamma }} \right),
\end{equation}
where $\gamma$ is the shape parameter, $S$ is the scale parameter and ${\eta _0}$ is the maximum value of $\eta$. The latter three parameters are given by
\begin{equation}\label{f2}
\begin{array}{l}
\gamma  = 8h\frac{{\exp \left( { - 4h} \right){I_1}\left[ {4h} \right]}}{{1 - \exp \left( { - 4h} \right){I_0}\left[ {4h} \right]}}{\left[ {\ln \left( {\frac{{2\eta _0^2}}{{1 - \exp \left( { - 4h} \right){I_0}\left[ {4h} \right]}}} \right)} \right]^{ - 1}},\\
\\
S = \beta{\left[ {\ln \left( {\frac{{2\eta _0^2}}{{1 - \exp \left( { - 4h} \right){I_0}\left[ {4h} \right]}}} \right)} \right]^{ - \left( {{1 \mathord{\left/
 {\vphantom {1 \gamma_s }} \right.
 \kern-\nulldelimiterspace} \gamma }} \right)}},\\
\\
\eta _0^2 = 1 - \exp \left( { - 2h} \right) ,
\end{array}
\end{equation}
where ${I_1}\left[ . \right]$ is the modified Bessel function, and where $h = {\left( {{\beta \mathord{\left/
 {\vphantom {a W}} \right.
 \kern-\nulldelimiterspace} W}} \right)^2}$, with $\beta$ being the receiver aperture radius and $W$ the beam-spot radius at the receiver aperture. Note,  $\beta$  and $W$ have the same units (meter). A schematic illustration of beam-wandering is shown in Fig.~\ref{fig:beam-wandering}. According to Eqs.~(\ref{Rice-d}) and (\ref{relation}), the probability distribution $p\left( \eta  \right)$ can be described by the log-negative Weibull distribution \cite{Wander2012}
\begin{equation}\
\begin{array}{l}
p(\eta ) = \frac{{2{S^2}}}{{\sigma _b^2\gamma \eta }}{\left( {2\ln \frac{{{\eta _0}}}{\eta }} \right)^{(\frac{2}{\gamma } - 1)}}{I_0}\left[ {\frac{{Sd}}{{\sigma _b^2}}{{\left( {2\ln \frac{{{\eta _0}}}{\eta }} \right)}^{\frac{1}{\gamma }}}} \right]\\
\\
 \times \exp \left( {\frac{{ - 1}}{{2\sigma _b^2}}\left[ {{S^2}{{\left( {2\ln \frac{{{\eta _0}}}{\eta }} \right)}^{\frac{2}{\gamma }}} + {d^2}} \right]} \right)
\end{array}
\label{f1}
\end{equation}
%\begin{equation}\
%p\left( \eta  \right) = \frac{{2{S^2}}}{{\sigma _b^2\gamma \eta }}{\left( {2\ln \frac{{{\eta _0}}}{\eta }} \right)^{\left( {\frac{2}{\gamma }}- 1 \right) }}\exp \left( { - \frac{{{S^2}}}{{2\sigma _b^2}}{{\left( {2\ln \frac{{{\eta _0}}}{\eta }} \right)}^{\left( {\frac{2}{\gamma }} \right)}}} \right)
%\label{f1}
%\end{equation}
for $\eta  \in \left[ {0,\,{\eta _0}} \right]$, with $p\left( \eta  \right) = 0$, otherwise. In some of the earlier literature, e.g. \cite{log-normal}, the log-normal distribution was used. However, we now know the log-negative Weibull distribution more accurately describes the operationally-important distribution tail \cite{Wander2012}. In Fig.~\ref{fig:fixed-sigma_b} the log-negative Weibull distribution is shown for fixed values of the beam-wandering standard deviation $\sigma_b$ and the receiver aperture radius $\beta$,  and for different values of the beam-spot radius at the receiver aperture $W$ (the mean fading loss increases with increasing $W$). In Fig.~\ref{fig:fixed-W} the log-negative Weibull distribution is shown for the fixed values of $W$ and $\beta$, with different values of $\sigma_b$ (the mean fading loss increases with increasing $\sigma_b$).

Now we analyse the influence of beam-width fluctuations (caused by atmospheric turbulence) to the beam-wandering model just given. We refer to this effect as turbulence-induced beam-spreading. In doing this analysis, we will assume beam deformation does not occur - meaning the beam shape remains circular as it traverses the atmospheric channel (beam-deformation  has been analysed in \cite{2016}). In turbulence-induced beam-spreading, the beam-spot radius $W$ randomly changes in the receiver aperture plane \cite{2016} with the probability distribution $p(W)$. Including this effect in our beam wandering model, the transmission coefficient of the channel, $\eta$, is now a function of the two random variables $l$ and $W$ according to Eqs.~(\ref{relation}) and (\ref{f2}). We define a new variable $\Theta$ by setting $\Theta  = 2\ln \left( {\frac{W}{{{w_0}}}} \right)$, where $w_0$ is the initial beam-spot radius at the radiation source. This is useful since $\Theta$ randomly changes according to a normal distribution with the mean value $\left\langle \Theta  \right\rangle $ and standard deviation $\sigma _\Theta$ \cite{2016}. That is
\begin{equation}\label{1D-Gaussian}
p(\Theta ) = \frac{1}{{\sqrt {2\pi \sigma _\Theta ^2} }}\exp \left( { - \frac{{{{\left( {\Theta  - \left\langle \Theta  \right\rangle } \right)}^2}}}{{2\sigma _\Theta ^2}}} \right).
\end{equation}
With the inclusion of beam-width fluctuations in beam wandering, the calculation of a closed-form solution for $p(\eta)$ is not straightforward. However, knowing the probability distribution of $p(l)$ of Eq.~(\ref{Rice-d}) and $p(\Theta)$ of Eq.~(\ref{1D-Gaussian}), we can calculate certain important quantities after averaging over all values of the channel's transmission coefficient. For instance, the mean fading loss in dB of a fading channel with the inclusion of beam-width fluctuations  is now given by $ - 10{\log _{10}}\left( {\int {{\eta ^2}(l,\Theta )p(l,\Theta )dld\Theta } } \right)$. Assuming that atmospheric turbulence is isotropic \cite{2016} and $d=0$, the mean fading loss in dB of a fading channel (after the inclusion of beam-width fluctuations in the beam-wandering model) is given by $ - 10{\log _{10}}\left( {\int {{\eta ^2}(l,\Theta )p(l)p(\Theta )dld\Theta } } \right)$. Note, with the inclusion of beam-width fluctuations, the maximum value of the channel's transmission coefficient $\eta_0$ is no longer fixed but rather randomly changes.

%For a fading channel dominated by beam wandering the mean fading loss in dB is given by $\int_0^{{\eta _0}} {{\eta ^2}p(\eta )} d\eta $, where $p(\eta )$ is given by Eq.~(\ref{f1}). %Note, for fixed values of $W$ and $\beta$ the mean fading loss decreases with increasing ${\sigma _b}$.

%Note, pointing error can also be included in beam wandering, however, in this chapter we assume pointing error is zero. The inclusion effect of pointing error in beam wandering will be discussed later in Chapter~\ref{C2.1:chapter2.1}. Note also, in this chapter we will ignore any fluctuations in beam width caused by atmospheric turbulence. The impact of such fluctuations will also be discussed later in Chapter~\ref{C2.1:chapter2.1}.

%As discussed in Sec.~\ref{Sources of loss to and from satellites}, photonic losses in the downlink are likely be dominated by diffraction effects. Nonetheless, for uniformity of presentation in this contribution we will ``parameterise'' losses in the downlink through $\sigma_b$, but set its value for ensuring that the anticipated losses in the downlink are met. Ultimately, the evolved final quantum states will be determined solely by the magnitude of the photonic losses.

\begin{figure}[t!]
    \begin{center}
      {\includegraphics[width=3 in]{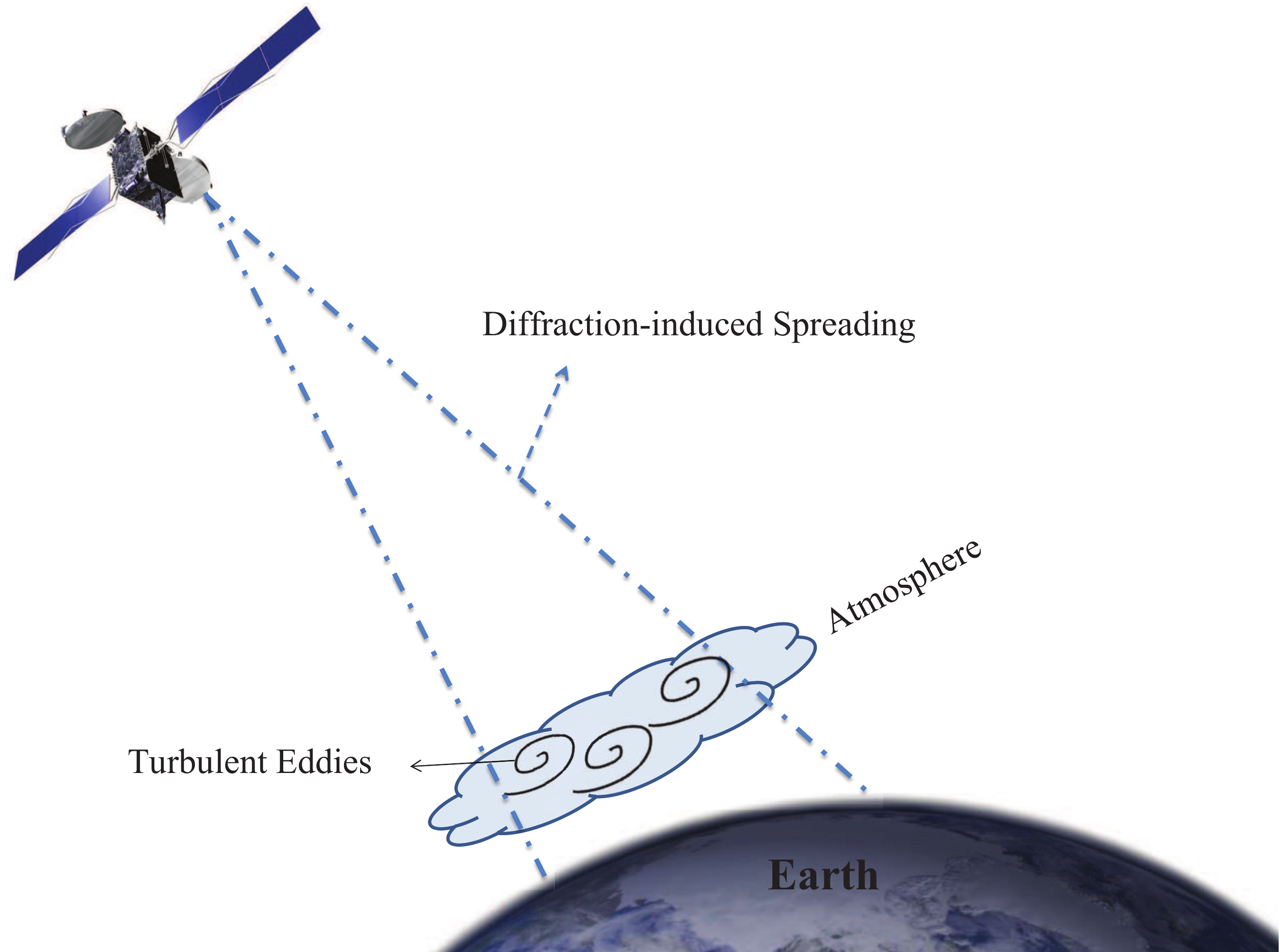}}
    \caption{Illustration of beam-spreading in downlink channels.}\label{fig:downlink}
    \end{center}
\end{figure}

Optical losses in the downlink are usually orders of magnitude lower relative to uplinks \cite{China1,China2,China3,satellite-survey}. This means that if the ``price'' is paid in terms of placing the critical quantum technology on board the satellite (rather than the easier case of maintaining the quantum technology in ground stations), then much better quantum communication channels can be obtained. As alluded to earlier, the principal reason for this improvement is that in the downlink, diffraction of the beam is the main contributor to photon losses - not beam-wandering as in the uplink (see Fig.~\ref{fig:downlink}). The important fact is that by the time the downward-link beam hits the main turbulence-inducing layers of the atmosphere (this layer commences at about 20~km from ground level) the beam is much closer to its target and therefore any induced beam-wandering is less effective.
%  Table X (do a similar table to table 1 from attached paper) indicates quantitatively the remarkable change in channel losses that can occur due to this simple fact for different transceiver diameters.
Clearly, as opposed to most communication channels, there will be no directional reciprocity in channel throughput for quantum communications with satellites. The recent experimental deployments of quantum communication in space have mostly exploited the more favourable downlink channel conditions \cite{China1,China2}. The losses in the downlink can then be modelled quite simply (to first order) through diffraction-only effects with the beam divergence following a $\lambda/D$ scaling, where $D$ is the diameter of the satellite telescope and $\lambda$ is the transmission wavelength \cite{satellite-survey}.

\subsection{Estimation of a FSO channel}

Note that the rate of atmospheric fluctuations we consider are the order of a few kHz, which is at least a thousand times slower than the typical transmission rates \cite{fso}. This means that the channel's transmission coefficient can be measured at the cost of additional (classical) transmission and receiver complexity \cite{Rf2012, Rf2015, sem, pp3}. These channel measurements may be carried out using several schemes, e.g., by transmitting coherent (classical) light pulses that are intertwined with the quantum information \cite{Rf2012, Rf2015} or by transmitting a local oscillator (i.e, a strong coherent laser pulse which is mixed with the signal field in the homodyne detection and serves as a phase reference) \cite{sem}. In  \cite{sem} measurement of the atmospheric channel's transmission coefficients was carried out in real time at the receiver by passing a local oscillator through the channel in a mode orthogonally polarized to the signal. The technique of measuring the atmospheric channel's transmission coefficient by an auxiliary classical laser beam was introduced in 2012 \cite{Rf2012}, and its practical employment was demonstrated for a one-way communication link in 2015 \cite{Rf2015}. The same technique based on the intensity of the signal itself was introduced and realized in \cite{pp3}.  %Let us consider the technique of real-time channel measurement in QKD. In QKD protocols Eve can have control of the entire channel, hence, she does have access to the coherent light pulse which is used to measure the channel transmission coefficient. However, this access would not help her extract information on Alice and Bob's encodings. Eve can also manipulate the coherent light pulse. This manipulation would provide her with no insight to the actual key. Such manipulation can only reduce the final key rate.

\section{Introduction to CV quantum systems}

%A quantum system is called a CV system when it has an infinite-dimensional Hilbert space described by observables with continuous eigenspectra(for review, see \cite{review-entanglement-2003, CV-review, rr1, rr2, review-CV-2010, Weedbrook2012, Adesso}).

One form of a  CV quantum system  is  that represented by $N$ bosonic modes, such as those corresponding to $N$ quantized radiation modes of the electromagnetic field \cite{review-entanglement-2003, CV-review, rr1, rr2, review-CV-2010, Weedbrook2012, Adesso}. A single photon has four degrees of freedom, helicity (polarization) and the three components of the momentum vector. In principle, quantum information can  be encoded into any one of these degrees of freedom. A single `mode' of an electromagnetic field refers to a specific combination of these photonic degrees of freedom. In many circumstances different modes can be simply  represented by  different frequencies (since frequency is related to momentum). For a beam of photons, the \emph{number} of photons in the beam constitutes another means to encode quantum information. Quantum information encoded into the quadratures of the electromagnetic field (formally defined below) are related to an encoding in this additional degree of freedom. Since the quadrature operators have continuous spectra, we can describe the values of such operators as CV variables.
%Here, a field mode refers to a single degree of freedom of the electromagnetic field, such as  polarization or frequency.

A single mode of a CV system can  be described  as a single quantum harmonic oscillator of a specific frequency, where the electric and magnetic fields play the `roles' of the position and momentum \cite{book-quantum-optics}. It will be useful to further illustrate this concept.
Consider the case of a single-frequency radiation field  confined to a one-dimensional cavity with walls that are perfectly conducting. Assume the $z$-axis  is parallel to the length of the cavity and the cavity walls are located at $z = 0$ and $z = L$. The electric field within the cavity will form  a standing wave. Without loss of generality, we can take the electric field to be polarized perpendicular to the $z$-axis, and in the positive $x$-direction (we take the $x$ and $z$ coordinates to be in same plane and the $y$ plane perpendicular to the $x$ plane). In terms of the distance vector {\emph{r} and time $t$ , the electric field can then be written as  $E(r, t) = e_x E_x (z, t)$, where $e_x$ is a unit-length polarization vector. Given our boundary conditions, and assuming a radiation source-free cavity, the electric field satisfying Maxwell's equations  can be written as \cite{book-quantum-optics}
\begin{eqnarray}\label{electric-field}
{E_x}(z,t) = \sqrt {(\frac{{2{\omega ^2}}}{{V_o{\varepsilon _0}}})} \,\,q(t)\sin (k z),
\end{eqnarray}
where $k=\omega/c$ is the wave number
($\omega$ is the frequency of the mode and $c$ is the speed of light in vacuum),  ${\varepsilon _0}$ is the  vacuum permittivity,  $q(t)$ is a time-dependent factor having the dimension of length (meters), and $V_o$ is the effective volume of the cavity.\footnote{To apply this formalism to the free field we calculate the physical observables we are interested in and then simply take the limit $V_0\rightarrow \infty$.} For the present purposes we will assume the frequency is one of those allowed by the boundary conditions, namely, $\omega_n = c(n \pi/L)$, where $n = 1,2,...$.

 Similarly, the magnetic field can be written $B(r, t) = e_y B_y (z, t)$, where $e_y$ is a unit-length polarization vector, and where\cite{book-quantum-optics}
\begin{eqnarray}\label{magnetic-field}
{B_y}(z,t) = \frac{{{\mu _0}{\varepsilon _0}}}{k}\sqrt {(\frac{{2{\omega ^2}}}{{V_o{\varepsilon _0}}})} \,\,p(t)\cos (k z).
\end{eqnarray}
Here $p(t) = \dot q(t)$, where the dot denotes the time derivative, and ${\mu _0}$ is the vacuum permeability. Based on these equations it is then straightforward to show that the Hamiltonian, $H_o$, of the  electromagnetic field can be written \cite{book-quantum-optics}
\begin{eqnarray}\label{Hamiltoniano1}
H_o = \frac{1}{2}\int {dV_o\left( {{\varepsilon _0}E_x^2(z,t) + \frac{1}{{{\mu _0}}}B_y^2(z,t)} \right)} .
\end{eqnarray}
Substituting $E_x(z,t)$ and $B_y(z,t)$ in $H_o$ from Eq.~(\ref{electric-field}) and Eq.~(\ref{magnetic-field}) respectively and exploiting that $\sin^2 (\frac{\omega }{c}z)+\cos^2 (\frac{\omega }{c}z)=1$ the Hamiltonian of the single-mode electromagnetic field can be written as
\begin{eqnarray}\label{Hamiltoniano2}
H_o = \frac{1}{{2}}( {{p^2} + {{( {\omega q} )}^2}} ).
\end{eqnarray} This equation can be compared with the Hamiltonian of the classical harmonic oscillator for a particle of mass $m$ viz., $H_o = \frac{1}{{2}}( {{p^2/m} + {{( {m \omega q} )}^2}} )$, where we have taken the generalised coordinate $q=x$ and set $p=m\dot x$, $x$ being the position.
Comparing these two Hamiltonians, it can be seen that a single-mode electromagnetic field is formally equivalent to a  harmonic oscillator of unity mass, where the electric and magnetic fields  play  roles similar to that of the position and momentum of a particle.\footnote{We emphasize that the terms `position' and `momentum' here  simply refer  to the similar roles played by the field quadratures and position and momentum of a particle - e.g. the  `position quadrature' does not in any manner refer to the position of a photon.}

In quantum systems we replace variables, such as $q$,  $p$ and $H$ of the classical system, by their corresponding operator\footnote{Note that operators can be regarded as matrices. In fact, the operator and matrix viewpoints turn out to be completely equivalent \cite{book-DV}.} equivalents, e.g.  $\hat q$, $\hat p$ and $\hat H$. Then the Hamiltonian of the single-mode electromagnetic field becomes $\hat H_o = \frac{1}{{2}}( {{\hat p^2} + {{( {\omega \hat q} )}^2}} )$. As such, we can now see how a single mode of a CV system can indeed  be described  as a single quantum harmonic oscillator. Furthermore, note that the operators $\hat q$ and $\hat p$ are Hermitian (or self-adjoint). In quantum physics Hermitian operators correspond to observable quantities, where an observable is an operator that corresponds to a physical quantity, such as position or momentum, that can be measured.

However, it will be useful to introduce non-Hermitian operators $\hat a$ (the annihilation operator) and $\hat a^\dagger$ (the creation operator). These can be written as,
\begin{eqnarray}\label{annih1}
 \hat a=(2\hbar\omega)^{(-1/2)}(\omega\hat q+i\hat p) , \\
 \hat a^\dagger=(2\hbar\omega)^{(-1/2)}(\omega\hat q-i\hat p) ,
\end{eqnarray}
where $\hbar  = h/2\pi $, with $h$ being  Planck's constant.
These  bosonic field operators satisfy the commutation relation
$[\hat a,\hat a^\dagger ] = 1$,
where  the commutator between two operators $\hat x$ and $\hat y$ is defined to be $[ {\hat x,\hat y} ] = \hat x \hat y - \hat y \hat x$.
 Note that since the annihilation and creation operators are non-Hermitian, they correspond to \emph{non-observable} quantities.

It can be easily shown that our new non-Hermitian operators have a time dependence, under free evolution, which can be expressed as $\hat a=\hat a(0)\exp(-i\omega t)$ and $\hat a^\dagger=\hat a^\dagger(0)\exp(i\omega t)$.
As such, the electric field operator can then be re-written as
\begin{eqnarray}\label{electric-field2a}
{E_x}(z,t) = \sqrt{(\frac{\hbar \omega}{ V_0 \varepsilon_0})}\sin(kz) [\hat a \exp(-i\omega t)+ \hat a^\dagger\exp(i\omega t)  ].
\end{eqnarray}
Removing the time dependence in the creation and annihilation operators by re-setting  $\hat a=\hat a(0)$ and $\hat a^\dagger=\hat a^\dagger(0)$,  we can in turn define the \emph{quadrature operators} (see later discussion on the freedom to choose the specific form of these)
\begin{eqnarray}\label{quad1}
 \hat X_1 =\frac{1}{2} (\hat a+\hat a^\dagger), \\
  \hat X_2 =\frac{1}{2i} (\hat a-\hat a^\dagger).
\end{eqnarray}
In terms of the quadrature operators we can then re-write ${E_x}(z,t)$ as
\begin{eqnarray}\label{electric-field2}
{E_x}(z,t) = 2 \sqrt{(\frac{\hbar \omega}{ V_0 \varepsilon_0})}\sin(kz) [ \hat X_1\cos(\omega t) + \hat X_2\sin(\omega t) ].
\end{eqnarray}
As such, we can see that the quadratures $\hat X_1$ and $\hat X_2$ can be considered as the   amplitudes of the electric field's time-dependent cos and sin components, respectively. Clearly, these components are $90^o$ out of phase with each other - hence the name, quadratures. The quadratures satisfy the commutation relation $[\hat X_1 , \hat X_2]=i/2$.\footnote{This can be derived from the constraint imposed by quantum mechanics that  $\left [\hat q, \hat p \right ]=i\hbar$.
Note,  that in contrast to classical physics where any two observables commute i.e., their commutator is zero (which means it is possible to know precisely the value of both
observables at the same time), in quantum mechanics the quadrature observables of the electromagnetic field do not commute.}

A CV system of $N$ modes follows a similar description to that we have just given for a single mode, except of course the Hilbert space containing the multimode system is larger. The $N$-mode system may be described by a Hilbert space given by the tensor product $\mathcal{H} =  \otimes _{k = 1}^N{\mathcal{H}_k}$, where ${\mathcal{H}_k}$ is a single-mode Hilbert space associated with the $k$-th mode. The creation and annihilation operators for each mode then satisfy the commutation relationships
\begin{eqnarray}\label{commutation-relation1}
[ {{{\hat a}_k},{{\hat a}_{k'}}} ] = [ {\hat a_k^\dag ,\hat a_{k'}^\dag } ] = 0,\,\,\,\,\,\,\,\,[ {{{\hat a}_k},\hat a_{k'}^\dag } ] = {\delta _{kk'}},
\end{eqnarray}
where ${\delta _{kk'}}$ is the Kronecker delta function.

Consider again the single-mode Hilbert space ${\mathcal{H}_k}$. This is spanned by the Fock, or number-state basis, $\left\{ {{{\left| n \right\rangle }_k}} \right\}_{n = 0}^\infty $, where the Fock state ${{{\left| n \right\rangle }_k}}$ is the eigenstate of the number operator ${{\hat n}_k} = \hat a_k^\dag {{\hat a}_k}$, i.e., ${{\hat n}_k}{\left| n \right\rangle _k} = n{\left| n \right\rangle _k}$. Put simply, ${\left| n \right\rangle _k}$ represents the state of the electromagnetic field containing exactly $n$ photons (quanta) of frequency $\omega_k$. Note that for each mode $k$ there exists a \emph{vacuum} state which contains \emph{no} quanta of the field, namely, ${\left| 0 \right\rangle _k}$, satisfying ${{\hat a}_k}{\left| 0 \right\rangle _k} = 0$.
%Note, a vacuum state ${\left| 0 \right\rangle _k}$ has a zero mean photon number, i.e., $\left\langle {{{\hat n}_k}} \right\rangle  = 0$, where $\left\langle . \right\rangle $ denotes the mean value.
The action of the bosonic field operators over the Fock states is given by \cite{CV-review,Weedbrook2012}
\begin{eqnarray}\label{action-on-Fock}
{{\hat a}_k}{\left| n \right\rangle _k} = \sqrt n {\left| {n - 1} \right\rangle _k},\,\,\,\,\hat a_k^\dag {\left| n \right\rangle _k} = \sqrt {n + 1} {\left| {n + 1} \right\rangle _k}.
\end{eqnarray}

Having now formally defined  the vacuum state, it is probably useful to note for the unwary that some \emph{apparent} inconsistency lies lurking in the literature (including the many references of this work). This applies to both the  constant value applied to $\hbar$, as well as the nomenclature itself.
 We note that our quadrature operators, as defined thus far, can be used to form $\hat q= \sqrt{2 \hbar/\omega}\hat X_1$ and $\hat p= \sqrt{2 \hbar\omega}\hat X_2$; from which we can easily show consistency with $[\hat q, \hat p]=i \hbar$.  In many works we will find that $\hat q$ and $\hat p$ written in this form (and also in `dimensionless' form with, say, $\hbar=\omega=1$)  are also referred to as the `quadratures.' Also, in many works the cofactor of $1/2$ in front of our definitions of $\hat X_1$ and $\hat X_2$ is replaced by some other constant, e.g., $1/\sqrt{2}$ or $1$ - allowable re-definitions of course.
 It is  straightforward to determine  the vacuum expectation value for any well-defined defined operator (or function of that operator), e.g. $\left\langle 0 \right | \hat X_1^2 \left|0 \right\rangle=1/4$, and
 $\left\langle 0 \right | \hat q^2 \left|0 \right\rangle=\hbar/(2\omega)$.
 It is common to set $\hbar$ to some numerical constant, usually $1/2, 1$ or $2$. However, no consistency exists in the literature on this either. Setting $\hbar=2$ has the convenience of setting the  vacuum-state variance of the  $\hat q$ and $\hat p$ operators to 1 (when $\omega$ set to unity).\footnote{Note the variance of $\hat q$ in the vacuum state is just $\left\langle 0 \right | \hat q^2 \left|0 \right\rangle$ since the vacuum expectation  of $\hat q$  is zero (variance $=\left\langle 0 \right | \hat q^2 \left|0 \right\rangle-
 \left\langle 0 \right | \hat q \left|0 \right\rangle^2$). Similarly $\hat p$.}

Bearing in mind the above discussion of inconsistency in nomenclature,
 we adopt henceforth that $\hbar=2$ and $\omega=1$ (unless  stipulated otherwise). We also redefine the `quadrature' operators to be
 $\hat q_k$ and $\hat p_k$, now given by the simpler form $\hat q_k = \hat a_k + \,{\hat a_k^\dag }$ and $\hat p_k = i({\hat a_k^\dag } - \hat a_k\,)$.  This will make the  notation to follow  less cluttered.

 % which satisfy the commutation relation $\left[ {\hat q,\,\hat p} \right] = 2i$ (here $\hbar=2$).
Defining the vector of quadrature operators for $N$ modes as ${\hat R} = ( {{{\hat q}_1},\,{{\hat p}_1}, \ldots ,{{\hat q}_N},{{\hat p}_N}\,} )$, the commutation relationship between the quadrature operators can be written as $[ {{{\hat R}_i},{{\hat R}_j}} ] = 2i{\Omega _{ij}}$, where $\hat R_i$ ($\hat R_j$) is the $i$-th ($j$-th) element of the vector $\hat R$, and ${\Omega _{ij}}$ is the element of the matrix
\begin{eqnarray}\label{ad2}
\bm{\Omega}  = \mathop  \oplus \limits_{k = 1}^N \,\,\bm{\Omega_0}  \,,\,\,\bm{\Omega_0} = \left( {\begin{array}{*{20}{c}}
0&1\\
{ - 1}&0
\end{array}} \right).
\end{eqnarray}

Since a Hermitian operator has an orthogonal set of eigenvectors with real-valued eigenvalues, the quadrature operator $\hat q$ ($\hat p$) (which is Hermitian) is an observable with continuous eigenspectra, i.e., $\hat q\left| q \right\rangle  = q\left| q \right\rangle $ ($\hat p\left| p \right\rangle  = p\left| p \right\rangle $), with orthogonal eigenvectors or eigenstates $\left| q \right\rangle $ ($\left| p \right\rangle $) having continuous eigenvalues $q \in \mathbb{R}$ ($p \in \mathbb{R}$). Note that the two sets of eigenstates ${\left| q \right\rangle }$ and ${\left| p \right\rangle }$ identify two different bases (i.e., two different sets of orthogonal and complete eigenstates), and each set constitutes a common basis for CV quantum information. A CV quantum state can be defined as a continuous-valued superposition of the field's eigenstates.

%The single-mode Hilbert space ${\mathcal{H}_k}$ is spanned by the Fock or number-state basis $\left\{ {{{\left| n \right\rangle }_k}} \right\}_{n = 0}^\infty $, where the Fock state ${{{\left| n \right\rangle }_k}}$ is the eigenstate of the number operator ${{\hat n}_k} = \hat a_k^\dag {{\hat a}_k}$, i.e., ${{\hat n}_k}{\left| n \right\rangle _k} = n{\left| n \right\rangle _k}$. Note that for each mode $k$ there exists a vacuum state ${\left| 0 \right\rangle _k}$ satisfying ${{\hat a}_k}{\left| 0 \right\rangle _k} = 0$.
%%Note, a vacuum state ${\left| 0 \right\rangle _k}$ has a zero mean photon number, i.e., $\left\langle {{{\hat n}_k}} \right\rangle  = 0$, where $\left\langle . \right\rangle $ denotes the mean value.
%The action of the bosonic field operators over the Fock states is given by \cite{CV-review,Weedbrook2012}
%\begin{eqnarray}\label{action-on-Fock}
%{{\hat a}_k}{\left| n \right\rangle _k} = \sqrt n {\left| {n - 1} \right\rangle _k},\,\,\,\,\hat a_k^\dag {\left| n \right\rangle _k} = \sqrt {n + 1} {\left| {n + 1} \right\rangle _k}.
%\end{eqnarray}

All the physical information about a CV system is contained in its quantum state, represented by a density operator $\hat \rho$, which is a trace-one positive operator. A quantum state $\hat \rho$ is said to be a pure state, when we have $\hat \rho^2=\hat \rho$. A pure state can be described as $\hat \rho  = \left| \psi  \right\rangle \left\langle \psi  \right|$, where $\left| \psi  \right\rangle$ is the  vector representation of the pure quantum state. A mixed quantum state is defined as a statistical ensemble of pure states, which cannot be described by a single vector. Instead, it is described by its associated density operator. The density operator describing a mixed state is in the form of $\hat \rho  = \sum\nolimits_i {{p_i}\left| {{\psi _i}} \right\rangle \left\langle {{\psi _i}} \right|} $, where $p_i$ is the specific fraction of the ensemble found in each pure state ${\left| {{\psi _i}} \right\rangle }$.

A quantum state $\hat \rho$ of a $N$-mode CV system can also be described in terms of a characteristic function ${\chi _c}\left( \xi  \right) = \rm{Tr}( {\hat \rho \hat D( \xi  )} )$, where $\rm{Tr}$ denotes trace, $\hat D( \xi  ) = \exp ( {i\hat R\bm{\Omega} \xi } )$ is the Weyl operator \cite{CV-review,Weedbrook2012}, and $\xi  \in {\mathbb{R}^{2N}}$. The quantum state $\hat \rho$ can also be described in terms of a Wigner function (quasi-probability distribution), which is given by the Fourier transform of the characteristic function ${\chi _c}$ as \cite{CV-review,Weedbrook2012}
%Then, the quantum state $\hat \rho$ is the Fourier transformed equivalent of a Wigner function (quasi-probability distribution) of \cite{CV-review,Weedbrook2012}
\begin{eqnarray}\label{general-wigner-function}
W\left( R \right) = \int_{{\mathbb{R}^{2N}}} {\frac{{{d^{2N}}\xi }}{{{{\left( {2\pi } \right)}^{2N}}}}} \exp \left( { - iR\bm{\Omega} \xi } \right){\chi _c}\left( \xi  \right),
\end{eqnarray}
where $R = \left( {{q_1},\,{p_1}, \ldots ,{q_N},{p_N}\,} \right)$ is the vector of quadrature variables, with the real-valued variables $q$ and $p$ being the eigenvalues of the quadrature operators. Note that for a single-mode quantum state the probability distribution of a quadrature measurement (marginal distribution) is obtained from the Wigner function of the quantum state by integration over the conjugate quadrature.

The CV quantum states can be visualized using their Wigner function in a phase-space representation, where the axes are defined by a pair of conjugate quadrature variables $q$ and $p$. In such a phase space, a classical optical field is represented by a single point corresponding to its complex-valued field amplitude. However, the quantum states of light cannot be represented by a single point, since conjugate quadrature variables cannot be measured simultaneously with arbitrary precision due to the Heisenberg uncertainty relationship.\footnote{According to quantum optics, any measurement of the complex amplitude of the light field can deliver different values within an uncertainty region. Furthermore, due to the Heisenberg uncertainty relationship for the quadrature components of the light field, the uncertainties in both quadrature components is at least some quantity times the Planck's constant. This fact is represented by the commutator relations for the quadratures that we have discussed earlier.} Hence the Wigner function is utilized to represent the quantum states in the phase space \cite{CV-review, rr2, review-CV-2010, Weedbrook2012}.

In a $N$-mode CV system the Heisenberg uncertainty relationship is defined for the quadrature operators of each mode $k$, and is given by $V\left( {{{\hat q}_k}} \right)V\left( {{{\hat p}_k}} \right) \ge 1$, where $V$ is the variance of the quadrature operator, and is given by $V( {{{\hat R}_i}} ) = \langle {\hat R_i^2} \rangle  - {\langle {{{\hat R}_i}} \rangle ^2}$, where $\langle . \rangle $ denotes the mean value. Note again, that the quadrature variance of the vacuum state of a single mode is one, i.e., we have $V\left( {\hat q} \right) = V\left( {\hat p} \right) = 1$, which is the lowest possible variance reachable symmetrically by the $\hat q$ and $\hat p$ quadratures according to the uncertainty relationship.%, and it is also known as vacuum noise.

\subsection{Gaussian quantum states}
Gaussian quantum states (for a detailed review, see \cite{rr2, Weedbrook2012, Adesso}) are completely characterized by the first moment (or the mean value) of the quadrature operators $\langle {{{\hat R}}} \rangle $ and a covariance matrix $\bm{M}$, i.e. a matrix of the second moments of the quadrature operators defined as
\begin{eqnarray}\label{ad1}
{M_{ij}} = \frac{1}{2}\langle {{{\hat R}_i}{{\hat R}_j} + {{\hat R}_j}{{\hat R}_i}} \rangle  - \langle {{{\hat R}_i}} \rangle \langle {{{\hat R}_j}} \rangle .
\end{eqnarray}
The CM of a $N$-mode quantum state is a $(2N \times 2N)$ real symmetric matrix, which must satisfy the uncertainty principle, \emph{viz.},
$\bm{M}  + i\,\bm{\Omega} \, \ge \,0$.
By definition, a Gaussian state having $N$ modes is a CV state whose Wigner function  is a Gaussian distribution of the quadrature variables. That is,
\begin{eqnarray}\label{Gaussian-Wigner-fun}
W\left( {{R}} \right) = \frac{{\exp \left( { - \frac{1}{2}\left( {{R} - \left\langle {{R}} \right\rangle } \right)\,{\bm{M}^{ - 1}}\,{{\left( {{R} - \left\langle {{R}} \right\rangle } \right)}^T}} \right)}}{{{{\left( {2\pi } \right)}^N}\sqrt {\det \left( \bm{M} \right)} }}.
\end{eqnarray}
Some important examples of Gaussian states are vacuum states \cite{CV-review, rr2,
Weedbrook2012,book-quantum-optics}, coherent states \cite{CV-review, rr2,Weedbrook2012,book-quantum-optics}, thermal states \cite{CV-review, rr2, Weedbrook2012,book-quantum-optics} and squeezed states \cite{CV-review, rr2, Weedbrook2012,book-quantum-optics}. We discuss some of these Gaussian states further.\hfill \break
%\emph{Vacuum state}: The vacuum state $\left| 0 \right\rangle $ of a single mode is the eigenstate of the annihilation operator having a zero eigenvalue, which is formulated as $\hat a\left| 0 \right\rangle  = 0$. The vacuum state has zero photons on average, i.e. we have $\bar n= 0$, where $\bar n=\left\langle {\hat n} \right\rangle$ is the average number of photons.
\emph{Vacuum state}: The Wigner function of the vacuum state with respect to the conjugate quadrature variables $q$ and $p$ is shown in Fig.~\ref{fig:Wignr-functions}(a), in which the Wigner function is centered at $(0,0)$, which means that the vacuum state has a zero mean. The covariance matrix of the vacuum state is the identity matrix, which means that a vacuum state has a symmetric distribution of the quadrature components (see Fig.~\ref{fig:Wignr-functions}(a)) with both the quadrature components having noise variance of one. This noise is usually termed the vacuum noise or quantum shot noise.  \hfill \break
\emph{Coherent state}: A coherent state is generated by applying the displacement operator $\hat D$ to the vacuum state formulated as $\left| \alpha  \right\rangle  = \hat D(\alpha )\left| 0 \right\rangle $, where $\hat D(\alpha ) = \exp (\alpha {{\hat a}^\dag } - {\alpha ^ * }\hat a)$ is the displacement operator and $\alpha=(q+ip)/2$ is the complex amplitude. Since the displacement operator does not change the variance of the quadratures, coherent states - similarly to vacuum states - exhibit the lowest possible variance reachable symmetrically by the $\hat q$ and $\hat p$ quadratures. The coherent state is the eigenstate of the annihilation operator, which is formulated as $\hat a\left| \alpha  \right\rangle  = \alpha \left| \alpha  \right\rangle $. To elaborate a little further, this state has a mean value of $\langle {\hat R} \rangle  = (q,p)$, and the covariance matrix is equal to the identity matrix, which means that a coherent state has a symmetric distribution of the quadrature components with both the quadrature components having noise variance equal to one. This symmetric distribution can be seen in Fig.~\ref{fig:Wignr-functions}(b), where the Wigner function of the coherent state with a mean value of $(3,5)$ (which is the centre of the Wigner function) is shown with respect to the conjugate quadrature variables $q$ and $p$. Note that coherent states are much easier to generate in the laboratory than any other Gaussian state. For example, the laser field is in a coherent state. As an important application in the context of quantum communication, coherent states are used to distribute secret keys in Gaussian CV-QKD protocols \cite{QKD-coh1, RR2002, RR, QKD-coh2}. \hfill \break
\emph{Thermal state}: Thermal states can be described as a mixture of coherent states. %A thermal state $\hat \rho_{th}$ having an average number of photons $\bar n>0$ in the Fock basis is given by
%\begin{equation}\label{M-i}
%{\hat \rho _{th}} = \sum\limits_{n = 0}^\infty  {\frac{{{{\bar n}^n}}}{{{{\left( {\bar n + 1} \right)}^{n + 1}}}}} \left| n \right\rangle \left\langle n \right|.
%\end{equation}
The thermal state has a zero mean and a covariance matrix $\bm{M_{th}} = v_t \bm{I}$ associated with $v_t=2\bar n + 1$, where $v_t$ is the noise variance of each quadrature component, $\bar n>0$ is the average number of photons and $\bm{I}$ is the $(2 \times 2)$-element identity matrix. This form of the covariance matrix means that a thermal state has a symmetric distribution of the quadrature components, which can be seen in Fig.~\ref{fig:Wignr-functions}(c) where the Wigner function of the thermal state with $v_t=5$ is shown with respect to the conjugate quadrature variables $q$ and $p$. Note that in the generic form of quantum communication the quantum noise of the channel is in a thermal state, called thermal noise.\hfill \break
\emph{Single-mode squeezed vacuum state}: According to the Heisenberg uncertainty relationship, the lowest possible variance reachable symmetrically by the $\hat q$ and $\hat p$ quadratures is one i.e., the noise variance of the vacuum state.
%which means there is a minimum noise amplitude of one for the amplitude and phase quadratures of light, which is also known as the vacuum noise.
A reduction in the variance of the $\hat q$ (or $\hat p$) quadrature below the vacuum noise is possible by \emph{squeezing}. In  squeezing, the variance of one continuous variable is in fact decreased below the vacuum noise, while the variance of the conjugate variable  is increased. For instance, in a $\hat q$-squeezed light, the variance of the $\hat q$ quadrature is reduced below the vacuum noise, while the variance of the $\hat p$ quadrature is increased above the vacuum noise. A single-mode squeezed vacuum state is generated by applying the single-mode squeezing operator of ${\hat S_s}({r_s}) = \exp [ {{r_s}({{\hat a}^2} - {{\hat a}^{{\dag ^2}}})/2} ]$ \cite{CV-review, rr2, Weedbrook2012,book-quantum-optics} to the vacuum state, where $r_s \in \left[ {0,\infty } \right)$ represents the single-mode squeezing parameter.\footnote{Note, in general, squeezing parameters are complex numbers. For simplicity (and to be consistent with most of the literature) we  limit them here to real numbers.} Such a squeezed state has  zero mean and a covariance matrix of $\bm{M} = diag[\exp (-2{r_s}),\exp (2{r_s})]$ when the quantum fluctuations of the $\hat q$ quadrature have been squeezed. In this case for the single-mode squeezing represented by $r_s>0$ we have $V({{\hat q}}) < 1$ and $V({{\hat p} }) > 1$. This means that a single-mode squeezed state does not have a symmetric distribution of the quadrature components, since the variance of one of the quadratures is reduced by squeezing at the expense of an increase in the variance of the conjugate quadrature by the counterpart operation of anti-squeezing. Note, the state still obeys the Heisenberg uncertainty relationship. Such an asymmetric distribution of quadrature components can be seen in Fig.~\ref{fig:Wignr-functions}(d), where the Wigner function of the single-mode squeezed vacuum state with $r_s=0.5$ is shown. Here,  the $\hat q$ quadrature is  squeezed. In terms of applications in quantum communications, single-mode squeezed vacuum states are also utilized to distribute secret keys in Gaussian CV-QKD protocols \cite{1st-gaussQKD,inefficient_homodyne}. Note that for $r_s=0$, the single-mode squeezed state corresponds to the vacuum state. \hfill \break
\emph{Two-mode squeezed vacuum state}: A two-mode squeezed vacuum (TMSV) state is generated by applying the two-mode squeezing operator of ${\hat S_t}(r) = \exp [ {r({{\hat a}_1}{{\hat a}_2} - \hat a_1^\dag \hat a_2^\dag )/2} ]$ \cite{CV-review, rr2, Weedbrook2012,book-quantum-optics} to a pair of vacuum states $\left| 0 \right\rangle \left| 0 \right\rangle $, where $r \in \mathbb{R}$ is the two-mode squeezing parameter, and the indices~1 and 2 represent the two modes. A TMSV state is described in the Fock basis as \cite{CV-review, rr2, Weedbrook2012,book-quantum-optics}
\begin{eqnarray}\label{TMSV-fock}
\begin{array}{l}
\left| {\rm{TMSV}} \right\rangle  = \sum\limits_{n = 0}^\infty  {{q_n}} {\left| n \right\rangle _1}{\left| n \right\rangle _2},\,{\rm where}\\
\\
{q_n} = \sqrt {1 - {\lambda ^2}} {\lambda ^n},
\end{array}
\end{eqnarray}
and $\lambda  = \tanh r $. The two-mode squeezing in dB is given by $ - 10{\log _{10}}\left[ {\exp ( - 2r)} \right]$. Such a squeezed state has a zero mean, and a covariance matrix in the following form \cite{CV-review, rr2, Weedbrook2012,book-quantum-optics}
\begin{eqnarray}\label{D1}
\bm{M} = \left( {\begin{array}{*{20}{c}}
{v\,\bm{I}}&{\sqrt {{v^2} - 1} \,\bm{Z}}\\
{\sqrt {{v^2} - 1} \,\bm{Z}}&{v\,\bm{I}}
\end{array}} \right) ,
\end{eqnarray}
where $v = \cosh \left( {2r} \right)$ is the quadrature variance of each mode, and $\bm{Z} = diag\left( {1, - 1} \right)$. Note that the two-mode squeezing operator $\hat S_t$ cannot be factorised into the product of the two single-mode squeezing operators ${\hat S_s}$. Hence, the TMSV state is not a product of the two single-mode squeezed vacuum states. In fact, the squeezing (anti-squeezing) operation applied to the quantum fluctuations does not squeeze (anti-squeeze) the variance of the individual modes, but rather that of the superposition of the two modes, so that we have $V({{\hat q}_ - }) = V({{\hat p}_ + }) = \exp ( - 2{r})$ and $V({{\hat q}_ + }) = V({{\hat p}_ - }) = \exp (  2{r})$, where ${{\hat q}_ - } = ({{\hat q}_1} - {{\hat q}_2})/\sqrt 2 $, ${{\hat p}_ + } = ({{\hat p}_1} + {{\hat p}_2})/\sqrt 2 $, ${{\hat q}_ + } = ({{\hat q}_1} + {{\hat q}_2})/\sqrt 2 $, and ${{\hat p}_ - } = ({{\hat p}_1} - {{\hat p}_2})/\sqrt 2 $. For a two-mode squeezing operation with $r>0$, we have $V({{\hat q}_ - }) = V({{\hat p}_ + }) < 1$ and $V({{\hat q}_ + }) = V({{\hat p}_ - }) > 1$. The correlations between the quadratures of the two modes are known as Einstein-Podolski-Rosen (EPR) correlations, which indicate the presence of bipartite entanglement. Hence, for the two-mode squeezing operation with $r>0$ the two modes are entangled, where the entanglement increases upon increasing $r$. The TMSV state associated with $r>0$ is the most commonly used Gaussian entangled state \cite{CV-review, review-entanglement-2003, rr1, rr2, Weedbrook2012, Adesso}. In the limit of $r \to \infty $ we have a maximally entangled state having perfect correlations, yielding ${{\hat q}_1} = {{\hat q}_2}$ and ${{\hat p}_1} = - {{\hat p}_2}$. Note that for $r=0$ the TMSV state corresponds to two (non-entangled) vacuum states.

The Gaussian entangled squeezed states can be generated by parametric down conversion in a non-degenerate optical parametric amplifier \cite{1987, Kwiat, 1997, 1997-2, PDC-2003}, where a crystal having an optical nonlinearity is pumped by a bright laser beam. A photon of the incoming pumping beam spontaneously transfigures in the non-linear crystal into a lower-energy pair of photons, termed as the signal and the idler \cite{1987, Kwiat, 1997, 1997-2, PDC-2003}.
%In type-I PDC (degenerate OPA regime), the signal and idler beams are identical, i.e., they have the same frequency and the same polarization. In this process, the pump photons of frequency $2\omega_p$ are split into pairs of photons (in the same frequency mode) of frequency $\omega_p$. In this case, the type-I PDC process corresponds to the single-mode squeezing operator.
%In type-II PDC (non-degenerate OPA regime), the signal and idler waves are physically separate beams with different frequencies and different polarizations.
In Type-II parametric down conversion, which is known as a source of entangled states in the CV domain, the signal and  idler are in orthogonal polarizations, forming a Gaussian entangled squeezed state \cite{1987, Kwiat, 1997, 1997-2, PDC-2003}. In this process, the pump photons of frequency $2\omega_p$ are converted into pairs of entangled photons having a pair of different-frequency modes, namely modes~1 and 2 of frequency $\omega_1$ and $\omega_2$, where $2\omega_p=\omega_1+\omega_2$. An alternative way of generating the Gaussian entangled squeezed state is by mixing two orthogonally single-mode squeezed vacuum states, where one of the states is squeezed in the $\hat q$ quadrature and the other one is squeezed in the $\hat p$ quadrature. This mixing can be achieved by a balanced (or 50:50) beam splitter. Note that the single-mode squeezed vacuum state can be generated by Type-I parametric down conversion in a degenerate optical parametric amplifier, where the pump photons of frequency $2\omega_p$ are split into pairs of photons having the same  frequency and polarization \cite{PDC-2003}.

%In this case, the type-II PDC process corresponds to the two-mode squeezing operator. Note, the type-II PDC process in a more realistic scenario where it is pumped by a broadband laser pulse e.g., \cite{1997-2} will be discussed in Chapter~\ref{C5:chapter5}.

%Pumping a nonlinear crystal in the nondegenerate optical parametric amplifier regime, we generate pairs of photons in two different modes, known as the signal and the idler
%Crystal materials lacking inversion symmetry can exhibit a so-called χ(2) nonlinearity. Apart from frequency doubling and sum and difference frequency generation, this allows for parametric amplification. Here, the signal beam propagates through the crystal together with a pump beam of shorter wavelength. are then converted into (lower-energy) signal photons and the same number of so-called idler photons

%In type-II PDC (which is known as a source of entangled squeezed states in CV domain). The signal and the idler are in orthogonal polarizations, forming an entangled squeezed state \cite{1987, 1997, 1997-2}. We have so far assumed that the PDC source is pumped by a narrowband laser pulse, which means that the resulting TMSV state

Finally, note that by invoking local unitary operators the first moment of every two-mode Gaussian state can be set to zero and the CM can be transformed into the following standard form \cite{rr2, Weedbrook2012, Adesso}
\begin{eqnarray}\label{p3}
\bm{M_{s}}= \left( {\begin{array}{*{20}{c}}
\bm{A}&\bm{C}\\
{{\bm{C}^T}}&\bm{B}
\end{array}} \right),\,
\end{eqnarray}
where we have $\bm{A} = a\bm{I}\,,\,\bm{B} = b\bm{I}\,,\,\bm{C} = diag\left ( {{c_ + },{c_ - }} \right )$,
 $a,b,{c_ + },{c_ - } \in \mathbb{R}$. % and $I$ is a $2 \times 2$ identity matrix.
%The diagonal elements $a$ and $b$ are nonnegative real scalars and ${c_ + },{c_ - }$ are real scalars.

\begin{figure}
    \begin{center}
   {\includegraphics[width=3.5 in]{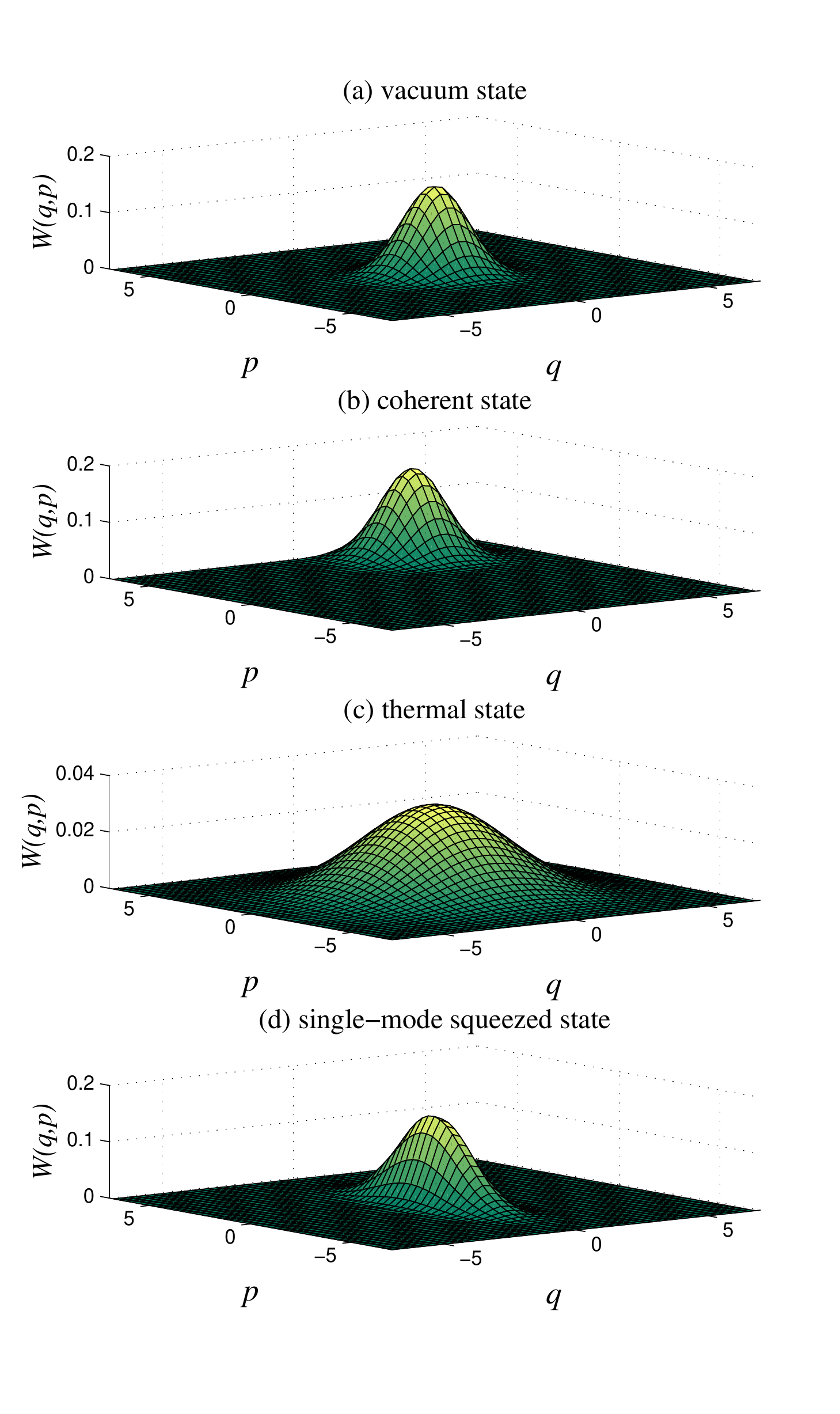}}
    \caption{The Wigner function of the important single-mode Gaussian states including vacuum state, coherent state with a mean value of $(3,5)$, thermal state with $\omega=5$, and single-mode squeezed vacuum state with $r_s=0.5$ and with $\hat q$ quadrature being squeezed.}\label{fig:Wignr-functions}
    \end{center}
\end{figure}

\subsection{Homodyne detection}

The homodyne detection of Fig.~\ref{fig:homodyne-heterodyne}(a) represents the most common measurement in CV quantum information processing \cite{CV-review, rr2, Weedbrook2012}. This detection scheme can be used for determining or observing the quadrature operator $\hat q$ (or $\hat p$ ) of a mode. The scheme of Fig.~\ref{fig:homodyne-heterodyne}(a) is experimentally implemented by combining the target mode (relying on the annihilation operator $\hat a$) with a local oscillator via a balanced beam splitter. The local oscillator is assumed to be in a bright coherent state $\left| {{\alpha _{LO}}} \right\rangle $. Since $\left| {{\alpha _{LO}}} \right\rangle $ is represented by a large number of photons, the local oscillator can be described by a classical complex amplitude $\alpha _{LO}$. The two output modes of the beam splitter can then be approximated by ${\hat a}_1=(\alpha _{LO}+\hat a)/\sqrt{2}$ and ${\hat a}_2=(\alpha _{LO}-\hat a)/\sqrt{2}$.

The intensity of each outgoing mode is then measured using a photodetector, which converts the photons of the electromagnetic mode into electrons, and hence into an electric current - which is termed as the photo-current $\hat i$. The photo-current is proportional to the number of photons in the electromagnetic mode. Hence, the pair of photodetectors of the two output modes of the beam splitter generate the photo-currents of
\begin{eqnarray}\label{BS-photocurrent}
\begin{array}{l}
{{\hat i}_1} \propto \hat n_1 = \hat a_1^\dag {{\hat a}_1} = (\alpha _{LO}^ *  + {{\hat a}^\dag })({\alpha _{LO}} + \hat a)/2,\\
\\
{{\hat i}_2} \propto \hat n_2 = \hat a_2^\dag {{\hat a}_2} = (\alpha _{LO}^ *  - {{\hat a}^\dag })({\alpha _{LO}} - \hat a)/2.
\end{array}
\end{eqnarray}
%A photodetector measuring an electromagnetic mode
Then the difference between the photo-currents ${{\hat i}_1}$ and ${{\hat i}_2}$ is measured, or more specifically,  ${{\hat i}_1} - {{\hat i}_2} \propto (\alpha _{LO}^ * \hat a + {\alpha _{LO}}{{\hat a}^\dag })$ is measured. Considering a local oscillator associated with ${\alpha _{LO}} = \left| {{\alpha _{LO}}} \right|\exp (i\Theta )$, where $\left| {{\alpha _{LO}}} \right|$ and $\Theta $ are the magnitude and phase of the local oscillator respectively, the quadrature operator $\hat q$ ($\hat p$) can be measured by setting the local oscillator's phase as $\Theta=0$ ($\Theta=\pi/2$).

\begin{figure}
    \begin{center}
   {\includegraphics[width=80mm]{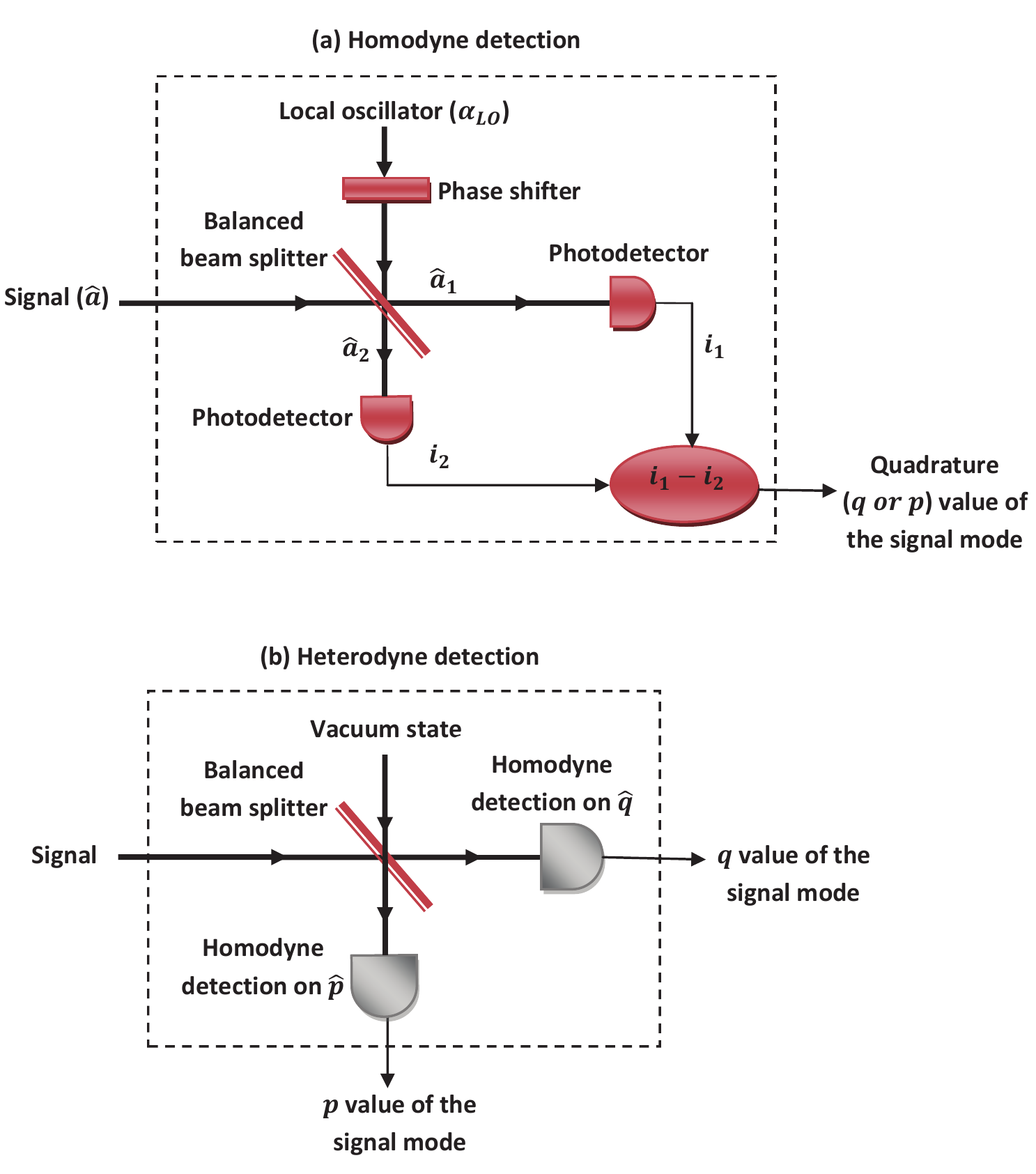}}
    \caption{Schematic of (a) homodyne detection and (b) heterodyne detection.}\label{fig:homodyne-heterodyne}
    \end{center}
\end{figure}

In contrast to homodyne detection, heterodyne detection allows us to measure both the quadrature operators $\hat q$ and $\hat p$ of a mode simultaneously \cite{CV-review, rr2, Weedbrook2012}. A heterodyne detector combines the target mode with a vacuum ancillary mode into a balanced beam splitter. Then, homodyne detection is applied to the conjugate quadratures of the two output modes, i.e., to $\hat q$ of one of the output modes and $\hat p$ of the other one, which are measured using homodyne detection. The `price' to pay for this simultaneous detection is the introduction of an additional noise term into the measurements (due to the mixing into the signal of the vacuum state). The implementation of heterodyne detection is shown in Fig.~\ref{fig:homodyne-heterodyne}(b).

\subsection{CV entanglement}

We have already discussed the notion of entanglement. Indeed, this property is one of the most important properties of quantum mechanics, and is widely recognized as a basic resource for quantum information processing and quantum communications (for review, see \cite{review-entanglement-2003, rr1, Adesso,Weedbrook2012}). We now attempt to quantify the entanglement property of CV states more carefully. We focus our attention on \emph{bipartite} CV entanglement, which relies on the entanglement between two CV quantum systems. Let us consider the pair of CV quantum systems $A$ and $B$ having Hilbert spaces $\mathcal{H}_A$ and $\mathcal{H}_B$, respectively. The Hilbert space of the composite system is given by the tensor product ${\mathcal{H}_A} \otimes {\mathcal{H}_B}$. By definition, a bipartite quantum state $\hat \rho_{AB}$ relying on the Hilbert space ${\mathcal{H}_A} \otimes {\mathcal{H}_B}$ is said to be separable, if it can be formulated as a probability distribution over a pair of uncorrelated states expressed as $\hat \rho_{AB}  = \sum\nolimits_i {{p_i}\hat \rho _i^A \otimes \hat \rho _i^B} $, where the quantum state $\hat \rho_i^A$ ($\hat \rho_i^B$) acts on the Hilbert space ${\mathcal{H}_A}$ (${\mathcal{H}_B}$), ${p_i} \ge 0$, and $\sum\nolimits_i {{p_i}}  = 1 $. %Note, $\otimes$ denotes the tensor product.
If a quantum state $\hat \rho_{AB}$ is separable, then its partial transpose ${\hat \rho_{AB} ^{PT}}$ with respect to either subsystem is positive \cite{Simon}. The partial transposition of $\hat \rho_{AB} $ represents the transposition with respect to only one of the two subsystems, for example to system $B$. By definition, a state is stated to be entangled, when it is not separable in the above-mentioned sense.

%\subsubsection{Entanglement measures}

The \emph{grade} (or quantifiable measure) of entanglement in a \emph{pure} bipartite quantum state $\left| \psi \right\rangle $ (with density operator ${{\hat \rho }_{AB}} = \left| \psi  \right\rangle \left\langle \psi  \right|$) can be quantified by the entropy of entanglement ${E_v}(\left| \psi  \right\rangle )$. The entropy of entanglement stipulates the number of entangled qubits (measured in ebits\footnote{An ebit  (entanglement bit) as the unit of bipartite entanglement is the amount of entanglement that is contained in a maximally entangled two-qubit state (Bell state). In fact, it is said that each of the Bell states contains one ebit of entanglement.}) that can be extracted from the state. It also can be considered as the amount of entanglement required to generate the state. The entropy of entanglement is given by the von Neumann entropy of the reduced density operators ${{\hat \rho }_A}$ or ${{\hat \rho }_B}$, where  ${{\hat \rho }_A} = \rm Tr_{B}({{\hat \rho }_{AB}})$ and ${{\hat \rho }_B} = \rm Tr_{A}({{\hat \rho }_{AB}})$, with $\rm Tr_{A}$ and $\rm Tr_{B}$ denoting the partial trace \cite{review-entanglement-2003, rr1, Adesso,Weedbrook2012}.

For a Gaussian state $\hat \rho$, the von Neumann entropy $S(\hat \rho)$  is given by $S(\hat \rho ) = \sum\nolimits_{k} {g({\nu _k})}  $, where we have $g(x) = \left[ {(x + 1)/2} \right]{\log _2} \left[ {(x + 1)/2} \right] - [(x - 1)/2]{\log _2} \left[ {(x - 1)/2} \right]$, and ${{\nu _k}}$ are the symplectic eigenvalues\footnote{For an arbitrary $N$-mode covariance matrix $\bm{M}$, there exists a symplectic matrix $\bm{S}$ such that $\bm{M} = \bm{S}\bm{M_d}{\bm{S}^T}$, where $\bm{M_d} = \mathop  \oplus \limits_{k = 1}^N \nu _k\bm{I}$ is a diagonal matrix, and the $N$ positive quantities $\nu _k$ are the symplectic eigenvalues of $\bm{M}$. Note that a symplectic matrix $\bm{S}$ is a matrix with real elements that satisfies the condition $\bm{S} \bm{\Omega} \bm{S}^T = \bm{\Omega}$ where $\bm{\Omega}$ is defined in Eq.~(\ref{ad2}) \cite{Weedbrook2012,Adesso}. For example, given a two-mode Gaussian state associated with a covariance matrix $\bm{M} = \left\{ {\bm{A},\bm{C};{\bm{C}^T},\bm{B}} \right\}$, where $\bm{A}=\bm{A}^T$, $\bm{B}=\bm{B}^T$, and $\bm{C}$ are $2 \times 2$ real matrices, the symplectic eigenvalues of $\bm{M}$ are given by $\nu _ \pm ^2 = \left( {\Delta  \pm \sqrt {{\Delta ^2} - 4\det (\bm{M})} } \right)/2$, where $\Delta = \det (\bm{A})+ \det (\bm{B}) + 2\det (\bm{C})$ \cite{Weedbrook2012, Adesso}.} of the covariance matrix of the state. For a pure two-mode entangled state in the form of $\left| \psi  \right\rangle  = \sum\nolimits_{n = 0}^\infty  {{q_n}} {\left| n \right\rangle _1}{\left| n \right\rangle _2}$, the entropy of entanglement is given by ${E_v}(\left| \psi  \right\rangle ) =  - \sum\nolimits_{n = 0}^\infty {q_n^2} {\log _2}q_n^2$.

Among the different quantifiable measures  used as a grade of entanglement for a \emph{mixed} bipartite quantum state ${{\hat \rho }_{AB}}  = \sum\nolimits_i {{p_i}\left| {{\psi _i}} \right\rangle \left\langle {{\psi _i}} \right|} $,  the most well-known  is perhaps the entanglement of formation \cite{EOF1,EOF2}, $E_f$. This is defined as ${E_f}\left( {\hat \rho_{AB} } \right) = \mathop {\min }\limits_{\left\{ {{p_i},\left| {{\psi _i}} \right\rangle } \right\}} \sum\nolimits_i {{p_i}{E_v}(\left| {{\psi _i}} \right\rangle )} $, where the minimum is taken over all the possible pure-state decompositions of the mixed state ${{\hat \rho }_{AB}}$. The entanglement of formation gives the minimal amount of entanglement of any ensemble of pure states realizing the given state $\hat \rho_{AB}$ - meaning it quantifies the minimum amount of entanglement needed to prepare the quantum state $\hat \rho_{AB}$ from a mix of pure entangled states. In fact, given an entangled state $\hat \rho_{AB}$, the entanglement of formation  expresses the number of maximally entangled states we need to create $\hat \rho_{AB}$. In general, this measure of entanglement is difficult to calculate.

%In the asymptotic generalization, entanglement of formation quantifies the minimum number of maximally entangled states (Bell states) which is needed to prepare the quantum state $\hat \rho$ through local operations and classical communications.

The distillable entanglement is another measure for entanglement, and is the amount of entanglement that can be distilled from a given mixed state \cite{review-entanglement-2003}. This quantity is also hard to calculate in general, since it would require optimization over all possible distillation protocols.
However,, there is an entanglement measure which is easy to compute, and  gives an upper bound on the amount of distillable entanglement.  This measure is the so-called logarithmic negativity \cite{LN, LN2}.

%Hence in this review we will adopt the logarithmic negativity, $E_{LN}$, in order to evaluate the entanglement \cite{LN, LN2}.
The logarithmic negativity exhibits the following properties. (i) $E_{LN}$ is a non-negative function, ${E_{LN}}\left( \hat \rho_{AB}  \right) \ge 0$. (ii) If $\hat \rho_{AB}$ is separable, ${E_{LN}}\left( \hat \rho_{AB}  \right) = 0$.  (iii) ${E_{LN}}\left( \hat \rho_{AB}  \right)$ does not increase on average under local (quantum) operations and classical communications. The logarithmic negativity of a bipartite state $\hat \rho_{AB} $ is defined as \cite{LN}
\begin{eqnarray}\label{E-LN-general}
{E_{LN}}(\hat \rho_{AB} ) = {\log _2}\left[ {1 + 2N(\hat \rho_{AB} )} \right],
\end{eqnarray}
where $N(\hat \rho_{AB} )$ is the negativity defined as the absolute value of the sum of the negative eigenvalues of ${\hat \rho_{AB} ^{PT}}$. The logarithmic negativity quantifies as to what degree the quantum state fails to satisfy the positivity of the partial transpose condition.% the partial transpose of $\hat \rho_{AB} $ with respect to either subsystem. %Partial transposition of $\hat \rho_{AB} $ means the transposition with respect to one of the two subsystems, e.g., system $B$.

%%The logarithmic negativity $E_{LN}(\hat \rho)$ in the  case of a Gaussian state $\hat \rho$, is given by ${E_{LN}}(\hat \rho ) = \sum\nolimits_{k} {f({{\tilde \nu }_k})} $, where $f(x) =  - {\log _2}x$ for $x<1$, and $f(x)=0$ for $x \ge 1$, and where ${{{\tilde \nu }_k}}$ are the symplectic eigenvalues of the partially transposed covariance matrix of the Gaussian state.

%Let us now consider a special case (which will be of much use in our later discussion of satellite-based quantum communications), where each of the CV quantum systems $A$ and $B$ in a bipartite quantum state $\hat \rho_{AB}$ has only a single mode. For a pure two-mode entangled state expressed in the form of $\left| \psi  \right\rangle  = \sum\nolimits_{n = 0}^\infty  {{q_n}} {\left| n \right\rangle _1}{\left| n \right\rangle _2}$, the logarithmic negativity can be calculated analytically as \cite{LN}
%\begin{eqnarray}\label{E-LN-pure}
%{E_{LN}}(\left| \psi  \right\rangle ) = 2{\log _2}\left( {\sum\nolimits_{n = 0}^\infty  {{q_n}} } \right).
%\end{eqnarray}
In the special case of two-mode Gaussian states, we are able to determine the logarithmic negativity  through the use of the covariance matrix \cite{rr1, Weedbrook2012, Adesso}. Given a two-mode Gaussian state associated with a covariance matrix $\bm{M} = \left\{ {\bm{A},\bm{C};{\bm{C}^T},\bm{B}} \right\}$
%$
%%\begin{eqnarray}\label{Atandard}
%{M}= \left( {\begin{array}{*{20}{c}}
%A&C\\
%{{C^T}}&B
%\end{array}} \right)
%%\end{eqnarray}
%$,
%where $A = aI\,,\,B = bI\,,\,C = diag\left ( {{c_ + },{c_ - }} \right )$, $a,b,{c_ + },{c_ - } \in \mathbb{R}$.
%Considering the standard form,
where $\bm{A}=\bm{A}^T$, $\bm{B}=\bm{B}^T$, and $\bm{C}$ are $2 \times 2$ real matrices, the logarithmic negativity is given by \cite{rr1, Adesso, Weedbrook2012}
\begin{eqnarray}\label{E-LN-Gaussian}
{E_{_{LN}}}\left( \bm{M} \right) = \max \left[ {0, - {{\log }_2}\left( {{\tilde \nu _ - }} \right)} \right],
\end{eqnarray}
 where ${\tilde \nu _ - }$ is the smallest symplectic eigenvalue of the partially transposed $\bm{M}$. This eigenvalue is given by \cite{rr1, Weedbrook2012, Adesso}
\begin{eqnarray}\label{symplectic eigenvalue}
\tilde \nu _ - ^2 = \left( {\Delta  - \sqrt {{\Delta ^2} - 4\det (\bm{M})} } \right)/2,
\end{eqnarray}
where $\Delta = \det (\bm{A})+ \det (\bm{B}) - 2\det (\bm{C})$.
%In the following when we refer to the grade of entanglement, we will mean the quantitative value of one of the above entanglement measures.

\subsection{Gaussian lossy quantum channel}

%Consider a single-mode Gaussian quantum state, described by the density operator ${\hat \rho}$, as the input

Consider a fixed-attenuation channel described by a transmissivity of $0 \le \tau  \le 1$ and thermal noise variance of $V_n  \ge 1$. Note that in the optical frequency domain the average number of photons is very low even at room temperature (300K), hence the thermal noise has a negligible impact on the signal. In fact, in the optical frequency domain the noise variance is effectively unity, simply representing the vacuum noise. However, in the millimeter-wave domain the thermal noise exhibits a  variance, $V_n$, which is much higher than unity. More specifically, we have $V_n  = 2\bar n + 1$ with $\bar n$ being the average number of photons \cite{MW-QKD1,MW-QKD2,MW-QKD3,Neda6}. In order to suppress the thermal noise, the system has to be operated at very low temperatures, e.g. $<100$mK. The average number of photons for a single mode is given by \cite{MW-QKD1,MW-QKD2,MW-QKD3,Neda6} $ \bar n = [{{\exp \left( {hf/{k_B}T_b} \right) - 1}}]^{-1}\,,$ where  $f$ is the frequency of the mode, $k_B$ is the Boltzmann's constant, and $T_b$ is the temperature.

A fixed-attenuation channel is a Gaussian channel, which transforms the Gaussian input states into Gaussian states. For example, if a single-mode Gaussian quantum state is transmitted through a fixed-attenuation channel, it will remain Gaussian at the output of the channel even though it has experienced channel loss. We can model the impact of a fixed-attenuation channel of transmissivity $\tau$ and thermal noise variance $V_n$ on the single-mode input Gaussian state ${\hat \rho}$ by a beam splitter transformation, with the transmissivity of the beam splitter being $\tau$ and reflectivity $ 1 - \tau $. In this channel representation shown in Fig.~\ref{fig:channel representation} the Gaussian input state is combined with the thermal noise in the beam splitter, such that one input mode of the beam splitter is the Gaussian input state ${\hat \rho }$ having the corresponding quadratures of ${{\hat q}_1},{{\hat p}_1}$ and the second input mode is the thermal noise with corresponding quadratures of ${{\hat q}_2},{{\hat p}_2}$. As a result of the beam splitter transformation we have the output modes~$1'$ (corresponding to the received quantum state ${\hat \rho' }$ at the output of the channel) and $2'$ with corresponding quadratures of ${{\hat q'}_1},{{\hat p'}_1}$ and ${{\hat q'}_2},{{\hat p'}_2}$ respectively. These output quadratures can be described by \cite{Weedbrook2012}

\begin{equation}\label{BS-transformation}
{{\hat R}_{out}} = \left( {\begin{array}{*{20}{c}}
{\sqrt \tau  \bm{I}}&{\sqrt {1 - \tau } \bm{I}}\\
{ - \sqrt {1 - \tau } \bm{I}}&{\sqrt \tau  \bm{I}}
\end{array}} \right){{\hat R}_{in}},
\end{equation}
where ${{\hat R}_{in}} = ({{\hat q}_1},{{\hat p}_1},{{\hat q}_2},{{\hat p}_2})$, and ${{\hat R}_{out}} = ({{\hat q'}_1},{{\hat p'}_1},{{\hat q'}_2},{{\hat p'}_2})$. As a result, the quadrature variance of the received quantum state at the output of the channel is given by $V({{\hat q'}_1}) = \tau V({{\hat q}_1}) + (1 - \tau )V_n$, and $V({{\hat p'}_1}) = \tau V({{\hat p}_1}) + (1 - \tau )V_n $.

\begin{figure}
    \begin{center}
   {\includegraphics[width=3.5 in]{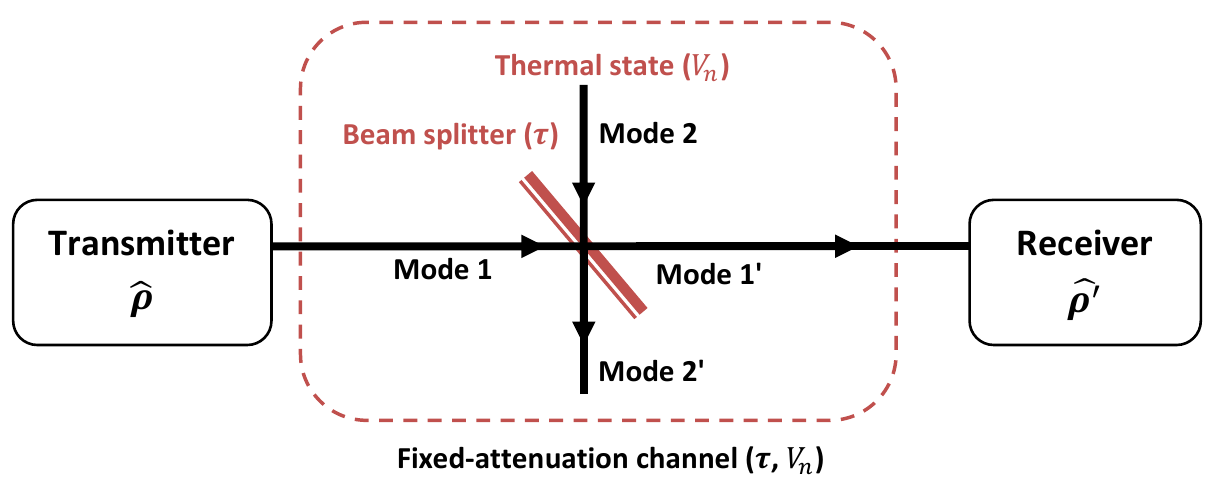}}
    \caption{The beam splitter representation of a fixed-attenuation channel with transmissivity $\tau$ and thermal noise variance $V_n$.}\label{fig:channel representation}
    \end{center}
\end{figure}

Let us now use such a channel representation to analyse the evolution of a two-mode Gaussian quantum state over a fixed-attenuation channel (the general multimode case can be significantly more complex, e.g. \cite {nedamult}). We consider a TMSV state with zero mean and covariance matrix in the form of Eq.~\eqref{D1} as the input quantum state of the channel. There are two settings for the transmission of a two-mode quantum state between two parties, namely, the single-mode transfer and the two-mode transfer \cite{Pirandola}. We discuss each of these in detail.

\emph{Single-mode transfer}: In this setting, the TMSV source is placed at one of the parties' site. In this case, only one mode (mode~2) is transmitted through a fixed-attenuation channel, with the other mode (mode~1) remaining unaffected. The Gaussian output state has a zero mean and CM in the following form \cite{Weedbrook2012,Pirandola}
\begin{equation}\label{M-asym}
\bm{M_{sm}} = \left( {\begin{array}{*{20}{c}}
{v\bm{I}}&{\sqrt \tau  \sqrt {{v^2} - 1} \bm{Z}}\\
{\sqrt \tau  \sqrt {{v^2} - 1} \bm{Z}}&{\left( {\tau v + (1 - \tau )V_n } \right)\bm{I}}
\end{array}} \right),
\end{equation}
where $v = \cosh ( {2r} )$ is the quadrature variance of each mode in the input TMSV state ($r$ being the two-mode squeezing parameter).% , and $\omega  = 2\bar n + 1$ and where $\bar n$ is the mean photon number of the thermal noise.

\emph{Two-mode transfer}: In this setting, the TMSV source is placed somewhere between the two parties. In this case, one mode (mode~1) of the TMSV state is transmitted through a fixed-attenuation channel with transmissivity $\tau_1$ and thermal noise variance $V_{n1}$, while the other mode (mode~2) being transmitted through another fixed-attenuation channel with transmissivity $\tau_2$ and thermal noise variance $V_{n2}$. The Gaussian output state has a zero mean and CM in the following form \cite{Weedbrook2012,Pirandola}
\begin{equation}\label{M-sym}
\bm{M_{tm}} = \left( {\begin{array}{*{20}{c}}
{\left( {\tau_1 v + (1 - \tau_1 )V_{n1} } \right)\bm{I}}&{\sqrt {\tau_1 \tau_2} \sqrt {{v^2} - 1} \bm{Z}}\\
{\sqrt {\tau_1 \tau_2} \sqrt {{v^2} - 1} \bm{Z}}&{\left( {\tau_2 v + (1 - \tau_2 )V_{n2} } \right)\bm{I}}
\end{array}} \right).
\end{equation}
Here, we have assumed  the two fixed-attenuation channels are independent and the two thermal noises are  uncorrelated.

\section{CV-QKD}

At the time of writing most of the classical cryptography schemes are based on the Rivest-Shamir-Adleman (RSA) protocol \cite{RSA} in which the encryption key is public. These cryptography schemes are based on the concept of one-way functions, i.e. on functions which are easy to compute but extremely difficult to invert. Hence, the security of these schemes cannot be proved in principle. In fact, the security of these schemes is not unconditional, since they are based on certain computational power assumptions. Thus, if quantum computers were available today with a substantial amount of computational power, RSA  cryptography schemes could be broken. However, unconditional security is indeed possible using the one-time pad scheme of \cite{one-time-pad}, where a symmetric, random secret key is shared between the transmitter and receiver. In the one-time pad scheme, the transmitter (Alice) encodes the message by applying a modular addition between the plaintext bits and an equal amount of random bits of the shared secret key. At the receiver, Bob decodes the received message by applying the same modular addition between the received ciphertext and the shared secret key. If Alice and Bob do not reuse their key, the one-time pad scheme of \cite{one-time-pad} cannot be broken, in principle. However, the main problem of this scheme is the generation of the secret key - a key which is as long as the message itself and must be used only once. This problem becomes severe, when a large amount of information has to be securely transmitted. Partially because of this limitation, public-key cryptography is more widely used than the one-time pad scheme.

QKD is the most well-developed and most widely known protocol of quantum communications. QKD, which is based on the laws of quantum physics, allows Alice and Bob to generate secret keys that can later be used to communicate with information-theoretic (unconditional) security, regardless of any future advances in computational power. A QKD protocol can be divided into two main stages. Firstly, a quantum communication part where a pair of distant and trusted parties, Alice and Bob, generate two sets of correlated data through the transmission of a significant number of quantum states over an insecure quantum channel. Secondly, by the use of a classical post-processing protocol \cite{Renner1,Renner2} operated over a public but authenticated (meaning that the transferred data is known to be unaltered) classical channel, Alice and Bob extract from their correlated data a secret key that is unknown to a potential eavesdropper, Eve. The final key, which is unconditionally secure  can then be used to transmit secret messages \cite{G, Scarani}. Note that in QKD the quantum channel is open to any possible manipulation from Eve, which means that Eve has full access to the quantum channel without any computational (classical or quantum) limitation other than those imposed by the laws of quantum physics. However, Eve can only \emph{monitor} the public classical channel, without \emph{modifying} the messages (since the channel is authenticated). A schematic of a QKD system is shown in Fig.~\ref{fig:QKD}.

\begin{figure}
    \begin{center}
   {\includegraphics[width=3.5 in]{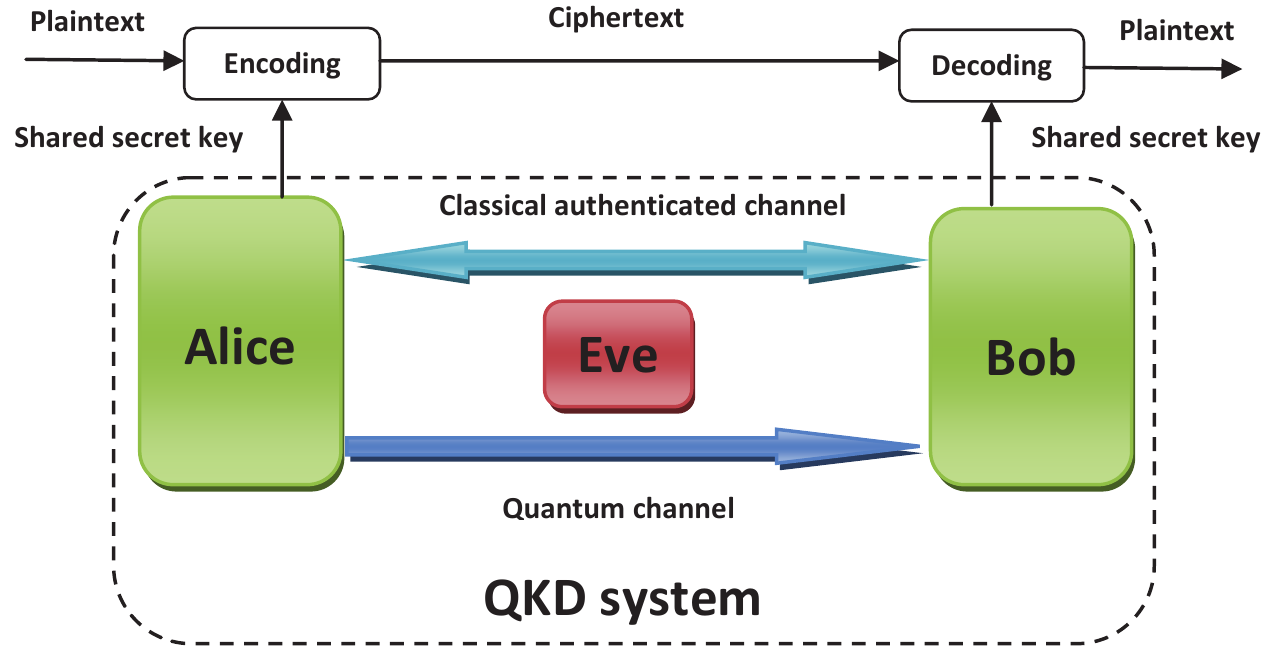}}
    \caption{A schematic of a QKD system: Alice and Bob are connected by a quantum channel, to which Eve has full access without any limitation (other than those constrained by the laws of physics). They are also connected by an authenticated classical channel, which Eve can only monitor. The final shared key between Alice and Bob, which is unconditionally  secure, can then be used to transmit (encode and decode) secret messages.}\label{fig:QKD}
    \end{center}
\end{figure}

The security of QKD is based on some of the fundamental principles of quantum physics. From an attack perspective we could consider that Eve's ultimate goal is to have a perfect copy of the quantum state sent by Alice to Bob. However, this outcome is impossible owing to the no-cloning theorem of quantum physics, which states that it is impossible to create an identical copy of an arbitrary unknown quantum state while keeping the original state intact\cite{wooters} \cite{Dieks}. This simple, but crucial,  observation can be traced back to the fact that quantum mechanics is a linear theory.

 %because copying or its observation destroys the quantum state, which collapses back into the classical domain. To elaborate a little further, any action by which Eve extracts some information from the quantum state (sent by Alice to Bob) is a generalized form of measurement; however a fundamental principle of quantum physics states that measurement in general changes the quantum state of the measured system. %, and then this modification can be detected by trusted parties.

There are two main techniques of implementing QKD, DV-QKD where the key information is mapped to a single photon's phase or polarization \cite{BB84, 144km, decoy}, and CV-QKD where the key information is mapped to the quadrature variables of the optical field \cite{1st-CVQKD,2nd-CVQKD,1st-gaussQKD,RR2002,RR,1st-EB}. In the DV-QKD technology detection is realized by single-photon detectors, while in the CV-QKD technology detection is realized by homodyne (or heterodyne) detectors. In this review we will focus our attention on the CV technology to implement QKD.

%CV-QKD is mostly implemented based on a point-to-point protocol, where Alice is the sender of quantum states and Bob is the receiver of the incoming quantum states e.g., \cite{RR, LDPC, exp-CVQKD2007-1, exp-CVQKD2007-2, exp-CVQKD2009-1, exp-CVQKD2009-2, exp-CVQKD2010, exp-CVQKD2012, exp-PM}. In Chapter~? we will study a non point-to-point protocol of CV-QKD where both Alice and Bob are the senders of quantum states. Note, from now on, unless stated otherwise we will consider the point-to-point protocols of CV-QKD.

CV-QKD is mostly implemented experimentally in a prepare-and-measure (PM) scheme
\cite{1st-gaussQKD, RR2002, RR, LDPC, exp-CVQKD2007-1, exp-CVQKD2007-2, exp-CVQKD2009-1, exp-CVQKD2009-2, inefficient_homodyne, exp-CVQKD2010, exp-CVQKD2012, exp-PM, exp-CVQKD-2016-fiber100}, where Alice prepares CV quantum states and encodes the key information onto the quantum states, which are then transmitted over an insecure quantum channel to Bob. At the output of the channel Bob receives the quantum states and measures them using homodyne or heterodyne detectors. As a result,  correlated data is created between Alice and Bob. Each PM scheme of CV-QKD can be represented by an equivalent entanglement-based (EB) scheme \cite{1st-EB, thesis, Weedbrook2012, exp-EB, Weedbrook2013, inefficient_homodyne}, where Alice generates a two-mode entangled state, with one mode being held by Alice and the other mode being transmitted through an insecure quantum channel to Bob. Alice and Bob then proceed by measuring their own modes using homodyne or heterodyne detectors in order to create correlated data. Following the generation of the correlated data, Alice and Bob proceed with classical post-processing over a public, but authenticate, classical channel (in both the PM scheme and EB scheme), so as to generate a secret key even in the presence of Eve.

CV-QKD protocols using Gaussian quantum states have been richly analysed in theory \cite{Weedbrook2012, 1st-gaussQKD, RR2002, 1st-EB, thesis, Weedbrook2013, inefficient_homodyne}, and they have also been implemented experimentally \cite{RR, LDPC, exp-CVQKD2007-1, exp-CVQKD2007-2, exp-CVQKD2009-1, exp-CVQKD2009-2, exp-CVQKD2010, exp-CVQKD2012, exp-PM, exp-CVQKD-2016-fiber100, exp-EB}. In the Gaussian PM scheme which is shown in Fig.~\ref{fig:Gaussian-PM}, the CV quantum states prepared by Alice are Gaussian states (squeezed states or coherent states) which are modulated by Gaussian distributions \cite{1st-gaussQKD, RR2002, RR, LDPC, exp-CVQKD2007-1, exp-CVQKD2007-2, exp-CVQKD2009-1, exp-CVQKD2012, exp-PM, exp-CVQKD-2016-fiber100, thesis, inefficient_homodyne}. In fact, Alice encodes a classical random variable drawn from a Gaussian distribution onto a Gaussian quantum state, which is transmitted to Bob, and then measured by him, thus extracting a classical random variable which is correlated to Alice's. In the Gaussian PM scheme, the measurements of the received quantum states are made by Gaussian measurements, namely by homodyne or heterodyne detection.

\begin{figure*}
    \begin{center}
   {\includegraphics[width=150mm]{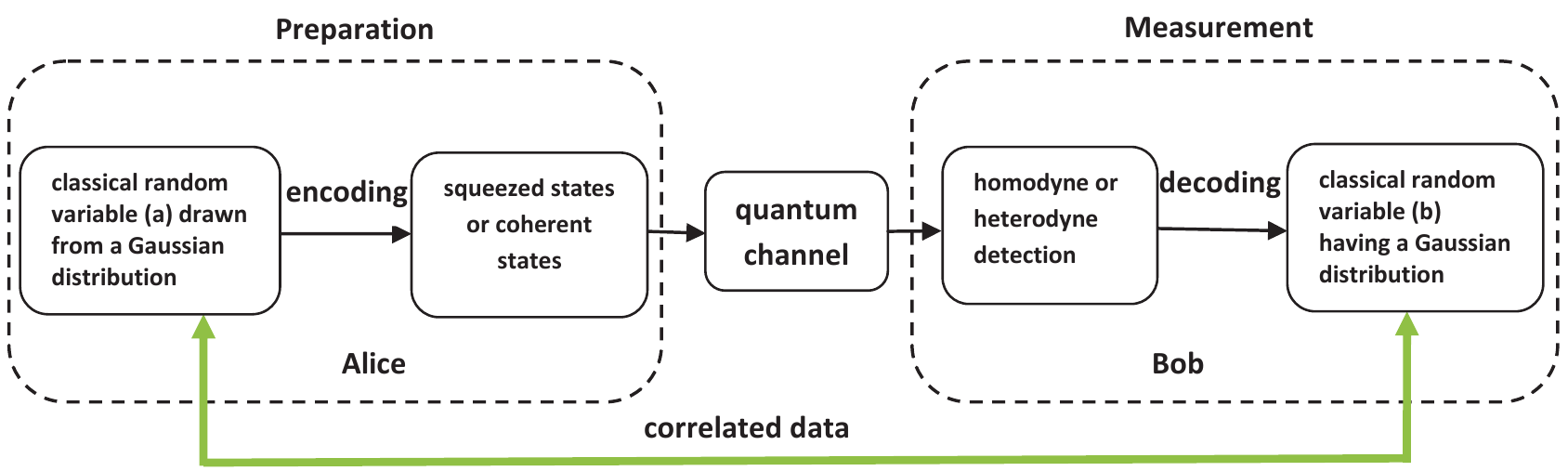}}
    \caption{Schematic of Gaussian CV-QKD protocols in a PM scheme.}\label{fig:Gaussian-PM}
    \end{center}
\end{figure*}

%CV-QKD  protocols using squeezed states can be described as follows: Alice generates a real random Gaussian-distributed variable $a$ with zero mean and variance $v_m$. Alice also generates a random bit $u$. Alice then prepares a single-mode squeezed vacuum state with CM ${M} = diag(1/v,v)$, where $v = \exp (2{r_s})$, and where $r_s$ is the single-mode squeezing. The prepared squeezed state is then modulated (displaced) by an amount $a$, where the modulation variance satisfies $v_m=v-1/v$. In fact, depending on the value of the random bit $u$ Alice sends a $q$-squeezed state with first moment $(a, 0)$ or a $p$-squeezed state with first moment $(0, a)$. Hence, Alice randomly chooses to squeeze and displace either the $\hat q$ or $\hat p$ quadrature. The prepared and modulated squeezed states are then transmitted over an insecure quantum channel towards Bob. For each incoming state, Bob depending on his own random bit $u'$, measures either $\hat q$ or $\hat p$ quadrature. Note, in order to warrant security, Alice and Bob must randomly choose different basis for preparation and measurement. When the quantum communication is finished and all the incoming states are measured by Bob, classical post-processing over a public channel starts by applying sifting where Alice and Bob reveal which quadrature (basis) they used to prepare and measure the information, thus discarding incompatible data. In fact, Alice reveals for each pulse the value of $u$ (whether she displaced $q$ or $p$ quadrature), and Bob keeps only the cases where he measured the right quadrature (i.e., $u=u'$).

CV-QKD protocols using squeezed states \cite{1st-gaussQKD} can be described as follows. Alice generates a real random Gaussian-distributed variable $a$ with zero mean and variance $v_m$. She also generates a random bit $u$, and then prepares a single-mode squeezed vacuum state having the covariance matrix $\bm{M} = diag(1/v,v)$, where $v = \exp (2{r_s})$, and where $r_s$ is the single-mode squeezing. The squeezed state prepared is then modulated (displaced) by an amount $a$, where the modulation variance satisfies $v_m=v-1/v$. In fact, depending on the value of the random bit $u$, Alice sends a $q$-squeezed state having a first moment of $(a_q, 0), a_q=a$ or a $p$-squeezed state associated with the first moment $(0, a_p), a_p=a$. Hence, Alice randomly chooses to squeeze and displace either the $\hat q$ or the $\hat p$ quadrature. The prepared and modulated squeezed states are then transmitted over an insecure quantum channel to Bob. For each incoming state, depending on his own random bit $u'$, Bob measures either the $\hat q$ or the $\hat p$ quadrature using homodyne detection, obtaining a real variable $b_q=b$ or $b_p=b$, respectively. Note that in order to warrant security, Alice and Bob choose different basis for preparation and measurement (in a random fashion). Following the measurement of all incoming states by Bob, classical post-processing over the public channel commences via a sifting operation. In this operation, Alice and Bob reveal to each other which of the two quadratures  they used for preparing (Alice) and measuring (Bob) the information,  discarding any incompatible data (i.e., $a \ne b$). In fact, Alice reveals for each pulse the value of $u$ (i.e., whether she displaced the $\hat q$ or the $\hat p$ quadrature), and Bob only retains the cases, where he measured the relevant quadrature (i.e., $u=u'$).

Another squeezed-state protocol was developed in \cite{inefficient_homodyne}, in which Bob uses heterodyne detection rather than homodyne detection and measures both the $\hat q$ and $\hat p$ quadratures for obtaining $(b_q, b_p)$. In the sifting step of this protocol, Bob then disregards one of his quadrature measurements, depending on Alice's specific choice of quadrature preparation. This protocol can be seen as a noisy version of the protocol with squeezed states and homodyne detection, since the heterodyne detection introduces a vacuum noise into the measurement. When Bob's data are the reference of error correction (see below) in the classical post-processing, the heterodyne detection protocol has a better robustness against the channel noise than the protocol associated with homodyne detection \cite{inefficient_homodyne}.

In contrast to the above CV-QKD protocols using squeezed states, CV-QKD  protocols using coherent states \cite{QKD-coh1, RR2002, RR} can be described as follows. Alice generates a pair of random real numbers, $a_q$ and $a_p$, chosen from two independent Gaussian distributions of variance $v'_m$. Alice then prepares a coherent state, which is then modulated (displaced) by the amounts of $a_q$ and $a_p$, where $(a_q, a_p)$ represents the mean value of the coherent state. The prepared and modulated coherent states are then transmitted over an insecure quantum channel to Bob. For each incoming state, depending on his own random bit $u'$, Bob measures either the $\hat q$ or the $\hat p$ quadrature using homodyne detection,  obtaining a real variable $b_q$ or $b_p$, respectively. When the quantum communication is finished and all the incoming states have been measured by Bob, classical post-processing over a public channel is commenced by applying sifting, where Bob reveals for each pulse the value of $u'$ (i.e., whether he measured the $\hat q$ or the $\hat p$ quadrature), and Alice keeps $a_q$ or $a_p$ depending on the value of $u'$. Note that in this protocol only one of the two real random variables generated by Alice is used for the key after the sifting stage.

Another coherent-state protocol was developed in \cite{QKD-coh2}, where Bob uses heterodyne detection rather than homodyne detection and measures both the $\hat q$ and $\hat p$ quadratures for obtaining $(b_q, b_p)$ at the cost of introducing a vacuum noise into the measurement. In this protocol, sifting is no longer needed, since both of the real random variables generated by Alice are used for the generation of the key, hence potentially resulting in higher secret key rates.

For all QKD protocols  parameter estimation is performed (in the classical post-processing stage, following the sifting step), where the two parties reveal a randomly chosen subset of their data. This allows them to estimate  parameters of the channel, such as the channel's transmissivity and the channel noise. This allows them to limit the maximum amount of information Eve can have about their values. This step is followed by a  reconciliation procedure - which encompasses error correction. As discussed more later, this procedure normally proceeds via the use of low density parity check (LDPC) codes \cite{LDPC}. QKD can be operated in two reconciliation scenarios, direct reconciliation \cite{DR-RR} and reverse reconciliation \cite{RR2002, RR}. In the direct reconciliation protocol Alice's data constitute the reference and she sends classical correction information to Bob which may be overheard by Eve. Then Bob corrects his key elements to arrive at the same values as Alice. By contrast, in the reverse reconciliation protocol Bob's data constitute the reference and must be estimated by Alice (also by Eve) \cite{RR2002, RR}. Since the upper bound on Eve's information is estimated during the parameter estimation stage, Alice and Bob apply a privacy amplification protocol (for discarding the information that may be known to Eve) to produce a shared binary secret key.

%Each Gaussian PM scheme can be represented by an equivalent EB scheme, in which Alice and Bob share a Gaussian two-mode entangled state \cite{1st-EB, thesis, Weedbrook2012, exp-EB, Weedbrook2013, inefficient_homodyne}.
Note that there are eight protocol choices for characterising Gaussian CV-QKD in a PM scheme.
This is because we must consider the type of quantum state (squeezed states or coherent states) which Alice prepares, and also the type of detection (homodyne or heterodyne detection) which Bob applies to the received states, as well as the specific type of reconciliation (direct reconciliation or reverse reconciliation). However, recalling all PM schemes have an equivalent EB scheme, we note all the PM protocols can be described in an unified way using the EB scheme \cite{thesis, Weedbrook2012} shown in Fig.~\ref{fig:EB-QKD}.
Here Alice generates a TMSV state, which we refer to as $\hat \rho_{AB}$. She keeps mode~$A$, and sends mode~$B$ to Bob. At some time later, Alice and Bob use an unbalanced beam splitter of transmissivity (${T_A}$ at Alice's side and ${T_B}$ at Bob's side),  to carry out \emph{generalized} \emph{heterodyne} detections. If Alice  applies homodyne detection (${T_A}=1$), the  prepared  state should be a squeezed state and if Alice makes a heterodyne detection (${T_A}=1/2$), the  prepared state should be a coherent state.
%Bob can invoke homodyne detection associated with ${T_B}=1$ and heterodyne detection with ${T_B}=1/2$.
The security of the CV-QKD protocols is mostly analysed using their equivalent EB scheme, where a two-mode entangled state is shared between Alice and Bob before their detection observations. Note, in the security analysis of CV-QKD discussed next we will assume that the  number of exchanges between Alice and Bob is considered to be infinite (the asymptotic regime). This assumption is adopted in most QKD security analyses since the ability to estimate some quantities (e.g. average values) exactly in the infinite sample-limit, greatly simplifies the analyses.

\begin{figure}
    \begin{center}
   {\includegraphics[width=3.5 in]{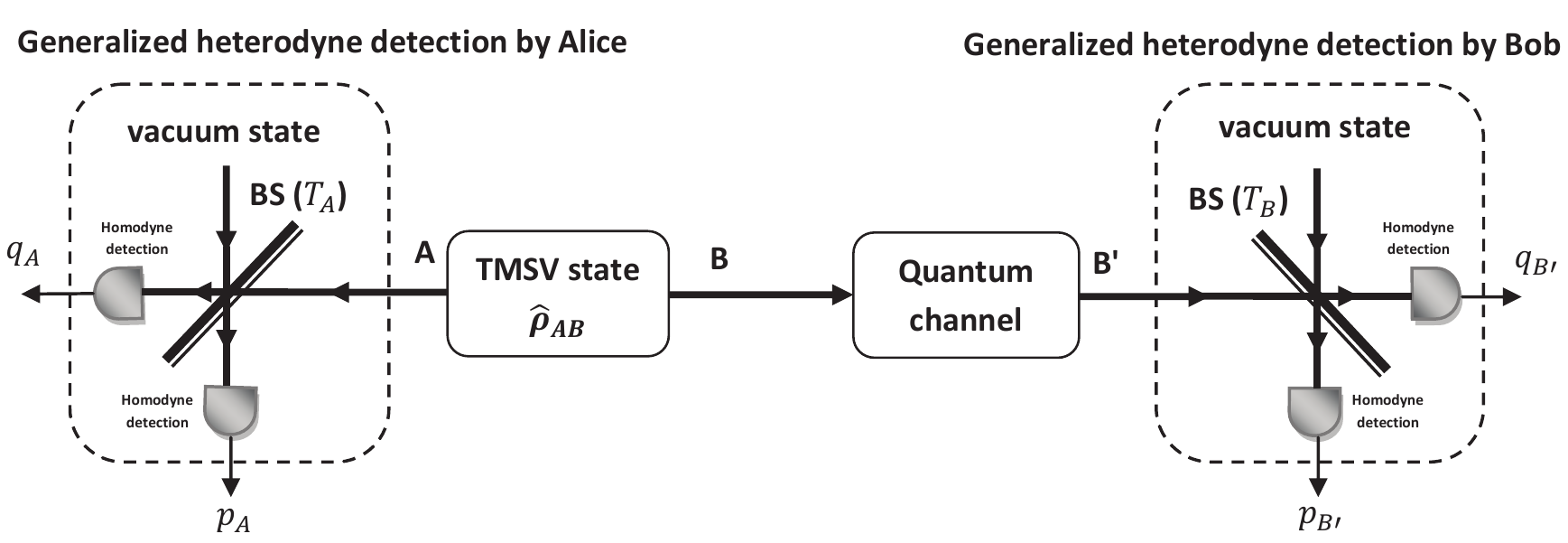}}
    \caption{The EB representation of Gaussian CV-QKD protocols.}\label{fig:EB-QKD}
    \end{center}
\end{figure}

\subsection{CV-QKD security analysis}
The most powerful, and most general, attack that Eve can implement against QKD is known as a coherent  attack \cite{Weedbrook2012,thesis}. In  this attack, Eve prepares her ancillary system in a global quantum state, which means she prepares an arbitrary joint (entangled) state of the ancillae. After the interaction of the global ancillary system with the signals sent by Alice, the output ancillary system is stored in a quantum memory. Once the classical post-processing relying on the public channel is finished, Eve applies an optimal joint measurement over the ancillary system stored in the quantum memory to maximize her knowledge on the quantum information of the trusted parties. The security analysis of CV-QKD in the face of coherent attacks is very complex. However, under some trivial constraint imposed on the classical post-processing protocol, collective attacks are just as detrimental as coherent attacks \cite{coherent-attack}. In a collective attack against QKD Eve prepares her ancillary system in a product state of identically prepared ancillae. After interaction of each ancilla with a single signal sent by Alice, the output ancilla is stored in a quantum memory. Once the classical post-processing is completed, Eve applies an optimal joint measurement over the ensemble of ancillae in the quantum memory.

For a realistic reconciliation algorithm, the asymptotic CV-QKD key rate (bits per pulse) against collective attacks is given by \cite{Weedbrook2012,thesis} $K = \xi {I_{AB}} - {I_E}$, where $I_{AB}$ is the mutual information between Alice and Bob (i.e., between Alice's variable, $a$, as well as Bob's variable, $b$), and $0<\xi<1$ is the reconciliation efficiency. This efficiency reflects that in a realistic reconciliation algorithm, Alice and Bob acquire not all of the maximum attainable mutual information. Note that for a perfect reconciliation algorithm we will have $\xi = 1$. Furthermore, $I_{E}$ is the Holevo bound defined in \cite{Weedbrook2012,thesis} as an upper bound on the quantum information stolen by Eve. In the reconciliation step, if we  assume that Alice's data represents the reference, then $I_{E} = I _{AE}$ is the Holevo bound on the mutual information between Eve's quantum memory and Alice's variable. By contrast, if we assume that Bob's data is the reference, then $I_{E} = I _{BE}$ is the Holevo bound on the mutual information between Eve's quantum memory and Bob's variable. Note that $I_{AB}$ remains the same, regardless of whose data represents the reference of reconciliation. It was also shown \cite{Gaussian-attack} that in the family of collective attacks, Gaussian attacks based on Gaussian operations\footnote{Gaussian operations are linear operations with respect to the quadrature amplitudes. Such operations maintain the Gaussian character of Gaussian states.} are the optimal attacks Eve can implement so as to minimize the secret key rate $K$\footnote{Gaussian collective attacks are as strong as coherent attacks in the limit of an infinite number of quantum states exchanged, however, it is not known this is the case for a realistic finite-length key protocols.}. %As a result, for the asymptotic security analysis of Gaussian CV-QKD protocols (which are based on the Gaussian modulation of Gaussian states) collective Gaussian attacks represent the fundamental benchmark.

Let us consider a Gaussian CV-QKD protocol in the EB scheme, where Alice generates a TMSV state $\hat \rho_{AB}$, and keeps mode~$A$ while sending mode~$B$ to Bob over an insecure quantum channel. In the optimal collective Gaussian attack (which is also referred to as the entangling-cloner attack \cite{RR}) shown in Fig.~\ref{fig:attack}, Eve models the quantum channel (with transmissivity of $0 \le \tau  \le 1$ and thermal noise variance of $\omega  \ge 1$) by a TMSV state $\hat \rho_{E_1E_2}$ having a quadrature variance of $\omega$ and a beam splitter of transmissivity $\tau$. In fact, the quadrature variance of $\hat \rho_{E_1E_2}$ and the transmissivity of the beam splitter in Fig.~\ref{fig:attack} are tuned in order to inject the same noise and to impose the same attenuation as in the original channel, respectively. In this beam splitter Eve combines the signal mode gleaned from Alice (mode~$B$) with one mode (mode~$E_1$) of the TMSV state. The first output of the beam splitter (mode $B'$) which is the quantum signal received by Bob is given by ${{\hat q}_{B'}} = \sqrt \tau  {{\hat q}_B} + \sqrt {1 - \tau } {{\hat q}_{{E_1}}}$, and ${{\hat p}_{B'}} = \sqrt \tau  {{\hat p}_B} + \sqrt {1 - \tau } {{\hat p}_{{E_1}}}$. The second output of the beam splitter (mode $E'_1$) and mode~$E_2$ of the TMSV state $\hat \rho_{E_1E_2}$ are stored by Eve in a quantum memory. Once the classical post-processing over the public channel is completed, this quantum memory is detected by means of an optimal joint measurement which estimates Alice's data (in direct reconciliation) or Bob's data (in reverse reconciliation). Note that in a Gaussian CV-QKD protocol, the asymptotic key rate against optimal collective Gaussian attacks can be calculated through the equivalent EB scheme based on the covariance matrix of the two-mode entangled state shared between Alice and Bob before their detection observations \cite{Weedbrook2012,Weedbrook2013,thesis}.

\begin{figure}
    \begin{center}
   {\includegraphics[width=3.5 in]{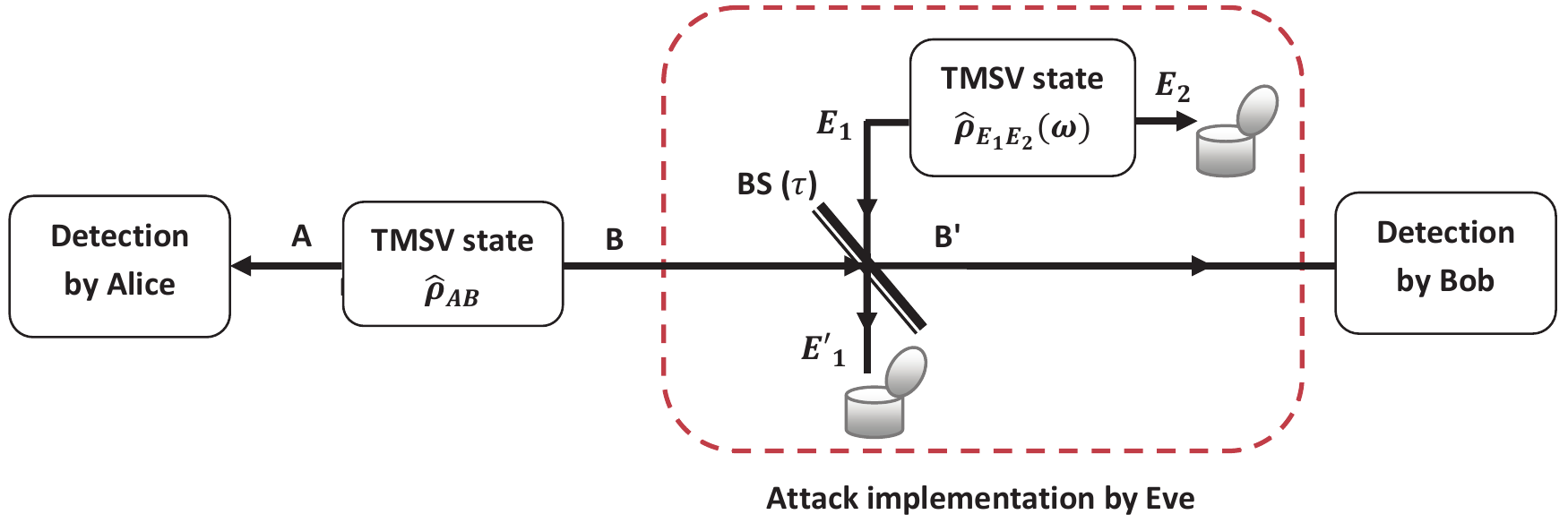}}
    \caption{Implementation of optimal collective Gaussian attack (entangling-cloner attack) by Eve.}\label{fig:attack}
    \end{center}
\end{figure}

\section{Entanglement distribution and CV-QKD implementation via satellite}

\subsection{Entanglement distribution and standard  QKD protocols}

Let us reconsider the quantum communication architectures of Fig.~\ref{fig:satellite-based-confs} for CV entanglement distribution and for CV-QKD implementation.
We assume that the source of quantum communication in the transmitter(s) is a two-mode entangled state associated with modes~1 and 2. In the scheme (a) (the scheme (b)) of Fig.~\ref{fig:satellite-based-confs}, a two-mode entangled state is generated by Alice at the ground station (satellite) with one mode, mode~1, kept by Alice, while the other mode, mode~2, is transmitted to Bob located at the satellite (ground station) over the uplink (downlink). In the scheme (c) of Fig.~\ref{fig:satellite-based-confs}, a two-mode entangled state is generated by Alice at the ground station transmitter with one mode, mode~1, held at the ground station transmitter and the other mode, mode~2, transmitted over the uplink to the  relay satellite. The received mode is then reflected in the satellite and transmitted through the downlink to Bob at the ground station receiver. In the scheme (d) of Fig.~\ref{fig:satellite-based-confs}, a two-mode entangled state is generated on board of the satellite with both modes then sent over the separate downlinks to Alice and Bob located at the separate ground stations. In the scheme (e) of Fig.~\ref{fig:satellite-based-confs}, Alice and Bob are located in the separate ground stations, both initially possessing a two-mode entangled state. One mode of each entangled state is kept by a ground station transmitter and the second mode of each state is transmitted over the uplink to the relay satellite, in which on-board entanglement swapping is performed on the arriving modes. To elaborate a little further, entanglement swapping \cite{1st-tele} is a standard quantum protocol conceived for establishing entanglement between distant quantum systems that have never interacted \cite{swap2005, swap2006, swap2014, Obermaier}. It is the central mechanism of quantum repeaters \cite{repeater}, enabling the distribution of entanglement over large distances. In the scheme (e) of Fig.~\ref{fig:satellite-based-confs}, the received modes are swapped at the satellite via a CV Bell measurement \cite{CV_Bell}, where the two modes are mixed through a balanced beam splitter. Explicitly, the $\hat q$ quadrature of one of the output modes of the beam splitter and the $\hat p$ quadrature of the output mode are separately measured by two homodyne detectors. This process is sometimes described by saying that the two output modes of the beam splitter are conjugately homodyned \cite{CV_Bell}. The classical outcome of the Bell measurement is then communicated to Alice and Bob so that they can optimally displace their modes, according to the measurement outcome, in order to maximize the resultant entanglement shared between the ground stations. This entanglement swapping scheme between two ground stations via satellite is shown more explicitly in Fig.~\ref{fig:swapping}.

\begin{figure}
    \begin{center}
   {\includegraphics[width=3.5 in]{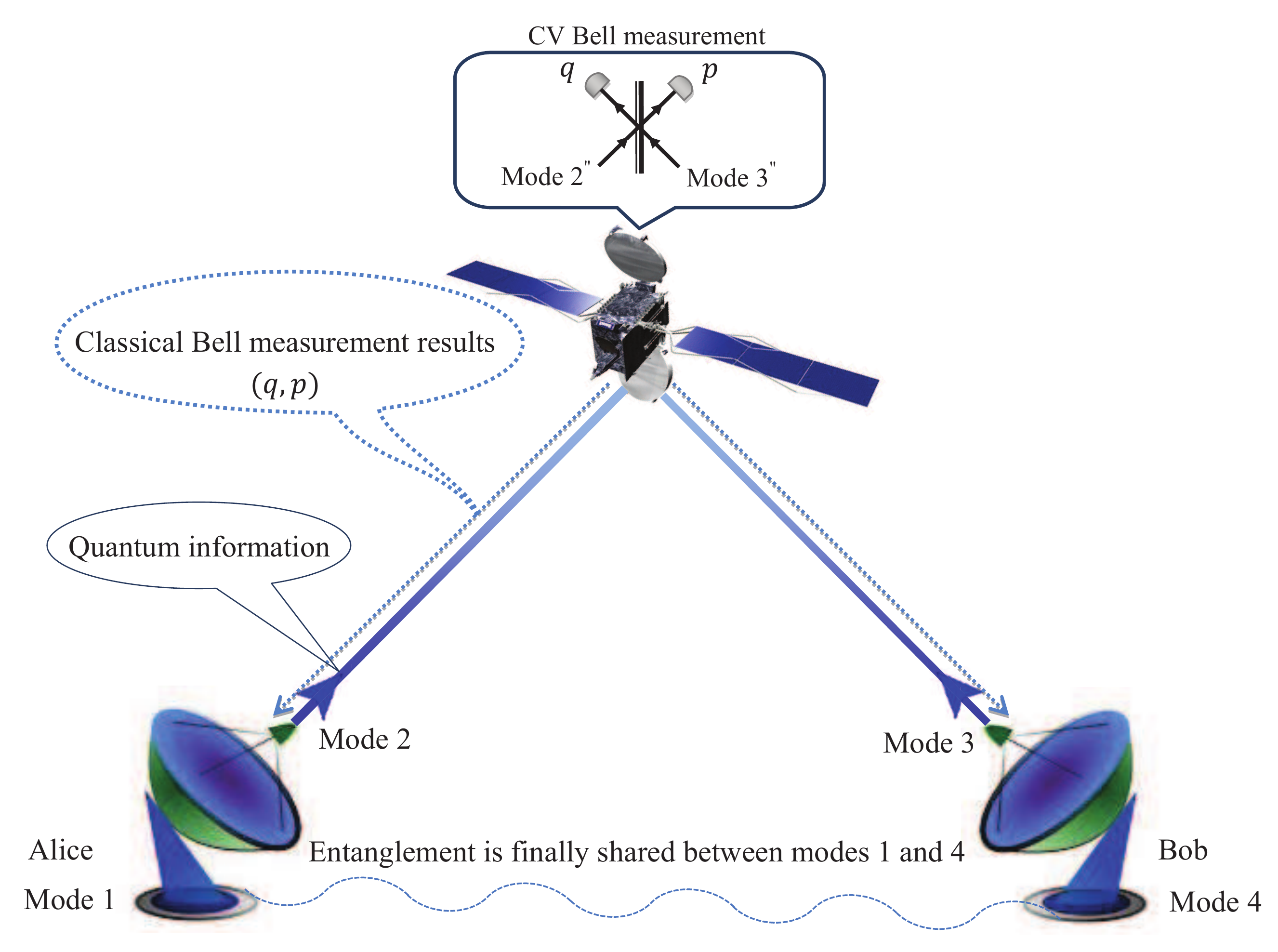}}
    \caption{Entanglement swapping between two ground stations via satellite: The two-mode entangled state of modes~1 and 2 (modes~3 and 4) is initially owned by Alice (Bob). Mode~1 (mode~4) is kept by Alice (Bob) and mode~2 (mode~3) is then transmitted over the uplink to the relay satellite. The received modes~$2''$ and $3''$ (where the $''$ indicates that the
modes have now incurred losses) are mixed through a balanced beam splitter and the $\hat q$ quadrature of one of the output modes and the $\hat p$ quadrature of the other one are measured by two homodyne detectors. The classical outcome of the Bell measurement is then communicated to Alice and Bob. As a result, there would exist an entangled state shared between modes~1 and 4.}\label{fig:swapping}
    \end{center}
\end{figure}

As a result of the entanglement distribution in each quantum communication scheme of Fig.~\ref{fig:satellite-based-confs}, there would exist an entangled state shared between Alice and Bob.
Once the entangled states have been shared between the stations, for each scheme of Fig.~\ref{fig:satellite-based-confs}, Alice and Bob are able to invoke CV-QKD protocols in the EB scheme by applying homodyne or heterodyne detection of their own modes. The level of entanglement produced by the quantum communication schemes considered here as well as the quantum key rates of the EB CV-QKD protocols in these schemes have recently been analyzed in \cite{Neda1, Neda2, Neda3, Neda4, Neda5}.

In the schemes (a), (b), and (c) of Fig.~\ref{fig:satellite-based-confs} the entangled source originates from one of the trusted parties (Alice). However, in the scheme (d) of Fig.~\ref{fig:satellite-based-confs} the entangled source originates from the satellite, which in some circumstances may be controlled by the eavesdropper, Eve. In \cite{Weedbrook2013}, it has been shown that in the context of the EB CV-QKD protocols Alice and Bob can still generate a secure key, even when Eve controls the entanglement source.

\subsection{Measurement-device-independent QKD protocols}

In the scheme (e) of Fig.~\ref{fig:satellite-based-confs} the entangled source originates from both trusted parties (Alice and Bob), however, the Bell measurement at the satellite may be controlled by Eve. In \cite{Nature}, it has been  demonstrated that in CV-QKD protocols the secret key to be shared between the two trusted parties can be generated by the measurement of an untrusted intermediate relay. In measurement-device-independent (MDI) protocols of QKD \cite{first_MDI1, 1st_DV_MDI_theory, Nature}, Alice and Bob are not connected by direct links, and an intermediate relay is used for completing the communication link. In MDI protocols the measurement device is the intermediate relay, whose operation may be controlled by an adversary.  Fig.~\ref{fig:swapping} is in fact one example of a scenario over which a MDI protocol may be implemented.

The security of CV-MDI protocols is usually analysed using  EB schemes that invokes CV entanglement swapping at the relay similar to that shown in Fig.~\ref{fig:swapping} Although CV-MDI protocols are practically implemented in a PM scheme (see below).

In the EB equivalent of the Gaussian MDI-QKD protocols, a pair of TMSV states associated with the quadrature variance of $v = \cosh \left( {2r} \right)$ (where $r$ is the two-mode squeezing), is initially owned by Alice and Bob. One mode of each entangled state is held by Alice and Bob, while the second mode of each state is transmitted to the intermediate relay over the insecure channel. The received modes are swapped via a CV Bell measurement at the intermediate relay. The swapping process continues by the relay communicating the Bell measurement result through a classical public channel to Alice and Bob. After receiving the Bell measurement outcome, Bob displaces his mode, while Alice keeps her mode unchanged. Then Alice and Bob measure their modes by homodyne (or heterodyne) detectors to create correlated data. After the establishment of a sufficiently large amount of correlated data, Alice and Bob proceed with the classical post-processing over an authenticated public channel to create a secret key.

In the EB scheme of the Gaussian MDI-QKD protocols, if Alice and Bob apply a homodyne detection of their modes, the scheme becomes equivalent to the PM scheme, in which Alice and Bob prepare squeezed states, and if Alice and Bob apply a heterodyne detection of their modes, the scheme becomes equivalent to the PM scheme in which Alice and Bob prepare coherent states. We discus these PM schemes next.

%\subsubsection{Gaussian MDI-QKD protocols}
The MDI implementation of Gaussian CV-QKD protocols in the PM scheme depends on whether the Gaussian resource is a squeezed or a coherent state. If a squeezed state, Alice  prepares here mode in a squeezed state with the quadrature variance $v = \exp (2{r_s})$, where $r_s$ is the single-mode squeezing. Which one of the two quadratures is to be squeezed   is based on a randomly generated bit. The chosen quadrature is then modulated by a random Gaussian-distributed variable with zero mean and variance $v_m$ conditioned on $v_m=v-1/v$.  The same procedure is applied independently at Bob's side. If the Gaussian resource is a coherent state, Alice  prepares her coherent-state mode with each quadrature  independently modulated by a random Gaussian-distributed variable having zero mean and variance of $v'_m$. Likewise Bob.

Following transmission to the satellite of the modes belong to Alice and Bob, and irrespective of the Gaussian resource used, the satellite makes a CV Bell measurement on each mode pair, announcing the results. Alice and Bob undertake some modification of their data based on these results and undergo some classical post-processing to end up with a shared key. More details of this process can be found in \cite{Neda4}.

Note the modulation variance $v'_m$ (in the protocol using coherent states) can reach very high values, e.g., $v'_m=60$ \cite{Nature}. With the use of squeezed states, however, achieving  high values of squeezing reamins experimentally challenging. As such, quadrature variance $v$ and of the modulation variance $v_m$ are limited in the range of values attained. Note that $v=5.05$ is equivalent to the two-mode squeezing of 10 dB \cite{sque10}. Note also that  vacuum squeezing at 15 dB is currently the highest obtainable in any experiment \cite{sque15}.

Previous contributions on MDI-QKD protocols have mainly been focussed on fixed-attenuation channels \cite{Nature, CV_MDI1, CV_MDI2, CV_MDI3, CV_MDI_sym, CV_MDI4, fiber_DV1, fiber_DV2, fiber_DV3, fiber_DV4, MDI2016, MDI2016-404}. In \cite{Neda4}, a MDI implementation has been investigated in order to establish Gaussian CV-QKD protocols between two ground stations, where the communication occurs between the ground stations via a LEO satellite over a pair of independent atmospheric channels. In this CV-MDI protocol the measurement device is the satellite itself, which can be controlled by an adversary. In \cite{Neda4}, it has been demonstrated that while the CV-MDI protocol is only feasible for low-loss fixed-attenuation channels, the protocol is capable of achieving a beneficial secure key rate even for transmission over high-loss atmospheric channels. Note that in MDI-QKD the devices of Alice and Bob have to be trusted \cite{Nature, CV_MDI1, CV_MDI2, CV_MDI3, CV_MDI_sym, CV_MDI4, fiber_DV1, fiber_DV2, fiber_DV3, fiber_DV4, MDI2016, MDI2016-404}. Nonetheless, it has recently been shown that QKD is possible even when the device of one of the parties is untrusted \cite{DI1, DI2, DI3}. The security of this one-sided device-independent protocol using CV quantum states has recently been investigated both theoretically and experimentally \cite{CV-DI1, CV-DI2}.

We note that MDI protocols represent a step closer to full device-independent protocols. These latter protocols are based on Bell violation measurements at the receivers, and represent the most robust form of QKD (the form that requires the least number of assumptions). Although some work has been carried out in relation to CV states in device independent QKD (e.g. \cite{violation}), practical progress is limited due to the very low key rates expected. CV MDI-QKD protocols, with their reduced assumptions on how the measurement device must operate, currently represent the most robust form of QKD that still lead to reasonable key rates. The MDI protocols remain unconditionally secure in their generation of keys - the best an adversary in charge of the measurement device can do is drive the key rate to zero (e.g. by broadcasting false Bell measurement results).

\subsection{Entanglement determination and quantum key rate computation}
The evolution of quantum states as they prorogate through atmospheric fading channels can be considered in two different scenarios. In the first scenario, the  transmission coefficient $\eta$ of the atmospheric fading channel is unknown, while in the second scenario it is known. In this latter scenario, it is assumed that the transmission coefficient can be measured in real time at the receiver.
%and we consider the ensemble-average state at the output of the atmospheric channel

\subsubsection{Scenario 1. The transmission coefficient of the fading channel is unknown}

Here, we consider the distribution of a two-mode entangled state over satellite-based atmospheric fading channels. In fact, we assume that the transmitter initially possesses a two-mode (mode~1 and mode~2) entangled state $\hat \rho$, with one (or more) of the modes transmitted to the receiving station(s) through atmospheric fading channels. This leads to two operational settings.

\emph{Single-mode transfer}: In this setting we assume that mode~1 of $\hat \rho$ remains at the ground station (satellite), while mode~2 of $\hat \rho$ is transmitted to the satellite (ground station) over the fading uplink (downlink) characterized by the probability distribution ${p\left( \eta  \right)}$ and the maximum transmission coefficient of $\eta _0$. The density operator of the two-mode state at the ground station and satellite for each realization of the transmission coefficient $\eta $ is given by ${\hat \rho' }(\eta )$. Since $\eta $ is a random variable, the elements of the total density operator of the resultant mixed state ${\hat \rho'_t}$ are calculated by averaging the elements of the density operator ${\hat \rho'}(\eta )$ over all possible transmission coefficients of the fading channel, giving the ensemble-averaged state of \cite{Neda3}
\begin{equation}\label{AA1}
\hat \rho' _t = \int_0^{{\eta _0}} {p\left( \eta  \right){\hat \rho' }\left( \eta  \right)\,d\eta }.
\end{equation}
Now, let us consider the initial two-mode entangled state $\hat \rho$ at the transmitter being a Gaussian state \cite{Dong, Usenko, Neda1, Neda2, Bohmann1}. In this case the resultant ensemble-averaged state ${\hat \rho'_t}$ is a non-Gaussian mixture of the Gaussian states $\rho'(\eta )$ obtained for each realization of $\eta $. Since the resultant ensemble-averaged state shared by the ground station and the satellite is a non-Gaussian state, it cannot be completely described by its first and second moments. Therefore, the final entanglement computed based on the covariance matrix of the resultant ensemble-averaged state will represent only the Gaussian entanglement between the ground station and the satellite, but not the total distributed entanglement \cite{Dong, Usenko, Neda1, Bohmann1}. In order to calculate the total shared entanglement between the stations, the entanglement has to be computed based on the density operator of the resultant ensemble-averaged state \cite{Neda3}.

Note that if we use the shared entanglement created for subsequent use in QKD, i.e. a EB CV-QKD protocols operating over atmospheric fading channels\footnote{Note that in \cite{fast-fading}, a fast-fading channel has been considered where the users are only able to estimate the probability distribution of the channel's transmission coefficient but not its instantaneous values, while the eavesdropper has full control of the fast-fading channel, so that she chooses the instantaneous transmission coefficient of the channel.}, then the same concept (use of ensemble averaged states) is invoked when the quantum key rate is calculated. Note that when the quantum key rate is in fact calculated based on the covariance matrix of the resultant ensemble-averaged state ${\hat \rho'_t}$, the key rate computed is only related to the Gaussian component of ${\hat \rho'_t}$ \cite{Neda2}.

\emph{Two-mode transfer}: In this setting we assume that the satellite initially possesses a two-mode entangled state $\hat \rho$, with mode~1 transmitted to ground station~1 over a fading downlink obeying the probability distribution of ${p_1}({\eta _1})$ and having the maximum transmission coefficient of $\eta _{01}$, while mode~2 is transmitted to ground station~2 over a different fading downlink characterized by the probability distribution ${p_2}({\eta _2})$ and having the maximum transmission coefficient of $\eta _{02}$. Here, the two fading downlinks are assumed to be independent. The density operator of the two-mode state at the ground stations for each realization of the transmission coefficients ${\eta _1}$ and ${\eta _2}$ is given by ${\hat \rho'}({\eta _1},{\eta _2})$. The elements of the total density operator of the resultant mixed state ${\hat \rho'_t}$ are calculated by averaging the elements of the density operator ${\hat \rho' }({\eta _1},{\eta _2})$ over all possible transmission coefficients of the two separate fading channels, giving the ensemble-averaged state of \cite{Neda3}
\begin{equation}\label{AA2}
\hat \rho '_t = \int_0^{{\eta _{01}}} {\int_0^{{\eta _{02}}} {{p_1}\left( {{\eta _1}} \right){p_2}\left( {{\eta _2}} \right){\hat \rho' }\left( {{\eta _1},{\eta _2}} \right)\,d{\eta _1}d{\eta _2}} }.
\end{equation}
 %Here again, when the initial two-mode entangled state $\hat \rho$ at the sending station is a Gaussian state, the resulting ensemble-average state ${\hat \rho'_t}$ is a non-Gaussian mixture of the Gaussian states obtained for each realization of ${\eta _1}$ and ${\eta _2}$.

\subsubsection{Scenario 2. The transmission coefficient of the fading channel can be measured}

Let us now assume a  modified scenario, in which the variable transmission coefficient of the atmospheric fading channel is measured with the aid of a separate coherent signal. For example, when a local oscillator in a polarized mode orthogonal to the signal is sent through the channel. Although this increases the complexity of the system, the grade of entanglement (and hence the quantum key rate of the EB CV-QKD protocols implemented based on this entanglement) generated between the stations will be increased.

When considering this  scenario in the single-mode transfer setting where the transmission coefficient $\eta$ is measured at the receiving station, the final entanglement can be calculated as \cite{Neda3}
\begin{equation}\label{total-E1}
E = \int_0^{{\eta _0}} {p\left( \eta  \right)} \,E\left[ \rho'(\eta ) \right]\,d\eta,
\end{equation}
where $E\left[ \rho'(\eta)  \right]$ is the grade of entanglement of a state received through the channel of transmission coefficient $\eta$.

In this scenario, when the initial two-mode entangled state $\hat \rho$ at the transmitter is a Gaussian state, the mixed states $\rho'(\eta )$ collected at the receiver during each transmission coefficient window remain Gaussian, because within each (small) fading bin we can assume that the transmission coefficient is constant and therefore the states during that particular bin remain Gaussian. In this case, the grade of entanglement of the mixed Gaussian state $\rho'(\eta) $ i.e., $E\left[ \rho'(\eta)  \right]$ can be calculated based on the covariance matrix of $\rho'(\eta) $, which results in $E$ of Eq.~(\ref{total-E1}) representing the total entanglement shared between the stations \cite{Neda3}.
%within each (small) bin we can assume the transmission coefficient is constant and therefore the states in that particular bin are Gaussian. In fact,

Considering this scenario in the EB CV-QKD protocols communicating over atmospheric fading channels, which are implemented based on the shared entangled states between the stations, the same concept is true when the quantum key rate is calculated. In fact, due to the relatively long coherence time of the atmospheric channel, it may be possible to devise a scheme, in which quantum key rates are derived for each realization of the fading (each fading bin realized), and summed \cite{Neda3,Neda4,Neda5,Pirandola-fundamentallimit}. Indeed, the quantum key rate $K\left[ \rho'(\eta)  \right]$ resulting from the mixed Gaussian state $\rho'(\eta) $ can be calculated based on the covariance matrix of $\rho'(\eta) $, and then the total key rate shared between the stations is calculated by $K = \int_0^{{\eta _0}} {p\left( \eta  \right)} \,K\left[ \rho'(\eta ) \right]\,d\eta$ \cite{Neda3,Neda4,Neda5}.

Similarly, considering this  scenario in the two-mode transfer setting, where the transmission coefficients $\eta_1$ and $\eta_2$ are measured at the two receiving stations, the final grade of entanglement can be calculated as \cite{Neda3}
\begin{equation}\label{total-E2}
E = \int_0^{{\eta _{01}}} {\int_0^{{\eta _{02}}} {{p_1}\left( {{\eta _1}} \right){p_2}\left( {{\eta _2}} \right)E[{\hat \rho' }\left( {{\eta _1},{\eta _2}} \right)]\,d{\eta _1}d{\eta _2}} },
\end{equation}
where $E[{\hat \rho' }\left( {{\eta _1},{\eta _2}} \right)]$ is the entanglement of a state that has traversed two channels having the transmission coefficients of $\eta_1$ and $\eta_2$ \cite{Neda3,Neda4,Neda5}.

%In the context of entanglement distribution, the two key performance indicators in this chapter will be $E_{LN}$, the entanglement (logarithmic negativity); and ${R_E}$, the entanglement-generation rate. $E_{LN}$ represents the entanglement generated between two stations following the transfer of a pulse through the lossy fading channel(s). When one mode is retained by the sender, the pulse refers to the second mode sent between the sender and receiver. When one mode is sent to one receiver and the other mode sent to a different receiver, the pulse refers to the two modes collectively. ${R_E}$  encapsulates directly the probability of creating the initial state. Introducing ${P_c}$, the creation probability of the initial state (we adopt $P_c=1$ for TMSV states), we have $R_E = {P_c}\,E_{LN}$.

\subsection{Enhancement of quantum communication performance}

Satellite-based communication channels tend to suffer from high uplink losses on the order of 25-30 dB (and beyond) for a LEO satellite receiver  \cite{fso,r7,satellite-survey}, while single downlink channels are anticipated to have losses of 5-10 dB for a LEO satellite transmitter \cite{fso,r7,satellite-survey}. Under such high losses, entanglement distribution and QKD via satellite will remain a fruitless endeavor without the beneficial intervention of the post-selection strategy \cite{Usenko} and entanglement distillation techniques \cite{Dong} detailed below. %The post-selection strategies which occur at the receiving station can be based on classical measurements of the channel transmission coefficient, or on quantum measurements.

\subsubsection{Post-selection}
Although atmospheric fading degrades both the  entanglement and the quantum key rate, its effects may be mitigated. Post-selection of high transmission-coefficient windows, as introduced in \cite{Usenko} for the case of a single point-to-point fading channel, is capable of improving both the entanglement and the quantum key rate. To elaborate a little further, in the post-selection strategy, a subset of the channel transmittance distribution, namely that associated with the high transmission coefficient, is selected to contribute to the resultant post-selected state and to the post-selected key rate.

To elaborate on the post-selection strategy, in addition to the quantum states, coherent (classical) light pulses are transmitted through the channel in order to estimate the channel's transmission coefficient $\eta$ at the receiver. The received quantum state is either retained or discarded, conditioned on the channel's transmission coefficient being higher or lower than the post-selection threshold ${\eta_{th}}$. Although this post-selection strategy can be invoked for enhancing the grade of entanglement and the quantum key rate between the transmitter and receiver, estimation of the channel's transmission coefficient will impose additional complexity on both the transmitter and receiver. The operation of this form of post-selection in the scheme (c) of Fig.~\ref{fig:satellite-based-confs} has been invoked in \cite{Neda1} for enhancing the grade of Gaussian entanglement and in \cite{Neda2} for increasing the quantum key rates between the ground stations.

\subsubsection{Entanglement distillation}

The other strategy, which can be used in order to enhance the grade of entanglement between the transmitter and receiver is entanglement distillation that is based on quantum measurement techniques without relying on channel estimation. Entanglement distillation represents the protocol of extracting a subset of states with a higher degree of entanglement from an ensemble of entangled states \cite{Distillation1}. In fact, entanglement distillation may be viewed as a purifying protocol that selects highly entangled pure states from a set of entangled states that have become mixed as a result of imperfect transmission \cite{Distillation2,Distillation3,Distillation4,Distillation5}. It has been shown that if the entangled states are Gaussian, entanglement distillation cannot be performed using only Gaussian operations carried out by linear optical components, such as beam splitters and phase shifters, homodyne detection and classical communication \cite{no_go1,no_go2,no_go3}. However, when the Gaussian entangled states are transmitted through a fading channel, the state at the output of the channel is a non-Gaussian mixed state (a non-Gaussian mixture of Gaussian states), and therefore the aforementioned no-go theorem does not apply. In \cite{Dong}, a method has been proposed for distilling entanglement from (initially) Gaussian entangled states received over a single point-to-point fading channel. This is achieved by carrying out a weak measurement (based on a beam splitter and a homodyne detector)  applied to the received non-Gaussian  mixed state. The entanglement distillation is implemented at the receiver by extracting a small portion of the received mixed state using a tap beam splitter. A single quadrature (for instance, the ${\hat q}$ quadrature) is then measured by applying homodyne detection to the tapped beam. If the measurement outcome is above the threshold value ${q_{th}}$, then the remaining state is retained, otherwise it is discarded. The operation of this form of entanglement distillation in the scheme (c) of Fig.~\ref{fig:satellite-based-confs} has been invoked in \cite{Neda1} for enhancing the Gaussian entanglement between the ground stations (which consequently leads to an improvement in the quantum key rates of the EB CV-QKD protocols).

Note that when  entangled states are conveyed over a fading channel, both the above-mentioned post-selection and entanglement distillation strategies act as ``Gaussification'' methods in the sense that the resultant conditioned states approach a Gaussian form due to the enhanced concentration of low-loss states in the final ensemble-averaged state. Note also that using the above-mentioned post-selection and entanglement distillation strategies, the entanglement established between the transmitter and receiver is only probabilistically increased.

Another entanglement distillation technique for is based on applying an initial non-Gaussian operation to the Gaussian entangled states (that again increases the entanglement probabilistically), which is followed by a Gaussification step that iteratively drives the output non-Gaussian state towards a Gaussian state. Non-deterministic noiseless linear amplification has been identified as a method of distilling Gaussian entanglement \cite{NLA1,NLA2,NLA-Distil1,NLA-Distil2,NLA-Distil3,error-correction7,NLA-QKD-F,NLA-QKD-B,NLA-QKD-W}. It has been shown that the non-deterministic noiseless linear amplification is capable of distilling improved CV entanglement \cite{NLA-Distil1,NLA-Distil2,NLA-Distil3} and enhancing CV-QKD performance \cite{NLA-QKD-W,NLA-QKD-F,NLA-QKD-B}, when applied after the lossy channel to the quantum states received. The non-Gaussian operations which result in the generation of non-Gaussian entangled states will be discussed in detail in the next section.

% Note that use of other entanglement distillation schemes (which are based on quantum measurement) such as the scheme of measurement-based noiseless linear amplification has been studied in e.g., \cite{Distil1, Distil2}, but we will not consider such distillation schemes here.

\section{Non-Gaussian CV quantum communication over atmospheric channels}

In the CV domain, previous efforts invested in entanglement distribution and QKD over atmospheric channels have been predominately focussed on Gaussian states \cite{Elser,Heim-2009,Usenko,Bohmann1,Heim,R4,Neda1,Neda2,Neda4}. Although Gaussian quantum states are well understood both from a theoretical and from an experimental perspective \cite{rr2,Weedbrook2012,Adesso}, the employment of CV non-Gaussian quantum states\footnote{Note that only pure states having a positive Wigner function are Gaussian states. However, the Wigner function of non-Gaussian pure states takes on negative values.} for quantum communication has also garnered interest \cite{1st_PSS, Kitagawa, 1st_PAS, telep-nG, Zhang-Loock, Navarrete, Oxford2013, added, ME, Seshadreesan, Oxford, Allegra, Adesso-proof, Sabapathy, Filippov, Amp1, Amp2, nG-modulation, nG1, nG2, nG-coherent}. Non-Gaussian quantum states are valuable resource for a range of protocols, including teleportation \cite{1st_PSS, Kitagawa, 1st_PAS, telep-nG, Zhang-Loock, ME, Seshadreesan, Oxford}, cloning \cite{Amp1, Amp2} and CV-QKD protocols \cite{nG-modulation, nG1, nG2, nG-coherent}. For two important reasons,  entangled non-Gaussian states are particularly interesting in the context of quantum communication via satellite. The first of these reasons is that the distillation of Gaussian entanglement is impossible using only Gaussian operations \cite{no_go1,no_go2,no_go3}. However, mixed non-Gaussian states can undergo entanglement distillation without any additional requirements. The second reason is that, relative to Gaussian entanglement,  non-Gaussian entanglement can be shown in some circumstances to be more robust against decoherence  \cite{Sabapathy, ME, Filippov}.

\subsection{Non-Gaussian entangled states}\label{non-Gaussian}

%For non-Gaussian entangled states we consider photon-subtracted squeezed (PSS) states \cite{1st_PSS, Kitagawa, telep-nG, Zhang-Loock, Seshadreesan, Oxford, Oxford2013, Navarrete}, photon-added squeezed (PAS) states \cite{1st_PAS, telep-nG, Oxford, Navarrete, added}, photon-replaced squeezed (PRS) states \cite{ME, Oxford}, and NOON states \cite{NOON0, NOON1, NOON2}. Such states cover a wide range of the non-Gaussian state possibilities, and represent the  non-Gaussian states most likely to be used in future quantum application and communication  deployment scenarios.

CV non-Gaussian states are mostly generated by applying non-Gaussian operations, such as photon subtraction \cite{1st_PSS, Kitagawa, telep-nG, Zhang-Loock, Seshadreesan, Oxford, Oxford2013, Navarrete}, photon addition \cite{1st_PAS, telep-nG, Oxford, Navarrete, added} and photon replacement \cite{ME, Oxford} to incoming Gaussian states.  We discuss here non-Gaussian entangled states which are created probabilistically by applying non-Gaussian operations to  (i.e at the receiver) Gaussian TMSV states. Note that a non-Gaussian operation can be applied to either a single mode, or to both modes, of the incoming Gaussian entangled state. Also note the non-Gaussian operation can be applied to the incoming mode at the sender (i.e. incoming from the local TMSV production site), or at the receiver side (after propagation through the atmosphere). Unless otherwise stated, we will consider  the former process in the following.

For the generation of an entangled photon-subtracted squeezed (PSS) state \cite{1st_PSS, Kitagawa, telep-nG, Zhang-Loock, Seshadreesan, Oxford, Oxford2013, Navarrete}, each mode of an incoming TMSV state interacts with a vacuum mode in a beam splitter. One of the outputs of each beam splitter feeds a photon number resolving detector. When both detectors simultaneously register $k$ photons, which are considered to be non-Gaussian measurements, a pure non-Gaussian state is heralded with a probability of $0<P_{sb}<1$. This photon-subtraction operation is shown in Fig.~\ref{fig:non-Gaussian-operations}(a) for $k=1$. A PSS state can also be generated by applying the photon subtraction technique described above to a single mode of the TMSV state \cite{Oxford}. The generation of non-Gaussian states via photon subtraction as described above has been experimentally demonstrated in \cite{exp0, exp1, exp2}. Note that in the photon-subtraction operation, other types of photon detectors such as on/off photon detectors (which only distinguish the presence and absence of photons, and are considered a non-Gaussian measurement) can also be used for generating a PSS state from a TMSV state \cite{Kitagawa, Zhang-Loock}. In this case the non-Gaussian output state is a mixed state.

An entangled photon-added squeezed (PAS) state \cite{1st_PAS, telep-nG, Oxford, Navarrete, added} is generated by adding a single photon to each mode of a TMSV state. This single-photon addition is performed at a beam splitter, as shown in Fig.~\ref{fig:non-Gaussian-operations}(b), with one of the outputs of each beam splitter being detected by an on/off photon detector. A pure non-Gaussian state is then generated (with a probability of $0<P_{ab}<1$) when a vacuum state is registered in both detectors simultaneously. Note that the final creation probability of a PAS state is obtained by multiplying $P_{ab}$ by the  probability of creating the two additional photons required. A PAS state can also be generated by applying the photon addition technique described above to a single mode of the TMSV state \cite{Oxford}. Note that the addition of single photons to coherent states and to thermal states of light has been experimentally realized in \cite{added_exp1, added_exp2}.

By contrast, an entangled photon-replaced squeezed (PRS) state \cite{ME, Oxford} is generated according to Fig.~\ref{fig:non-Gaussian-operations}(c), where each mode of a TMSV state interacts with a single photon in a beam splitter, with one of the outputs of each beam splitter being detected by a photon number resolving detector. When both detectors register a single photon simultaneously, a pure non-Gaussian state is heralded with a probability of $0<P_{rb}<1$. The final creation probability of a PRS state is obtained by multiplying $P_{rb}$ by the  probability of creating the two additional photons required. A PRS state can also be generated by applying the photon replacement process described above to a single mode of the TMSV state \cite{Oxford}.

\begin{figure}
    \begin{center}
   {\includegraphics[width=3 in]{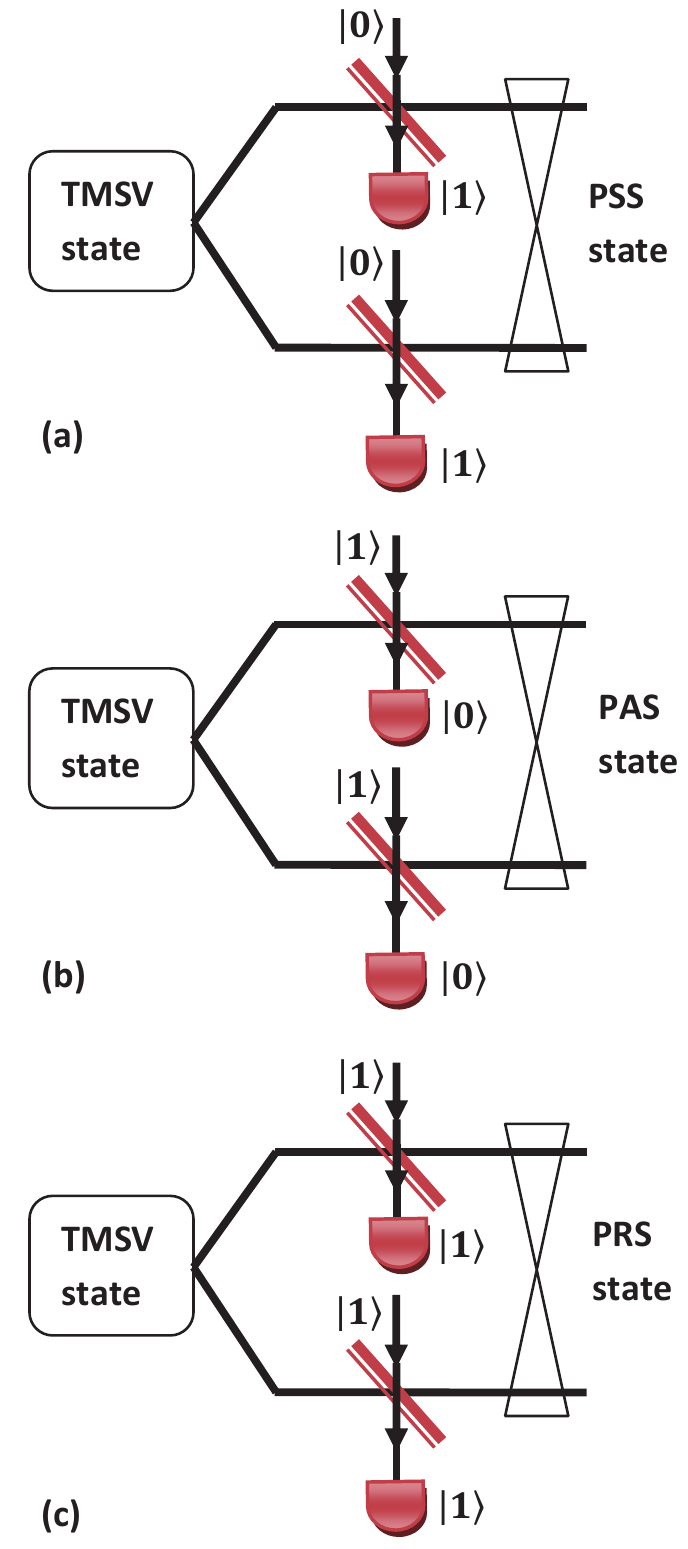}}
    \caption{Implementation of non-Gaussian operations on the Gaussian TMSV state; (a) photon subtraction, (b) photon addition, and (c) photon replacement.}\label{fig:non-Gaussian-operations}
    \end{center}
\end{figure}

\subsection{Evolution of non-Gaussian entangled states over a lossy channel}

Unlike Gaussian states, the evolution of non-Gaussian states cannot be analysed solely through the covariance matrix. Previous contributions have analysed the evolution of non-Gaussian states for transmission over fixed-attenuation channels relying on the so-called Master equation approach of \cite{Allegra}, the characteristic function approach of \cite{ME} or the Kraus operator approach of \cite{Sabapathy}. Here we discuss the general approach of Kraus representation \cite{Kraus} of the channel in order to directly analyze the evolution of the entangled states (Gaussian or non-Gaussian) through the channel. Considering a quantum state associated with the density operator ${\hat \rho _{in}}$ as the input of a trace-preserving\footnote{In a trace-preserving channel, the trace of the density operator is preserved, which means the trace of the output density operator of the channel remains one.} completely positive channel, the output density operator of the channel can be described in an operator-sum representation of the form
%\begin{eqnarray}%\label{k1}
${\hat \rho _{out}} = \sum\nolimits_{\ell  = 0}^\infty  {{G_\ell }{\hat \rho _{in}}\,G_\ell ^\dag }$,
%\end{eqnarray}
where the Kraus operators ${G_\ell }$ satisfy $\sum\nolimits_{\ell  = 0}^\infty  {{G_\ell }\,G_\ell ^\dag }  = I$, with $I$ being the identity operator. In \cite{Kraus}, the Kraus operators of a wide range of channels including a fixed-attenuation channel subject to vacuum noise (i.e., $V_n=1$ in Fig.~\ref{fig:channel representation}) are given. In \cite{Sabapathy}, the Kraus operators of a fixed-attenuation channel subject to vacuum noise but with additional Gaussian noise is given. The results of \cite{Kraus} have been generalized to a fixed-attenuation channel subject to thermal noise (i.e., $V_n>1$ in Fig.~\ref{fig:channel representation}) in \cite{Neda6}.

\subsection{Entanglement determination and quantum key rate computation}

Following the evolution of pure non-Gaussian states over the lossy channel(s), the quantum state of the channel output is a non-Gaussian mixed state. In general it is not possible to analytically determine the total grade of entanglement of the mixed non-Gaussian states after transmission over a lossy channel. Since the grade of entanglement is determined by the output density operator $\hat \rho _{out}$, which possesses an infinite number of elements, a numerical method is required for approximating the matrix $\hat \rho _{out}$ by its truncated-dimensional version, as discussed in \cite{Kitagawa, Neda3,Neda5,Neda6}, whilst ensuring that the trace of the truncated matrix is close to 1.

%limiting its size, i.e., creating a truncated $\hat \rho _{out}$. This matrix truncation (i.e., size limitation) has been conducted by setting some appropriate cutoffs in \cite{Kitagawa, Neda3,Neda5,Neda6}, where the final check on the truncated matrix (the cutoff values) is that the trace of the truncated matrix is very close to 1.

Given the non-deterministic nature of the non-Gaussian operations, in the context of non-Gaussian entanglement distribution, there are two key performance indicators, namely the grade of entanglement $E$ between two stations following the transmission of a pulse through the lossy channel(s), and the entanglement-generation rate $R_E$, where we have $R_E = {P_c}\,E$, with ${P_c}$ being the creation probability of the initial non-Gaussian state. The evolution of a wide range of non-Gaussian entangled states in both single-mode and two-mode transfer over atmospheric fading channels has been investigated both when the transmission coefficient of the atmospheric fading channel is unknown and when it is estimated in real time \cite{Neda3}.
The work of \cite{Neda3} considered operational scenarios where the non-Gaussian entangled states transmitted through the atmospheric channel are created ``just-in-time'' via non-Gaussian operations applied to the Gaussian entangled input states that would otherwise be transmitted directly over the communication channel. In this scenario transmitting the incoming Gaussian state directly over the atmospheric channel would be the best option in terms of maximizing the \emph{entanglement-generation rate}. However, if the transmission rates of all the states through the channel could be equalized for example with the aid of quantum memory (see \cite{Neda3} for more details), some non-Gaussian states lead to enhanced \emph{entanglement} transfer relative to that obtained by Gaussian state transfer.

The performance of CV-QKD protocols has been analysed in \cite{Neda5} for transmission over atmospheric fading channels, where the source is constituted by PSS states in the context of EB CV-QKD protocols. In \cite{Neda5}, one mode of the PSS state remains at the ground station (satellite), while the other photon-subtracted mode is transmitted to the satellite (ground station) over the fading uplink (downlink) channel characterized by the probability distribution $p\left( \eta  \right)$ and maximum transmission coefficient of $\eta_0$. When the transmission coefficient of the atmospheric channel can be measured in real time, after acquiring each realization of $\eta $, the key rate ${K(\eta )}$ is calculated based on the covariance matrix of the mixed non-Gaussian state at the output of the channel. The final key rate is then computed as $K = P_c\int_0^{{\eta _0}} {K(\eta )} p(\eta )\,d\eta $ in units of bits per pulse, with $P_c$ being the creation probability of the initial non-Gaussian entangled state. The resultant key rate represents a lower bound on the actual key rate of the CV-QKD protocol. However, to determine the actual resultant key rates (not just its lower bounds), ${K(\eta )}$ must be computed based on the density operator of the mixed non-Gaussian output state.% which is an open problem in the security analysis of non-Gaussian CV-QKD protocols.

In \cite{Neda3,Neda5} the non-Gaussian operations are first applied to the initial Gaussian states, with the resultant non-Gaussian states being transmitted through the atmospheric fading channel. An alternative approach would be to transmit the initial Gaussian states through the atmospheric channel, and then apply the non-Gaussian operations after the atmospheric channel to the quantum states received. In \cite{ME}, the distillation of CV entanglement using a coherent superposition-based non-Gaussian operation has been studied, where the non-Gaussian operation is the superposition of the photon subtraction and of the photon addition operations, and where the non-Gaussian operation is applied either before or after a fixed-attenuation channel. %In \cite{Distil2,After2} it has been shown that non-deterministic noiseless linear amplification \cite{Distil1} is capable of distilling improved CV entanglement, when applied after the lossy channel to the quantum states received.

\section{Comparison with Discrete-variable technologies}\label{CV and DV}
The family of DV systems invoked for satellite-based quantum communications constitutes an alternative technology, which has been deployed in Micius \cite{China1,China2,China3}. In space-based deployment, a range of pragmatic issues comes into play when considering the pros and cons of DV \emph{vs.} CV implementations. Perhaps the strongest argument in favour of DV systems in the space-based context is that photon losses have a less grave impact on quantum information processing in DV systems. In CV systems the photon losses in the channel introduce vacuum noise, leading to a  reduction in the correlation between Alice and Bob's data. By contrast, in DV systems, photon losses reduce the communication efficiency, but they do not trigger a false single-photon detection event. A photon is either lost in the channel, in which case Bob does not register anything, or it is simply detected at Bob's detector. In high-loss scenarios, this effect can lead to advantages for DV systems. However, this benefit may by outweighed by other considerations, as discussed briefly below. More details on  satellite-based DV quantum communication can be found elsewhere, for example in \cite{satellite-survey}.

The performance of DV-QKD~\cite{[8]} is limited both by the difficulty of single-photon generation, as well as by the expense of single-photon detectors. It is a challenge to construct a true single-photon source owing to  implementation challenges. Alternatively, single-photon sources can be approximated using an attenuated laser (weak coherent state pulses) \cite{WCP1,WCP2}. By contrast, CV-QKD systems rely on low-cost implementations and are potentially capable of supporting  higher key rates than DV-QKD systems. Recall that CV-QKD can be implemented by modulating both the amplitude and phase quadratures of a coherent laser and can be subsequently measured in the receivers using homodyne detectors, which operate faster and more efficiently than the single-photon detectors. Moreover, CV-QKD systems are more compatible with  standard telecommunication encoding, transmission and detection techniques. All these advantages potentially allow CV-QKD protocols to achieve higher secret key rates than DV-QKD systems. %and to be implemented in a relatively easier fashion.

Furthermore, the single-photon detectors of DV systems are very
sensitive to background light sources. By contrast, the homodyne
detectors used for CV systems offer beneficial robustness to
background light. Indeed, an explicit advantage of using a local
oscillator is that it has an `automatic' spectral-domain filtering
effect. Consequently, homodyne detectors remain to a large extent
unimpaired in daylight conditions without the extra filtering that are
needed by the single-photon detectors \cite{Elser}. Furthermore, in CV
systems, a tapped component of the local oscillator can be simply
obtained and measured, thereby allowing for \emph{direct} monitoring
of atmospheric fluctuations effects, such as beam wandering (which can
then be compensated for using adaptive optics
\cite{Elser,Heim-2009,Heim}).

Nonetheless, the issue of whether DV or CV systems should be deployed as the quantum information carrier in space-based quantum communications remains very much an open issue at the time of writing. Ultimately, it could well be that hybrid DV$+$CV architectures, accommodating time-variant atmospheric conditions, turn out to be the most beneficial in many circumstances. The employment of such hybrid architectures has been extensively studied for example in \cite{hybrid}. %It could well be that future sophisticated deployments of quantum technology in space are based on similar ideas.

\section{Future directions}

Quantum communication via satellite is in its infancy. Building on the early work and verification studies (both experimental and theoretical) of many researchers e.g. \cite{aircraft,moving-patforms,truck,144km,decoy,f2,freespace-entanglement-2005,Peng-2005,freespace-QKD-2006,freespace-QKD-2009, freespace-entanglement-2009,freespace-teleportation100-2012,freespace-teleportation143-2012,daylight-QKD,pp1,pp2,R2, r6,r7,r8,r9,r10,r11, r12,r13,r14,r15,r16,r17,r18, nano-satellite,micro-satellite,China1,China2,China3,proposal-2014,proposal-2008,Boone, IEEE1,IEEE2,IEEE3,IEEE4,satellite-survey,IEEE-s,Bacsardi2,Semonov2016-DV,Semonov2010-DV,Neda1,Neda2,Neda3,Neda4,Neda5, Elser,Heim-2009,Heim,R4,Usenko,Semenov2009,Wander2012,2016,Bohmann1,Bohmann2,Geo-2017},
%\cite{Neda1,Neda2,Neda3,Neda4,Neda5,Semonov2016-DV,Semonov2010-DV,144km,freespace-QKD-2009,freespace-entanglement-2009,Rf2012, freespace-teleportation100-2012,freespace-teleportation143-2012,aircraft,moving-patforms,truck,r10,r17,Elser,Heim2009,Heim,R4},
the pioneering experimental result of the Micius \cite{China1,China2,China3} collaboration has now provided us with the first glimpse of what is truly achievable via space-based platforms. However, there remains much to do before quantum communications via satellites can be considered mainstream. This is especially so in the CV quantum domain, where no space-based deployments have yet been achieved, despite the numerous theoretical studies e.g. \cite{Neda1,Neda2,Neda3,Neda4,Neda5,Elser,Heim-2009,Heim,R4,Usenko,Semenov2009,Wander2012,2016,Bohmann1,Bohmann2}. We briefly mention here some of the research topics within space-based CV quantum communications that we consider of particular interest to any multi-disciplinary engineering community.

\subsection{Channel transmissivity measurements}
The  Micius \cite{China1,China2,China3} data provides us with our first real insight into the channel conditions experienced by quantum states, as they traverse through the turbulent atmosphere, to and from Earth. The measured photonic losses in the downlink \cite{China1,China2} and in the uplink \cite{China3} are now available (the losses in the latter case were a minimum of 41 dB). Leveraging this data for better understanding the channel conditions experienced by CV states as they travel to and from Earth would be an insightful, but costly endeavour. As discussed earlier in Sec.~\ref{CV and DV}, the loss of photons in the CV context fundamentally affects any subsequent information processing, as opposed to the DV case, where photons not received can be simply ignored. Ultimately, the study of how the CV states are affected by the atmosphere reduces to a determination of the statistical distribution of the channel transmissivity. Detailed knowledge of this distribution has wide ranging implications for studies pertaining to non-classical signatures of CV states traversing through atmospheric channels \cite{Bohmann2}, as well as for a host of CV-based applications. The latter outcome is due to the fact that many applications are very sensitive to the channel's transmissivity \cite{Neda1,Neda2,Neda3,Neda4,Neda5}. As discussed previously, beyond the dominant effects of beam wandering and beam broadening, other more subtle effects induced by the atmosphere can play a non-negligible role. These effects include beam deformation, attenuation, absorption and scattering. Sophisticated theoretical studies of these effects are now becoming available, and in general these models are found to be consistent with  terrestrial experiments carried out over a wide range of turbulence conditions \cite{2016,scintillation2016,Bohmann3}. Experimental confirmations of existing turbulence models in the realm of  Earth-to-satellite (and vice versa) channels would be very important. Of particular importance would be a robust validation of the beam-wandering models used for the transmissivity statistics in the Earth-to satellite channels \cite{Wander2012,Usenko,2016}, and the validation of the beam-broadening models expected to dominate the satellite-to-Earth channels \cite{r10}.
\subsection{LDPC codes}
The reconciliation phase of any QKD protocol is perhaps the area of quantum communications most closely associated with classical communications. In the DV scenario, long LDPC codes can be used to correct transmission errors. For scenarios, where DV quantum measurements are mapped directly to binary outcomes, the transmission of bits via a classical binary symmetric channel can be adopted as the underlying model. A range of high-performance LDPC codes which approach reconciliation factors close to 1 in the large key length limit are known for such channels \cite{LDPC-DV1,LDPC-DV2,LDPC-CV1}. However, in the CV setting the extraction of binary information is substantially more involved. Currently, there are two main techniques that are widely adopted in this regard, namely, slice reconciliation \cite{LDPC-CV2, LDPC}, and multi-dimensional reconciliation \cite{LDPC-CV3, LDPC-CV4}. For the low signal to noise ratios (SNRs) routinely anticipated for satellite communications, the  multi-dimensional reconciliation technique is likely to be more appropriate. In this context, multi-dimensional reconciliation via multi-edge LDPC codes is considered by many as the most appropriate path due to the high performance of such codes at low SNRs \cite{LDPC-CV4}.

Nonetheless, numerous open research issues remain. Perhaps the most important of these is constituted by the finite key effects. Much of the work in formally determining the security of a key within QKD systems assumes having an infinite key length. However, in reality, this assumption is never satisfied and the consideration of the finite-length key effects must be analysed. This is an issue that affects both the DV \cite{Finite0} and CV security analyses \cite{Finite1,Finite2,Finite3,Finite4,CV-DI1}. This problem is of particular concern for space-based QKD due to the short transit times of LEO satellites. Hence, the finite-length key processing invoked in the context of CV-QKD conceived for satellites has to be considered. Naturally, this analysis will be strongly dependent on the specific CV-QKD protocol adopted. Finite-length key based analyses of standard coherent state protocols \cite{Finite-coherent}, of MDI protocols \cite{Finite-MDI1,Finite-MDI2} and of full device-independent protocols \cite{Finite-DI} follow quite distinct paths.

Beyond the finite-length effects within the reconciliation decoding phase, the construction of near-capacity adaptive-rate LDPC codes for CV space-based implementations would be useful. Again, these issues are particularly relevant to satellite-based communication due to the time-variant properties of the channel. For LEO satellites we can expect the SNR to exhibit quite rapid variations versus time, as the satellite appears above the horizon and disappears again. Furthermore, for a given set of orbital parameters, we could anticipate the SNR's evolution versus time to be reasonably predictable. Adaptive-rate LDPC codes well suited for counteracting the SNR vs time evolution should be constructed.
The employment of puncturing techniques \cite{adaptive-codes1} used for multi-edge LDPC codes appears to be an appropriate pathway to achieving this \cite{adaptive-codes2}. These studies are only in their early phases of development, hence further research into the design of adaptive-rate codes as a path to low-complexity CV-QKD via satellites is expected to be fruitful. An important focus of such future studies should be the maintenance of high reconciliation efficiencies over the anticipated range of SNRs \cite{adaptive-codes3}.

Finally, we note that in principle other codes beyond LDPC codes  could be used in the CV-QKD reconciliation phase. Currently, however, limited work has been reported in this area. Nonetheless,  we do note some work on turbo codes  \cite{tc_teq_st_2:book} applied to the CV domain, as reported in \cite{DBLP:journals/corr/cs-IT-0406001} (for use of such codes in  the DV domain see \cite{turbo-dv-qkd-2014,turbo-dv-qkd-2017}). Furthermore, polar codes \cite{5075875} have recently been invoked for CV-QKD  in \cite{Jouguet:2014:HPE:2600508.2600516}. These contributions suggest that further performance comparisons using various error correction codes for the CV-QKD reconciliation phase may become fruitful.

%As such, future research efforts on non-LDPC codes in the context of CV QKD may lead to reconciliation efficiencies that, in some scenarios, outperform current offerings.

\subsection{CV quantum error correction codes}
Of special importance for CV quantum communications are the non-Gaussian operations that form the basis of quantum error correction. Such operations are required due to the no-go theorem, stipulating that Gaussian errors cannot be corrected by purely Gaussian operations \cite{no-go-error-correction}. It is possible to build a pathway from standard classical LDPC codes to qubit error correction codes, and then to CV error correction codes. Following on from the original CV error correction protocols of \cite{error-correction1,error-correction2,error-correction3}, there are several examples of CV quantum error correction codes appearing in the recent literature \cite{error-correction4,error-correction5,error-correction6,error-correction7,error-correction8,error-correction9,error-correction10 ,error-correction11}. However, in the context of space-based implementations there is evidence to suggest that direct non-Gaussian measurement at the receiver is likely to be the most fruitful pathway to CV error correction - at least in the short term.

In Sec.~\ref{non-Gaussian} we have discussed a host of non-Gaussian operations in the form of photon subtraction and addition techniques that were used to form our non-Gaussian states, as seen in Fig.~\ref{fig:non-Gaussian-operations}. Such operations can also be used for producing CV entanglement distillation - a form of quantum error correction for CV variables. Photon subtraction and addition techniques are becoming mainstream in laboratories throughout the world and the imminent integration of such techniques directly into future satellite communications is expected. In QKD implementations though,  a balance must be struck between the relatively low probabilities of success for the subtraction/addition operations required and the resultant degradation of the key rates. More detailed studies of these design options for space-based communications are warranted.

\subsection{The interface with classical terrestrial networks}

Although fundamentally a breakthrough, the birth of space-based quantum communications can be seen from a more pragmatic perspective - it will allow for the creation of the global ``Quantum Internet''. This new Internet will interconnect a vast range of devices, from mobile devices all the way through to the much anticipated quantum computers. These devices will be able to transfer quantum information and communicate with each other in an unconditionally secure manner. Importantly, this new network will consist of not only quantum communication channels but also of classical communication channels. As such, consideration of how best to accommodate integration of the quantum information received via satellites into a wider integrated network will be required. Currently, very little detailed thought has been given to this ambitious enterprise, and therefore there is much opportunity for high-impact future research in the context of the integrated system-oriented vision of Fig.~\ref{moores}.

In the CV setting, perhaps the integration of CV quantum information into the microwave setting is the most important example. The implementation of quantum communication protocols in the  optical frequency domain is usually preferred, which is an explicit benefit of the negligible background thermal radiation at optical frequencies, hence all of our discussions have been in this domain. However, the advent of super-conducting microwave quantum circuits have led to an increasing interest in the implementation of quantum communication protocols in the microwave regime \cite{MW-QKD1, MWE1, MWE2, MWE3, MW-T1, MW-QKD2, MW-QKD3, MW-T2, MW-S, MW-R}. These interests are further fuelled by advances in macro electro-optomechanical resonators that are capable of coupling quantum information with the microwave-optical interface \cite{MW-T1,MW-S,MW-R}. With the advent of this technology, quantum information created via super-conducting circuits may be readily upconverted to the optical regime for direct transfer to an overhead satellite. The satellite could then communicate that information optically to a second terrestrial receiver with subsequent conversion back to the microwave regime for storage, error correction or further information processing. Such a scenario could well represent how future quantum computers will share information globally through the quantum Internet. We also note that it is even possible to directly transmit quantum information via microwave carriers to nearby wireless receivers \cite{Neda6}. The development of such integration techniques for the quantum Internet is still in its infancy. % and much research is still required.

\section{Conclusions}
We have  discussed  the recent research advances that are most relevant to CV quantum communication via low-Earth-orbit satellites. Recent experimental results gleaned from the Micius satellite on a range of DV-based quantum communication protocols indicate that CV quantum communication via large distances over the ether has become entirely plausible. We have outlined many of the technical advances in the field of CV quantum communication encompasses and highlighted a range of technical challenges it faces. However, the many advantages of this intriguing technology warrant its experimental deployment to make the vision of the perfectly secure future quantum-communications scenario portrayed in Fig.~\ref{moores} a reality.

{\em Our hope is valued Colleague that you would join this community-effort...}


\begin{thebibliography}{1}
\bibitem{6191306}
L.~Hanzo, H.~Haas, S.~Imre, D.~O'Brien, M.~Rupp, and L.~Gyongyosi, ``Wireless
  myths, realities, and futures: From {3G/4G} to optical and quantum
  wireless,'' \emph{Proceedings of the IEEE}, vol. 100, Invited vision
  paper in the special centennial issue, pp. 1853--1888  (2012).

\bibitem{6515077}
P.~Botsinis, S.~X. Ng, and L.~Hanzo, ``Quantum search algorithms, quantum
  wireless, and a low-complexity maximum likelihood iterative quantum
  multi-user detector design,'' \emph{IEEE Access}, vol.~1, pp. 94--122 (2013).

\bibitem{BB84} C. H. Bennett and G. Brassard, ``Quantum cryptography: Public key distribution and coin tossing,'' \emph{Proceedings of IEEE International Conference on Computers, Systems and Signal Processing}, Bangalore, India, pp. 175-179 (1984).



 \bibitem{teleB}C. H. Bennett, G. Brassard, C. Cr\'epeau, R. Jozsa, A. Peres, and W. K. Wootters, ``Teleporting an unknown quantum State via dual classical and Einstein-Podolsky-Rosen channels,'' \emph{Physical Review Letters} 70, pp. 1895-1899 (1993).

 \bibitem{vaid}  L. Vaidman,  ``Teleportation of quantum states,''
 \emph{Physical Review A} 49, pp. 1473-1476 (1994).

\bibitem{OP-Telep} A. Furusawa, J. L. Sorensen, S. L. Braunstein, C. A. Fuchs, H. J. Kimble, and E. S. Polzik, ``Unconditional quantum teleportation,'' \emph{Science} 282, pp. 706-709 (1998).

\bibitem{1st-tele} P. van Loock and S. L. Braunstein, ``Unconditional teleportation of continuous-variable entanglement,'' \emph{Physical Review A} 61, 010302 (1999).

\bibitem{book-DV} M. A. Nielsen and I. L. Chuang, \emph{Quantum Computation and Quantum Information} (Cambridge University Press, Cambridge, England, 2000).

\bibitem{CV-review} S. L. Braunstein and P. van Loock, ``Quantum information with continuous variables,'' \emph{Reviews of Modern Physics} 77, pp. 513-577 (2005).

\bibitem{1st-CVQKD} C. H. Bennett ``Quantum cryptography using any two nonorthogonal states,'' \emph{Physical Review Letters} 68, pp. 3121-3124 (1992).

\bibitem{2nd-CVQKD} T. C. Ralph ``Continuous variable quantum cryptography,'' \emph{Physical Review A} 61, 010303(R) (1999).

\bibitem{1st-gaussQKD} N. J. Cerf, M. Levy, and G. Van Assche, ``Quantum distribution of Gaussian keys using squeezed states,'' \emph{Physical Review A} 63, 052311 (2001).

\bibitem{RR2002} F. Grosshans and P. Grangier, ``Reverse reconciliation protocols for quantum cryptography with continuous variables,'' Proceedings of the 6th International Conference on Quantum Communications, Measurement, and Computing,  Massachusetts Institute of Technology (MIT), Cambridge, Ma. USA (2002).

\bibitem{RR} F. Grosshans, G. V. Assche, J. Wenger, R. Brouri, N. J. Cerf, and P. Grangier, ``Quantum key distribution using Gaussian-modulated coherent states,'' \emph{Nature} 421, pp. 238-241 (2003).

\bibitem{1st-EB}  F. Grosshans, N. J. Cerf, J. Wenger, R. Tualle-Brouri, and P. Grangier, ``Virtual entanglement and reconciliation protocols for quantum cryptography with continuous variables,'' \emph{Quantum Information and Computation} 3, pp. 535-552 (2003).

\bibitem{Elser} D. Elser, T. Bartley, B. Heim, C. Wittmann, D. Sych, and G. Leuchs, ``Feasibility of free space quantum key distribution with coherent polarization states,'' \emph{New Journal of Physics} 11 045014 (2009).

\bibitem{sem} A. A. Semenov, F. Toppel, D. Yu. Vasylyev, H. V. Gomonay, and W. Vogel, ``Homodyne detection for atmosphere channels,'' \emph{Physical Review A} 85, 013826 (2012).

\bibitem{Rf2016} C. Croal, C. Peuntinger, B. Heim, I. Khan, C. Marquardt, G. Leuchs, P. Wallden, E. Andersson, and N. Korolkova, ``Free-space quantum signatures using heterodyne measurements,'' \emph{Physical Review Letters} 117, 100503 (2016).


\bibitem{fiber-QKD-2006} P. A. Hiskett, D. Rosenberg, C. G. Peterson, R. J. Hughes, S. Nam, A. E. Lita, A. J. Miller, and J. E. Nordholt, ``Long-distance quantum key distribution in optical fibre,'' \emph{New Journal of Physics} 8, 193 (2006).

\bibitem{LDPC} J. Lodewyck, M. Bloch, R. Garcia-Patron, S. Fossier, E. Karpov, E. Diamanti, T. Debuisschert, N. J. Cerf, R. Tualle-Brouri, S. W. McLaughlin, and P. Grangier, ``Quantum key distribution over 25 km with an all-fiber continuous-variable system,'' \emph{Physical Review A} 76, 042305 (2007).

\bibitem{exp-CVQKD2007-1} B. Qi, L.-L. Huang, L. Qian, and H-K. Lo, ``Experimental study on the Gaussian modulated coherent-state quantum key distribution over standard telecommunication fibers,'' \emph{Physical Review A} 76, 052323 (2007).

\bibitem{fiber-QKD-2009} D. Rosenberg, C. G. Peterson, J. W. Harrington, P. R. Rice, N. Dallmann, K. T. Tyagi, K. P. McCabe, S. Nam, B. Baek, R. H. Hadfield, R. J. Hughes, and J. E. Nordholt, ``Practical long-distance quantum key distribution system using decoy levels,'' \emph{New Journal of Physics} 11, 045009 (2009).

\bibitem{exp-CVQKD2009-2} Q. Dinh Xuan, Z. Zhang, and P. Voss, ``A 24 km fiber-based discretely signaled continuous variable quantum key distribution system,'' \emph{Optics Express} 17, pp. 24244-24249 (2009).

\bibitem{exp-PM} P. Jouguet, S. Kunz-Jacques, A. Leverrier, P. Grangier, and E. Diamanti, ``Experimental demonstration of long-distance continuous-variable quantum key distribution,'' \emph{Nature Photonics} 7, pp. 378-381 (2013).

\bibitem{exp-CVQKD-2016-fiber100} D. Huang, P. Huang, D. Lin, and G. Zeng, ``Long-distance continuous-variable quantum key distribution by controlling excess noise,'' \emph{Scientific Reports} 6, 19201 (2016).


\bibitem{Takesue} H. Takesue, S. W. Nam, Q. Zhang, R. H. Hadfield , T. Honjo, K. Tamaki, and Y. Yamamoto, ``Quantum key distribution over a 40-dB channel loss using superconducting single-photon detectors,'' \emph{Nature Photonics} 1, pp. 343-348 (2007).

\bibitem{Stucki} D. Stucki, N. Walenta, F. Vannel, R. T. Thew , N. Gisin, H. Zbinden, S. Gray, C. R. Towery, and S. Ten, ``High rate, long-distance quantum key distribution over 250 km of ultra low loss fibres,'' \emph{New Journal of Physics} 11, 075003 (2009).

\bibitem{fiber2014} H.-K. Lo, M. Curty, and K. Tamaki, ``Secure quantum key distribution,'' \emph{Nature Photonics} 8, pp. 595-604 (2014).

\bibitem{fiber2015} B. Korzh, C. C. W. Lim, R. Houlmann, N. Gisin, M. J. Li, D. Nolan, B. Sanguinetti, R. Thew, and H. Zbinden, ``Provably secure and practical quantum key distribution over 307 km of optical fibre,'' \emph{Nature Photonics} 9, pp. 163-168 (2015).

\bibitem{MDI2016-404} H.-L. Yin, T.-Y. Chen, Z.-W. Yu, H. Liu, L.-X. You, Y.-H. Zhou, S.-J. Chen, Y. Mao, M.-Q. Huang, W.-J. Zhang, H. Chen, M. Jun Li, D. Nolan, F. Zhou, X. Jiang, Z. Wang, Q. Zhang, X.-B. Wang, and J.-W. Pan, ``Measurement-device-independent quantum key distribution over a 404 km optical fiber,'' \emph{Physical Review Letters} 117, 190501 (2016).


\bibitem{repeater} H.-J. Briegel, W. Dur, J. I. Cirac, and P. Zoller, ``Quantum repeaters: The role of imperfect local operations in quantum communication,'' \emph{Physical Review Letters} 81, pp. 5932-5935 (1998).


\bibitem{Peng-2005} C.-Z. Peng, T. Yang, X.-H. Bao, J. Zhang, X.-M. Jin, F.-Y. Feng, B. Yang, J. Yang, J. Yin, Q. Zhang, N. Li, B.-L. Tian, and J.-W. Pan, ``Experimental free-space distribution of entangled photon pairs over 13 km: towards satellite-based global quantum communication,'' \emph{Physical Review Letters} 94, 150501 (2005).

\bibitem{freespace-entanglement-2005} K. Resch, M. Lindenthal, B. Blauensteiner, H. Bohm, A. Fedrizzi, C. Kurtsiefer, A. Poppe, T. Schmitt-Manderback, M. Taraba, R. Ursin, P.Walther, H. Weier, H. Weinfurter, and A. Zeilinger, ``Distributing entanglement
and single photons through an intra-city free-space quantum channel,'' \emph{Optics Express} 13, 202 (2005).

\bibitem{freespace-QKD-2006} I. Marcikic, A. Lamas-Linares, and C. Kurtsiefer, ``Free-space quantum key distribution with entangled photons,'' \emph{Applied Physics Letters} 89, 101122 (2006).

\bibitem{f2} C. Erven, C. Couteau, R. Laamme, and G. Weihs, ``Entangled quantum key distribution over two free-space optical links,'' \emph{Optics Express} 16, pp. 16840-16853 (2008).


\bibitem{aircraft} S. Nauerth, F. Moll,	M. Rau,	C. Fuchs, J. Horwath, S. Frick, and H. Weinfurter, ``Air-to-ground quantum communication,'' \emph{Nature Photonics} 7, pp. 382-386 (2013).

\bibitem{moving-patforms} J.-Y. Wang, B. Yang, S.-K. Liao, L. Zhang, Q. Shen, X.-F. Hu, J.-C. Wu, S.-J. Yang, H. Jiang, Y.-L. Tang, B. Zhong, H. Liang, W.-Y. Liu, Y.-H. Hu, Y.-M. Huang, B. Qi, J.-G. Ren, G.-S. Pan, J. Yin, J.-J. Jia, Y.-A. Chen, K. Chen, C.-Z. Peng, and J.-W. Pan, ``Direct and full-scale experimental verifications towards ground-satellite quantum key distribution,'' \emph{Nature Photonics} 7, pp. 387-393 (2013).

\bibitem{truck} J.-P. Bourgoin, B. L. Higgins, N. Gigov, C. Holloway, C. J. Pugh, S. Kaiser, M. Cranmer, and T. Jennewein, ``Free-space quantum key distribution to a moving receiver,'' \emph{Optics Express} 23, pp. 33437-33447 (2015).


\bibitem{R2} D. E. Bruschi, T. C. Ralph, I. Fuentes, T. Jennewein, and M. Razavi, ``Spacetime effects on satellite-based quantum communications,'' \emph{Physical Review D} 90, 045041 (2014).

% litrature review

\bibitem{proposal-2008} J. M. Perdigues Armengol, B. Furch, C. J. de Matos, O. Minster, L. Cacciapuoti, M. Pfennigbauer, M. Aspelmeyer, T. Jennewein, R. Ursin, T. Schmitt-Manderbach, G. Baister, J. Rarity, W. Leeb, C. Barbieri, H. Weinfurter, and A. Zeilinger, ``Quantum communications at ESA: Towards a space experiment on the ISS,'' \emph{Acta Astronautica} 63, pp. 165-178 (2008).

\bibitem{r11} T. Scheidl, E. Wille, and R. Ursin, ``Quantum optics experiments using the International Space Station: a proposal,'' \emph{New Journal of Physics} 15, 043008 (2013).

\bibitem{proposal-2014} T. Jennewein, J. P. Bourgoin, B. Higgins, C. Holloway, E. Meyer-Scott, C. Erven, B. Heim, Z. Yan, H. Hubel, G. Weihs, E. Choi, I. D'Souza, D. Hudson, R. Laflamme, ``QEYSSAT: A mission proposal for a quantum receiver in space,'' \emph{Proc. of SPIE} 8997, 89970A (2014).

\bibitem{r12} H. Xin, ``Chinese academy takes space under its wing,'' \emph{Science} 332, pp. 904 (2011).

\bibitem{r13} R. Ursin, T. Jennewein, J. Kofler, J. M. Perdigues, L. Cacciapuoti, C. J. de Matos, M. Aspelmeyer, A. Valencia, T. Scheidl, A. Acin, C. Barbieri, G. Bianco, C. Brukner, J. Capmany, S. Cova, D. Giggenbach, W. Leeb, R. H. Hadfield, R. Laflamme, N. Lütkenhaus, G. J. Milburn, M. Peev, T. C. Ralph, J. Rarity, R. Renner, E. Samain, N. Solomos, W. Tittel, J. P. Torres, M. Toyoshima, A. Ortigosa-Blanch, V. Pruneri, P. Villoresi, I. Walmsley, G. Weihs, H. Weinfurter, M. Zukowski, and A. Zeilinger, ``Space-quest, experiments with quantum entanglement in space,'' \emph{Europhysics News} 40, pp. 26-29 (2009).

\bibitem{r14} T. Jennewein and B. Higgins, ``The quantum space race,'' \emph{Physics World} 26, pp. 52-56 (2013).

\bibitem{r15} P. Villoresi , T. Jennewein, F. Tamburini, M. Aspelmeyer, C. Bonato, R. Ursin, C. Pernechele, V. Luceri, G. Bianco, A. Zeilinger, and C. Barbieri, ``Experimental verification of the feasibility of a quantum channel between space and Earth,'' \emph{New Journal of Physics} 10, 033038 (2008).

\bibitem{r16} J. Yin, Y. Cao, S.-B. Liu, G.-S. Pan, J.-H. Wang, T. Yang, Z.-P. Zhang, F.-M. Yang, Y.-A. Chen, C.-Z. Peng, and J.-W. Pan, ``Experimental quasi-single-photon transmission from satellite to earth,'' \emph{Optics Express} 21, pp. 20032-20040 (2013).

\bibitem{r17} G. Vallone, D. Bacco, D. Dequal, S. Gaiarin, V. Luceri, G. Bianco, and P. Villoresi, ``Experimental satellite quantum communications,'' \emph{Physical Review Letters} 115, 040502 (2015).

\bibitem{r18} M. Er-long, H. Zheng-fu, G. Shun-sheng, Z. Tao, D. Da-sheng, and G. Guang-can, ``Background noise of satellite-to-ground quantum key distribution,'' \emph{New Journal of Physics} 7, 215 (2005).

\bibitem{r6} J. G. Rarity, P. R. Tapster, P. M. Gorman, and P Knight, ``Ground to satellite secure key exchange using quantum cryptography,'' \emph{New Journal of Physics} 4, 82 (2002).

\bibitem{r7} M. Aspelmeyer, T. Jennewein, M. Pfennigbauer, W. R. Leeb, and A. Zeilinger, ``Long-distance quantum communication with entangled photons using satellites,'' \emph{IEEE Journal of Selected Topics in Quantum Electronics} 9, pp. 1541-1551 (2003).

\bibitem{r8} C. Bonato, M. Aspelmeyer, T. Jennewein, C. Pernechele, P. Villoresi, and A. Zeilinger, ``Influence of satellite motion on polarization qubits in a Space-Earth quantum communication link,'' \emph{Optics Express} 14, pp. 10050-10059 (2006).

\bibitem{pp1} C. Bonato, A. Tomaello, V. D. Deppo, G. Naletto and P. Villoresi, ``Feasibility of satellite quantum key distribution,'' \emph{New Journal of Physics} 11, 045017 (2009).

\bibitem{r9} A. Tomaello, C. Bonato, V. Da Deppo, G. Naletto, P. Villoresi, ``Link budget and background noise for satellite quantum key distribution,'' \emph{Advances in Space Research} 47, pp. 802-810 (2011).

\bibitem{pp2} E. Meyer-Scott, Z. Yan, A. MacDonald, J.-P. Bourgoin, H. Huebel, and T. Jennewein, ``How to implement decoy state quantum key distribution for a satellite uplink with 50-db channel loss,'' \emph{Physical Review A} 84, 062326 (2011).

\bibitem{r10} J.-P. Bourgoin, E. Meyer-Scott, B. L. Higgins, B. Helou, C. Erven, H. Hubel, B. Kumar, D. Hudson, I. D'Souza, R. Girard, R. Laflamme, and T. Jennewein, ``A comprehensive design and performance analysis of low Earth orbit satellite quantum communication,'' \emph{New Journal of Physics} 15, 023006 (2013).

\bibitem{IEEE1} L. Moli-Sanchez, A. Rodrıguez-Alonso, and G. Seco-Granados, ``Performance analysis of quantum cryptography protocols in optical Earth-satellite and intersatellite Links,'' \emph{IEEE Journal on Selected Areas in Communications} 27, pp. 1582-1590 (2009).

\bibitem{IEEE2} Z. Yan, E. Meyer-Scott, J.-P. Bourgoin, B. L. Higgins, N. Gigov, A. MacDonald, H. Hubel, and T. Jennewein, ``Novel high-speed polarization source for Decoy-state BB84 quantum key distribution over free space and satellite links,'' \emph{Journal of Lightwave Technology} 31, pp. 1399-1408 (2013).

\bibitem{IEEE3} C. Cheng, R. Chandrasekara, Y. Chuan Tan, and A. Ling, ``Space-qualified nanosatellite electronics platform for photon pair experiments,'' \emph{Journal of Lightwave Technology} 33, pp. 4799-4804 (2015).

\bibitem{IEEE4} B. Qi, S. Liu, Q. Shen, S. Liao, W. Cai, Z. Lin, W. Liu, C. Peng, and Q. An, ``A compact readout electronics for the ground station of a quantum communication satellite,'' \emph{IEEE Transactions on Nuclear Science} 62, pp. 883-888 (2015).

\bibitem{IEEE-s} L. Bacsardi, ``On the way to quantum-based satellite communication,'' \emph{IEEE Communications Magazine} 51, pp. 50-55 (2013).

\bibitem{Bacsardi2} L. Bacsardi, ``Satellite communication over quantum channel,'' \emph{Acta Astronautica} 61, pp. 151-159 (2007).

\bibitem{satellite-survey} R. Bedington, J. M. Arrazola, and A. Ling, ``Progress in satellite quantum key distribution,'' \emph{npj Quantum Information} 3(30) (2017).

\bibitem{[8]}
H.~Nguyen, P.~Trinh, A.~Pham, Z.~Babar, D.~Alanis, P.~Botsinis, D.~Chandra,
  S.-X. Ng, and L.~Hanzo, ``Network coding aided cooperative quantum key
  distribution over free-space optical channels,'' \emph{IEEE Access} (2017).

\bibitem{nano-satellite} Z. Tang, R. Chandrasekara, Y. C. Tan, C. Cheng, L. Sha, G. C. Hiang, D. K. L. Oi, and A. Ling, ``Generation and analysis of correlated pairs of photons aboard a nanosatellite,'' \emph{Physical Review Applied} 5, 054022 (2016).

\bibitem{micro-satellite} H. Takenaka, A. Carrasco-Casado, M. Fujiwara, M. Kitamura, M. Sasaki, and M. Toyoshima, ``Satellite-to-ground quantum-limited communication using a 50-kg-class microsatellite,'' \emph{Nature Photonics} 11, pp. 502-508 (2017).

\bibitem{China1} S.-K. Liao, W.-Q. Cai, W.-Y. Liu, L. Zhang, Y. Li, J.-G. Ren, J. Yin, Q. Shen, Y. Cao, Z.-P. Li, F.-Z. Li, X.-W. Chen, L.-H. Sun, J.-J. Jia, J.-C. Wu, X.-J. Jiang, J.-F. Wang, Y.-M. Huang, Q. Wang, Y.-L. Zhou, L. Deng, T. Xi, L. Ma, T. Hu, Q. Zhang, Y.-A. Chen, N.-L. Liu, X.-B. Wang, Z.-C. Zhu, C.-Y. Lu, R. Shu, C.-Z. Peng, J.-Y. Wang, and J.-W. Pan, ``Satellite-to-ground quantum key distribution,'' \emph{Nature} 549, pp. 43-47 (2017).

\bibitem{China2} J. Yin, Y. Cao, Y.-H. Li, S.-K. Liao, L. Zhang, J.-G. Ren, W.-Q. Cai, W.-Y. Liu, B. Li, H. Dai, G.-B. Li, Q.-M. Lu, Y.-H. Gong, Y. Xu, S.-L. Li, F.-Z. Li, Y.-Y. Yin, Z.-Q. Jiang, M. Li, J.-J. Jia, G. Ren, D. He, Y.-L. Zhou, X.-X. Zhang, N. Wang, X. Chang, Z.-C. Zhu, N.-L. Liu, Y.-A. Chen, C.-Y. Lu, R. Shu, C.-Z. Peng, J.-Y. Wang, and J.-W. Pan, ``Satellite-based entanglement distribution over 1200 kilometers,'' \emph{Science} 356, pp. 1140-1144 (2017).

\bibitem{China3} J.-G. Ren, P. Xu, H.-L. Yong, L. Zhang, S.-K. Liao, J. Yin, W.-Y. Liu, W.-Q. Cai, M. Yang, L. Li, K.-X. Yang, X. Han, Y.-Q. Yao, J. Li, H.-Y. Wu, S. Wan, L. Liu, D.-Q. Liu, Y.-W. Kuang, Z.-P. He, P. Shang, C. Guo, R.-H. Zheng, K. Tian, Z.-C. Zhu, N.-L. Liu, C.-Y. Lu, R. Shu, Y.-A. Chen, C.-Z. Peng, J.-Y. Wang, and J.-W. Pan, ``Ground-to-satellite quantum teleportation,'' \emph{Nature} 549, pp. 70-73 (2017).

\bibitem{Boone} K. Boone, J.-P. Bourgoin, E. Meyer-Scott, K. Heshami, T. Jennewein, and C. Simon, ``Entanglement over global distances via quantum repeaters with satellite links,'' \emph{Physical Review A} 91, 052325 (2015).

 \bibitem{SECOQC-Vienna} M Peev, \emph{et al}., ``The SECOQC quantum key distribution network in Vienna,'' \emph{New Journal of Physics} 11, 075001 (2009).

\bibitem{Tokyo} M. Sasaki, \emph{et al}., ``Field test of quantum key distribution in the Tokyo QKD Network,'' \emph{Optics Express} 19, pp. 10387-10409 (2011).

\bibitem{quantum-access-network} B. Frohlich, J. F. Dynes, M. Lucamarini, A. W. Sharpe, Z. Yuan, and A. J. Shields, ``A quantum access network,'' \emph{Nature} 501, pp. 69-72 (2013).

\bibitem{144km} R. Ursin, F. Tiefenbacher, T. Schmitt-Manderbach, H. Weier, T. Scheidl, M. Lindenthal, B. Blauensteiner, T. Jennewein, J. Perdigues, P. Trojek, B. Omer, M. Furst, M. Meyenburg, J. Rarity, Z. Sodnik, C. Barbieri, H. Weinfurter, and A. Zeilinger, ``Entanglement-based quantum communication over 144km,'' \emph{Nature Physics} 3, pp. 481-486 (2007).

\bibitem{decoy} T. Schmitt-Manderbach, H. Weier, M. Furst, R. Ursin, F. Tiefenbacher, T. Scheidl, J. Perdigues, Z. Sodnik, C. Kurtsiefer, J. G. Rarity, A. Zeilinger, and H. Weinfurter, ``Experimental demonstration of free-space decoy-state quantum key distribution over 144 km,'' \emph{Physical Review Letters} 98, 010504 (2007).

\bibitem{freespace-QKD-2009} T. Scheidl, R. Ursin, A. Fedrizzi, S. Ramelow, X.-S. Ma, T. Herbst, R. Prevedel, L. Ratschbacher, J. Kofler, T. Jennewein, and A. Zeilinger, ``Feasibility of 300 km quantum key distribution with entangled states,'' \emph{New Journal of Physics} 11, 085002 (2009).

\bibitem{freespace-entanglement-2009} A. Fedrizzi, R. Ursin, T. Herbst, M. Nespoli, R. Prevedel, T. Scheidl, F. Tiefenbacher, T. Jennewein, and A. Zeilinger, ``High-fidelity transmission of entanglement over a high-loss free-space channel,'' \emph{Nature Physics} 5, pp. 389-392 (2009).

\bibitem{freespace-teleportation100-2012} J. Yin, J.-G. Ren, H. Lu, Y. Cao, H.-L. Yong, Y.-P. Wu, C. Liu, S.-K. Liao, F. Zhou, Y. Jiang, X.-D. Cai, P. Xu, G.-S. Pan, J.-J. Jia, Y.-M. Huang, H. Yin, J.-Y. Wang, Y.-A. Chen, C.-Z. Peng, and J.-W. Pan, ``Quantum teleportation and entanglement distribution over 100-kilometre free-space channels,'' \emph{Nature} 488, pp. 185-188 (2012).

\bibitem{freespace-teleportation143-2012} X.-S. Ma, T. Herbst, T. Scheidl, D. Wang, S. Kropatschek, W. Naylor, B. Wittmann, A. Mech, J. Kofler, E. Anisimova, V. Makarov, T. Jennewein, R. Ursin, and A. Zeilinger, ``Quantum teleportation over 143 kilometres using active feedforward,'' \emph{Nature} 489, pp. 269-273 (2012).

\bibitem{daylight-QKD} S.-K. Liao, H.-L. Yong, C. Liu, G.-L. Shentu, D.-D. Li, J. Lin, H. Dai, S.-Q. Zhao, B. Li, J.-Y. Guan, W. Chen, Y.-H. Gong, Y. Li, Z.-H. Lin, G.-S. Pan, J. S. Pelc, M. M. Fejer, W.-Z. Zhang, W.-Y. Liu, J. Yin, J.-G. Ren, X.-B. Wang, Q. Zhang, C.-Z. Peng, and J.-W. Pan, ``Long-distance free-space quantum key distribution in daylight towards inter-satellite communication,'' \emph{Nature Photonics} 11, pp. 509-513 (2017).



%% CV-free-space


\bibitem{Heim-2009} B. Heim, D. Elser, T. Bartley, M. Sabuncu, C. Wittmann, D. Sych, C. Marquardt, G. Leuchs, ``Atmospheric channel characteristics for quantum communication with continuous polarization variables,'' \emph{Applied Physics B} 98 pp. 635-640 (2010).

\bibitem{Semenov2009} A. A. Semenov, and W. Vogel, ``Quantum light in the turbulent atmosphere,'' \emph{Physical Review A} 80, 021802(R) (2009).

\bibitem{Wander2012} D. Yu. Vasylyev, A. A. Semenov, and W. Vogel, ``Towards global quantum communication: Beam wandering preserves nonclassicality,'' \emph{Physical Review Letters} 108, 220501 (2012).

\bibitem{2016} D. Vasylyev, A. A. Semenov, and W. Vogel, ``Atmospheric quantum channels with weak and strong turbulence,'' \emph{Physical Review Letters} 117, 090501 (2016).


\bibitem{Usenko} V. C. Usenko, B. Heim, C. Peuntinger, C. Wittmann, C. Marquardt, G. Leuchs, and R. Filip, ``Entanglement of Gaussian states and the applicability to quantum key distribution over fading channels,'' \emph{New Journal of Physics} 14, 093048 (2012).

\bibitem{Bohmann1} M. Bohmann, A. A. Semenov, J. Sperling, and W. Vogel, ``Gaussian entanglement in the turbulent atmosphere,'' \emph{Physical Review A} 94, 010302(R) (2016).

\bibitem{Bohmann2} M. Bohmann, J. Sperling, A. A. Semenov, and W. Vogel, ``Higher-order nonclassical effects in fluctuating-loss channels,'' \emph{Physical Review A} 95, 012324 (2017).

\bibitem{Neda1} N. Hosseinidehaj and R. Malaney, ``Gaussian entanglement distribution via satellite,'' \emph{Physical Review A} 91, 022304 (2015).

\bibitem{Neda2} N. Hosseinidehaj and R. Malaney,  ``Quantum key distribution over combined atmospheric fading channels,'' \emph{Proceedings of IEEE International Conference on Communications (ICC)}, London, UK, pp. 7413-7419 (2015).

\bibitem{Neda3} N. Hosseinidehaj and R. Malaney, ``Entanglement generation via non-Gaussian transfer over atmospheric fading channels,'' \emph{Physical Review A} 92, 062336 (2015).

\bibitem{Neda4} N. Hosseinidehaj and R. Malaney, ``CV-MDI quantum key distribution via satellite,'' \emph{Quantum Information and Computation} 17, pp. 0361-0379 (2017).

\bibitem{Neda5} N. Hosseinidehaj and R. Malaney,  ``CV-QKD with Gaussian and non-Gaussian entangled states over satellite-based channels,'' \emph{Proceedings of IEEE Global Communications Conference (GLOBECOM)}, Washington, DC, USA, pp. 1-7 (2016).

\bibitem{Heim} B. Heim, C. Peuntinger, N. Killoran, I. Khan, C. Wittmann, Ch. Marquardt, and G. Leuchs, ``Atmospheric continuous-variable quantum communication,'' \emph{New Journal of Physics} 16, 113018 (2014).

\bibitem{R4} C. Peuntinger, B. Heim, C. R. Muller, C. Gabriel, C. Marquardt, and G. Leuchs, ``Distribution of squeezed states through an atmospheric channel,'' \emph{Physical Review Letters} 113, 060502 (2014).

\bibitem{Geo-2017} K. Gunthner, I. Khan, D. Elser, B. Stiller, O. Bayraktar, C. R. Muller, K. Saucke, D. Trondle, F. Heine, S. Seel, P. Greulich, H. Zech, B. Gutlich, S. Philipp-May, C. Marquardt, and G. Leuchs, ``Quantum-limited measurements of optical signals from a geostationary satellite,'' \emph{Optica} 4, pp. 611-616 (2017).

%%



\bibitem{exp-EB} L. S. Madsen, V. C. Usenko, M. Lassen, R. Filip, and U. L. Andersen, ``Continuous variable quantum key distribution with modulated entangled states,'' \emph{Nature Communications} 3, 1083 (2012).

%% surveys


\bibitem{Pirandola-survey1} S. Pirandola and S. Mancini, ``Quantum teleportation with continuous variables: a survey,'' \emph{Laser Physics} 16, pp. 1418-1438 (2006).

\bibitem{Pirandola-survey2} S. Pirandola, J. Eisert, C. Weedbrook, A. Furusawa, and S. L. Braunstein, ``Advances in quantum teleportation,'' \emph{Nature Photonics} 9, pp. 641-652 (2015).

\bibitem{rr1} G. Adesso and F. Illuminati, ``Entanglement in continuous-variable systems: recent advances and current perspectives,'' \emph{Journal of Physics A: Mathematical and Theoretical} 40, pp. 7821-7880 (2007).

\bibitem{Gisin-survey} N. Gisin and R. Thew, ``Quantum communication,'' \emph{Nature Photonics} 1, pp. 165-171 (2007).

\bibitem{Scarani} V. Scarani, H. Bechmann-Pasquinucci, N.J. Cerf, M. Dusek, N. Lutkenhaus, and M. Peev, ``The security of practical quantum key distribution,'' \emph{Reviews of Modern Physics} 81, pp. 1301-1350 (2009).


\bibitem{review-CV-2010} U. L. Andersen, G. Leuchs, and C. Silberhorn, ``Continuous-variable quantum information processing,'' \emph{Laser and Photonics Reviews} 4, pp. 337-354 (2010).

\bibitem{rr2} X.-B. Wang, T. Hiroshima, A. Tomita, and M. Hayashi, ``Quantum information with Gaussian states,'' \emph{Physics Reports} 448, pp. 1-111 (2007).

\bibitem{Weedbrook2012} C. Weedbrook, S. Pirandola, R. Garcia-Patron, N. J. Cerf, T. C. Ralph, J. H. Shapiro, and S. Lloyd, ``Gaussian quantum information,'' \emph{Reviews of Modern Physics} 84, pp. 621-669 (2012).

\bibitem{QKD-survey} H.-K. Lo, M. Curty, and K. Tamaki, ``Secure quantum key distribution,'' \emph{Nature Photonics} 8, pp. 595-604 (2014).

\bibitem{QKD-survey2} E. Diamanti, H.-K. Lo, B. Qi, and Z. Yuan, ``Practical challenges in quantum key distribution,'' \emph{npj Quantum Information} 2, 16025 (2016).

\bibitem{QKD-survey3} E. Diamanti, and A. Leverrier, ``Distributing secret keys with quantum continuous variables: principle, security and implementations,'' \emph{Entropy} 17, pp. 6072-6092 (2015).


\bibitem{swapping-survey} K. Marshall, and C. Weedbrook, ``Continuous-variable entanglement swapping,'' \emph{Entropy} 17, pp. 3152-3159 (2015).

\bibitem{China-CVQKD-survey} Y.-M. Li, X.-Y. Wang, Z.-L. Bai, W.-Y. Liu, S.-S. Yang, and K.-C. Peng, ``Continuous variable quantum key distribution,'' \emph{Chinese Phys. B} 26, 040303 (2017).

\bibitem{India-CVQKD-survey} A. Shenoy-Hejamadi, A. Pathak, and S. Radhakrishna, ``Quantum cryptography: Key distribution and beyond,'' \emph{Quanta} 6 pp. 1-47 (2017).

%% atmospheric channels

\bibitem{Shaik} K. S. Shaik, ``Atmospheric propagation effects relevant to optical communications,'' \emph{TDA Progress Report} 42-94, pp. 180-200 (1988).

\bibitem{Scintillation2001} L. C. Andrews, R. L. Phillips, and C. Y. Hopen, \emph{Laser Beam Scintillation with Applications}, (SPIE, Bellingham, WA, 2001).

\bibitem{fso} L. C. Andrews and R. L. Phillips, \emph{Laser Beam Propagation Through Random Media}, Vol. PM152. 2nd ed. (SPIE, Bellingham, WA, 2005).

\bibitem{Scintillation&wander} F. Dios, J. A. Rubio, A. Rodrıguez, and A. Comeron, ``Scintillation and beam-wander analysis in an optical ground station–satellite uplink,'' \emph{Applied Optics} 43, pp. 3866-3873 (2004).

\bibitem{Rf2012} I. Capraro, A. Tomaello, A. DallArche, F. Gerlin, R. Ursin, G. Vallone, and P. Villoresi, ``Impact of turbulence in long range quantum and classical communications,'' \emph{Physical Review Letters} 109, 200502 (2012).

\bibitem{Rf2015} G. Vallone, D. G. Marangon, M. Canale, I. Savorgnan, D. Bacco, M. Barbieri, S. Calimani, C. Barbieri, N. Laurenti, and P. Villoresi, ``Adaptive real time selection for quantum key distribution in lossy and turbulent free-space channels,'' \emph{Physical Review A} 91, 042320 (2015).

\bibitem{pinting-error} X. Yi and M. Yao, ``Free-space communications over exponentiated Weibull turbulence channels with nonzero boresight pointing errors,'' \emph{Optics Express} 23, pp. 2904-2917 (2015).


\bibitem{eddies} J. B. Pors, Ph.D. Thesis, Leiden University (2011).

%\bibitem{log-normal1} P. Diament and M. C. Teich, ``Photodetection of low-level radiation through the turbulent atmosphere,'' \emph{Journal of the Optical Society of America} 60, pp. 1489-1494 (1970).

\bibitem{log-normal} P. W. Milonni, J. H. Carter, Ch. G. Peterson, and R. J. Hughes, ``Effects of propagation through atmospheric turbulence on photon statistics,'' \emph{Journal of Optics B} 6, pp. S742-S745 (2004).

\bibitem{pp3} C. Erven, B. Heim, E. Meyer-Scott, J.-P. Bourgoin, R. Laflamme, G. Weihs, and T. Jennewein, ``Studying free-space transmission statistics and improving free-space quantum key distribution in the turbulent atmosphere,'' \emph{New Journal of Physics} 14, 123018 (2012).

\bibitem{review-entanglement-2003} J. Eisert and M. B. Plenio, ``Introduction to the basics of entanglement theory in continuous-variable systems,'' \emph{International Journal of Quantum Information} 1, pp. 479-506 (2003).

\bibitem{Adesso} G. Adesso, Ph.D. Thesis, University of Salerno  (2007).

\bibitem{book-quantum-optics} C. C. Gerry and P. L. Knight, \emph{Introductory Quantum Optics} (Cambridge University Press, Cambridge, 2005).

%\bibitem{book-quantum-mechanics} D. J. Griffiths, \emph{Introduction to Quantum Mechanics: International edition.} (Pearson: Prenctice Hall, 2005).

\bibitem{QKD-coh1} F. Grosshans and P. Grangier, ``Continuous variable quantum cryptography using coherent states,'' \emph{Physical Reveiw Letters} 88, 057902 (2002).

\bibitem{QKD-coh2} C. Weedbrook, A. M. Lance, W. P. Bowen, T. Symul, T. C. Ralph, and P. K. Lam, ``Quantum cryptography without switching,'' \emph{Physical Review Letters} 93, 170504 (2004).


\bibitem{inefficient_homodyne} R. Garcia-Patron and N. J. Cerf, ``Continuous-Variable quantum key distribution protocols over noisy channels,'' \emph{Physical Review Letters} 102, 130501 (2009).

\bibitem{1987} J. G. Rarity, P. R. Tapster, and E. Jakeman, ``Observation of sub-Poissonian light in parametric downconversion,'' \emph{Optics Communications} 62, pp. 201-206 (1987).

\bibitem{Kwiat} P. G. Kwiat, K. Mattle, H. Weinfurter, A. Zeilinger, A. V. Sergienko, and Y. Shih,  ``New high-intensity source of polarization-entangled photon pairs,'' \emph{Physical Review Letters} 75, pp. 4337-4341 (1995).

\bibitem{1997} M. E. Anderson, D. F. McAlister, M. G. Raymer, and M. C. Gupta, ``Pulsed squeezed-light generation in ${\chi ^2}$ nonlinear waveguides,'' \emph{Journal of the Optical Society of America B} 14, pp. 3180-3190 (1997).

\bibitem{1997-2} W. P. Grice, and I. A. Walmsley, ``Spectral information and distinguishability in type-II down-conversion with a broadband pump,'' \emph{Physical Review A} 56, pp. 1627-1634 (1997).

\bibitem{PDC-2003} Y. Shih, ``Entangled biphoton source - property and preparation,'' \emph{Reports on Progress in Physics} 66, pp. 1009-1044 (2003).

\bibitem{Simon} R. Simon, ``Peres-Horodecki separability criterion for continuous variable systems,'' \emph{Physical Review Letters} 84, pp. 2726-2729 (2000).

\bibitem{EOF1} G. Giedke, M. M. Wolf, O. Kruger, R. F. Werner, and J. I. Cirac, ``Entanglement of formation for symmetric Gaussian states,'' \emph{Physical Review Letters} 91, 107901 (2003).

\bibitem{EOF2} M. M. Wolf, G. Giedke, O. Kruger, R. F. Werner, and J. I. Cirac, ``Gaussian entanglement of formation,'' \emph{Physical Review A} 69, 052320 (2004).

\bibitem{LN} G. Vidal and R. F. Werner, ``Computable measure of entanglement,'' \emph{Physical Review A} 65, 032314 (2002).

\bibitem{LN2} M. B. Plenio, ``Logarithmic negativity: A full entanglement monotone that is not convex,'' \emph{Physical Review Letters} 95, 090503 (2005).

\bibitem{MW-QKD1} C. Weedbrook, S. Pirandola, S. Lloyd, and T. C. Ralph, ``Quantum cryptography approaching the classical limit,'' \emph{Physical Review Letters} 105, 110501 (2010).

\bibitem{MW-QKD2} C. Weedbrook, S. Pirandola, and T. C. Ralph, ``Continuous-variable quantum key distribution using thermal states,'' \emph{Physical Review A} 86, 022318 (2012).

\bibitem{MW-QKD3} C. Weedbrook, C. Ottaviani, and S. Pirandola, ``Two-way quantum cryptography at different wavelengths,'' \emph{Physical Review A} 89, 012309 (2014).

\bibitem{Neda6} N. Hosseinidehaj and R. Malaney,  ``Quantum entanglement distribution in next-generation wireless communication systems,'' \emph{85th IEEE Vehicular Technology Conference (VTC)}, Spring, Sydney, Australia (2017).

\bibitem{nedamult} N. Hosseinidehaj and R. Malaney,
``Multimode entangled states in the lossy channel,'' \emph{IEEE VTC International Workshop on Quantum Communications for Future Networks (QCFN)}, Sydney, Australia (2017).

\bibitem{Pirandola} S. Pirandola, ``Entanglement reactivation in separable environments,'' \emph{New Journal of Physics} 15, 113046 (2013).

\bibitem{RSA} R. L. Rivest, A. Shamir, and L. Adleman, ``A method for obtaining digital signatures and public-key cryptosystems,'' \emph{Communications of the ACM} 21, pp. 120-126  (1978).

\bibitem{one-time-pad} G. S. Vernam, ``Cipher printing telegraph systems: For secret wire and radio telegraphic communications,'' \emph{Journal of the American Institute of Electrical Engineers} 45, pp. 109-115 (1926).



\bibitem{Renner1} R. Renner, Ph.D. Thesis, ETH Zurich (2005).

\bibitem{Renner2} R. Renner, N. Gisin, and B. Kraus, ``Information-theoretic security proof for quantum-key-distribution protocols,'' \emph{Physical Review A} 72, 012332 (2005).



\bibitem{G} N. Gisin, G. Ribordy, W. Tittel, and H. Zbinden, ``Quantum cryptography,'' \emph{Reviews of Modern Physics} 74, pp. 145-195 (2002).


    \bibitem{wooters}   W. K. Wootters and  W. H. Zurek,  (1982), ``A single quantum cannot be cloned,'' \emph{Nature}, 299, pp. 802-803 (1982).

   \bibitem{Dieks}    D. Dieks, ``Communication by EPR devices,'' \emph{Physics Letters A}. 92 (6), pp. 271-272, (1982).

\bibitem{exp-CVQKD2007-2} T. Symul, D. J. Alton, S. M. Assad, A. M. Lance, C. Weedbrook, T. C. Ralph, and P. K. Lam, ``Experimental demonstration of post-selection-based continuous variable quantum key distribution in the presence of Gaussian noise,'' \emph{Physical Review A} 76, 030303 (2007).

\bibitem{exp-CVQKD2009-1} S. Fossier, E. Diamanti, T. Debuisschert, A. Villing, R. Tualle-Brouri, and P. Grangier, ``Field test of a continuous-variable quantum key distribution prototype,'' \emph{New Journal of Physics} 11, 045023 (2009).


% free space below
\bibitem{exp-CVQKD2010} Y. Shen, H. Zou, L. Tian, P. Chen, J. Yuan, ``Experimental study on discretely modulated continuous-variable quantum key distribution,'' \emph{Physical Review A} 82, 022317 (2010).

\bibitem{exp-CVQKD2012} P. Jouguet, S. Kunz-Jacques, T. Debuisschert, S. Fossier, E. Diamanti, R. Alleaume, R. Tualle-Brouri, P. Grangier, A. Leverrier, P. Pache, and P. Painchault, ``Field test of classical symmetric encryption with continuous variables quantum key distribution,'' \emph{Optics Express} 20, pp. 14030-14041 (2012).

\bibitem{thesis} R. Garcia-Patron, Ph.D. Thesis, Universite Libre de Bruxelles (2007).

\bibitem{Weedbrook2013} C. Weedbrook, ``Continuous-variable quantum key distribution with entanglement in the middle,'' \emph{Physical Review A} 87, 022308 (2013).

\bibitem{DR-RR} S. Pirandola, R. Garcia-Patron, S. L. Braunstein, and S. Lloyd, ``Direct and reverse secret-key capacities of a quantum channel,'' \emph{Physical Review Letters} 102, 050503 (2009).

\bibitem{coherent-attack} R. Renner and J. I. Cirac, ``de Finetti representation theorem for infinite-dimensional quantum systems and applications to quantum cryptography,'' \emph{Physical Review Letters} 102, 110504 (2009).

\bibitem{Gaussian-attack} R. Garcıa-Patron, and N. J. Cerf, ``Unconditional optimality of Gaussian attacks against continuous-variable quantum key distribution,'' \emph{Physical Review Letters} 97, 190503 (2006).

\bibitem{swap2005} H.-R. Li, F.-L. Li, Y. Yang, and Q. Zhang, ``Entanglement swapping of two-mode Gaussian states in a thermal environment,'' \emph{Physical Review A} 71, 022314 (2005).

\bibitem{swap2006} S. Pirandola, D. Vitali, P. Tombesi, and S. Lloyd, ``Macroscopic entanglement by entanglement swapping,'' \emph{Physical Review Letters} 97, 150403 (2006).

\bibitem{swap2014} M. Abdi, S. Pirandola, P. Tombesi, and D. Vitali, ``Continuous-variable-entanglement swapping and its local certification: Entangling distant mechanical modes,'' \emph{Physical Review A} 89, 022331 (2014).

\bibitem{Obermaier} J. Hoelscher-Obermaier and P. van Loock, ``Optimal Gaussian entanglement swapping,'' \emph{Physical Review A} 83, 012319 (2011).

\bibitem{CV_Bell} S. Pirandola, J. Eisert, C. Weedbrook, A. Furusawa, and S. L. Braunstein, ``Advances in quantum teleportation,'' \emph{Nature Photonics} 9, pp. 641-652 (2015).

%%% MDI

\bibitem{Nature} S. Pirandola, C. Ottaviani, G. Spedalieri,	C. Weedbrook, S. L. Braunstein,	S. Lloyd, T. Gehring, C. S. Jacobsen, and U. L. Andersen, ``High-rate measurement-device-independent quantum cryptography,'' \emph{Nature Photonics} 9, pp. 397-402 (2015).

\bibitem{first_MDI1} S. L. Braunstein, and S. Pirandola, ``Side-channel-free quantum key distribution,'' \emph{Physical Review Letters} 108, 130502 (2012).

\bibitem{1st_DV_MDI_theory} H.-K. Lo, M. Curty, and B. Qi, ``Measurement-device-independent quantum key distribution,'' \emph{Physical Review Letters} 108, 130503 (2012).

\bibitem{sque10} T. Eberle, V. Handchen, and R. Schnabel, ``Stable control of 10 dB two-mode squeezed vacuum states of light,'' \emph{Optics Express} 21, pp. 11546-11553 (2013).

\bibitem{sque15} H. Vahlbruch, M. Mehmet, K. Danzmann, and R. Schnabel, ``Detection of 15 dB squeezed states of light and their application for the absolute calibration of photoelectric quantum efficiency,'' \emph{Physical Review Letters} 117, 110801 (2016).

\bibitem{CV_MDI1} X.-C. Ma, S.-H. Sun, M.-S. Jiang, M. Gui, and L.-M. Liang, ``Gaussian-modulated coherent-state measurement-device-independent quantum key distribution,'' \emph{Physical Review A} 89, 042335 (2014).

\bibitem{CV_MDI2} Z. Li, Y.-C. Zhang, F. Xu, X. Peng, and H. Guo, ``Continuous-variable measurement-device-independent quantum key distribution,'' \emph{Physical Review A} 89, 052301 (2014).

\bibitem{CV_MDI3} Y.-C. Zhang, Z. Li, S. Yu, W. Gu, X. Peng, and H. Guo, ``Continuous-variable measurement-device-independent quantum key distribution using squeezed states,'' \emph{Physical Review A} 90, 052325 (2014).

\bibitem{CV_MDI_sym} C. Ottaviani, G. Spedalieri, S. L. Braunstein, and S. Pirandola, ``Continuous-variable quantum cryptography with an untrusted relay: Detailed security analysis of the symmetric configuration,'' \emph{Physical Review A} 91, 022320 (2015).

\bibitem{CV_MDI4} Y. Zhang, Z. Li, C. Weedbrook, K. Marshall, S. Pirandola, S. Yu, and H. Guo, ``Noiseless linear amplifiers in entanglement-based continuous-variable quantum key distribution,'' \emph{Entropy} 17, pp. 4547-4562 (2015).

\bibitem{fiber_DV1} A. Rubenok, J. A. Slater, P. Chan, I. Lucio-Martinez, and W. Tittel, ``Real-world two-photon interference and proof-of-principle quantum key distribution immune to detector attacks,'' \emph{Physical Review Letters} 111, 130501 (2013).

\bibitem{fiber_DV2} T. Ferreira da Silva, D. Vitoreti, G. B. Xavier, G. C. do Amaral, G. P. Temporao, and J. P. von der Weid, ``Proof-of-principle demonstration of measurement-device-independent quantum key distribution using polarization qubits,'' \emph{Physical Review A} 88, 052303 (2013).

\bibitem{fiber_DV3} Y. Liu, , T.-Y. Chen, L.-J. Wang, H. Liang, G.-L. Shentu, J. Wang, K. Cui, H.-L. Yin, N.-L. Liu, L. Li, X. Ma, J. S. Pelc, M. M. Fejer, C.-Z. Peng, Q. Zhang, and J.-W. Pan, ``Experimental measurement-device-independent quantum key distribution,'' \emph{Physical Review Letters} 111, 130502 (2013).

\bibitem{fiber_DV4} Y.-L. Tang, H.-L. Yin, S.-J. Chen, Y. Liu, W.-J. Zhang, X. Jiang, L. Zhang, J. Wang, L.-X. You, J.-Y. Guan, D.-X. Yang, Z. Wang, H. Liang, Z. Zhang, N. Zhou, X. Ma, T.-Y. Chen, Q. Zhang, and J.-W. Pan, ``Measurement-device-independent quantum key distribution over 200 km,'' \emph{Physical Review Letters} 113, 190501 (2014).

\bibitem{MDI2016} Y.-L. Tang, H.-L. Yin, Q. Zhao, H. Liu, X.-X. Sun, M.-Q. Huang, W.-J. Zhang, S.-J. Chen, L. Zhang, L.-X. You, Z. Wang, Y. Liu, C.-Y. Lu, X. Jiang, X. Ma, Q. Zhang, T.-Y. Chen, and J.-W. Pan, ``Measurement-device-independent quantum key distribution over untrustful metropolitan network,'' \emph{Physical Review X} 6, 011024 (2016).

%%%%% Added Refs DI

\bibitem{DI1} M. Tomamichel, and R. Renner, ``Uncertainty relation for smooth entropies,'' \emph{Physical Review Letters} 106, 110506 (2011).

\bibitem{DI2} C. Branciard, E. Cavalcanti, S. Walborn, V. Scarani, and H. Wiseman, ``One-sided device-independent quantum key distribution: security, feasibility, and the connection with steering,'' \emph{Physical Review A} 85, 010301(R) (2012).

\bibitem{DI3} M. Tomamichel, S. Fehr, J. Kaniewski, and S. Wehner, ``A monogamy-of-entanglement game with applications to device-independent quantum cryptography,'' \emph{New Journal of Physics} 15, 103002 (2013).



\bibitem{CV-DI1} T. Gehring, V. Handchen, J. Duhme, F. Furrer, T. Franz, C. Pacher, R. F. Werner, and R. Schnabel, ``Implementation of continuous-variable quantum key distribution with composable and one-sided-device-independent security against coherent attacks,'' \emph{Nature Communications} 6, 8795 (2015).

\bibitem{CV-DI2} N. Walk, S. Hosseini, J. Geng, O. Thearle, J. Y. Haw, S. Armstrong, S. M. Assad, J. Janousek, T. C. Ralph, T. Symul, H. M. Wiseman, and P. K. Lam, ``Experimental demonstration of Gaussian protocols for one-sided device-independent quantum key distribution,'' \emph{Optica} 3, pp. 634-642 (2016).

    \bibitem{violation}  K. Marshall and C. Weedbrook, ``Device-independent quantum cryptography for continuous variables,'' \emph{Phys. Rev. A }90, 042311 (2014).

\bibitem{fast-fading} P. Papanastasiou, C. Weedbrook, and S. Pirandola, ``Continuous-variable quantum key distribution in fast fading channels,'' arXiv:1710.03525 (2017).

\bibitem{Dong} R. Dong, M. Lassen, J. Heersink, C. Marquardt, R. Filip, G. Leuchs, and U. L. Andersen, ``Continuous-variable entanglement distillation of non-Gaussian mixed states,'' \emph{Physical Review A} 82, 012312 (2010).

\bibitem{Pirandola-fundamentallimit} S. Pirandola, R. Laurenza, C. Ottaviani, and L. Banchi, ``Fundamental limits of repeaterless quantum communications,'' \emph{Nature Communications} 8, 15043 (2017).


%% Distillation
\bibitem{Distillation1} C. H. Bennett, G. Brassard, S. Popescu, B. Schumacher, J. A. Smolin, and W. K. Wootters, ``Purification of noisy entanglement and faithful teleportation via noisy channels,'' \emph{Physical Review Letters} 76, pp. 722-725 (1996).

\bibitem{Distillation2} D. E. Browne, J. Eisert, S. Scheel, and M. B. Plenio, ``Driving non-Gaussian to Gaussian states with linear optics,'' \emph{Physical Review A} 67, 062320 (2003).

\bibitem{Distillation3} J. Eisert, D. E. Browne, S. Scheel, and M. B. Plenio, ``Distillation of continuous-variable entanglement with optical means,'' \emph{Annals of Physics} 311 pp. 431-458 (2004).

\bibitem{Distillation4} J. Fiurasek, P. Marek, R. Filip, and R. Schnabel, ``Experimentally feasible purification of continuous-variable entanglement,'' \emph{Physical Review A} 75, 050302(R) (2007).

\bibitem{Distillation5} A. P. Lund, and T. C. Ralph, ``Continuous-variable entanglement distillation over a general lossy channel,'' \emph{Physical Review A} 80, 032309 (2009).

\bibitem{no_go1} J. Eisert, S. Scheel, and M. B. Plenio, ``Distilling Gaussian states with Gaussian operations is impossible,'' \emph{Physical Review Letters} 89, 137903 (2002).

\bibitem{no_go2} G. Giedke and J. I. Cirac, ``Characterization of Gaussian operations and distillation of Gaussian states,'' \emph{Physical Review A} 66, 032316 (2002).

\bibitem{no_go3} J. Fiurasek, ``Gaussian transformations and distillation of entangled Gaussian states,'' \emph{Physical Review Letters} 89, 137904 (2002).

\bibitem{NLA1} T. C. Ralph and A. P. Lund,  ``Nondeterministic noiseless linear amplification of quantum systems,'' \emph{Proceedings of the 9th International Conference on Quantum Communication Measurement and Computing (QCMC)}, (ed. A. Lvovsky) pp. 155-160 (2009).

\bibitem{NLA-Distil1} G. Y. Xiang, T. C. Ralph, A. P. Lund, N. Walk, and G. J. Pryde, ``Heralded noiseless linear amplification and distillation of entanglement,'' \emph{Nature Photonics} 4, pp. 316-319 (2010).

\bibitem{error-correction7} T. C. Ralph, ``Quantum error correction of continuous-variable states against Gaussian noise,'' \emph{Physical Review A} 84, 022339 (2011).

\bibitem{NLA2} N. Walk, A. P. Lund, and T. C. Ralph, ``Nondeterministic noiseless amplification via non-symplectic phase space transformations,'' \emph{New Journal of Physics} 15, 073014 (2013).

\bibitem{NLA-Distil2} H. M. Chrzanowski, N. Walk, S. M. Assad, J. Janousek, S. Hosseini, T. C. Ralph, T. Symul, and P. K. Lam, ``Measurement-based noiseless linear amplification for quantum communication,'' \emph{Nature Photonics} 8, pp. 333-338 (2014).

\bibitem{NLA-Distil3} A. E. Ulanov, I. A. Fedorov, A. A. Pushkina, Y. V. Kurochkin, T. C. Ralph, and A. I. Lvovsky, ``Undoing the effect of loss on quantum entanglement,'' \emph{Nature Photonics} 9, pp. 764-768 (2015).

\bibitem{NLA-QKD-F} J. Fiurasek and N. Cerf, ``Gaussian postselection and virtual noiseless amplification in continuous-variable quantum key distribution,'' \emph{Physical Review A} 86, 060302 (2012).

\bibitem{NLA-QKD-B} R. Blandino, A. Leverrier, M. Barbieri, J. Etesse, P. Grangier, and R. Tualle-Brouri, ``Improving the maximum transmission distance of continuous-variable quantum key distribution using a noiseless amplifier,'' \emph{Physical Review A} 86, 012327 (2012).

\bibitem{NLA-QKD-W} N. Walk, T. C. Ralph, T. Symul, P. K. Lam, ``Security of continuous-variable quantum cryptography with Gaussian postselection,'' \emph{Physical Review A} 87, 020303 (2013).


\bibitem{1st_PSS} T. Opatrny, G. Kurizki, and D.-G. Welsch, ``Improvement on teleportation of continuous variables by photon subtraction via conditional measurement,'' \emph{Physical Review A} 61, 032302 (2000).

\bibitem{Kitagawa} A. Kitagawa, M. Takeoka, M. Sasaki, and A. Chefles, ``Entanglement evaluation of non-Gaussian states generated by photon subtraction from squeezed states,'' \emph{Physical Review A} 73, 042310 (2006).

\bibitem{1st_PAS} F. Dell'Anno, S. De Siena, L. Albano, and F. Illuminati, ``Continuous-variable quantum teleportation with non-Gaussian resources,'' \emph{Physical Review A} 76, 022301 (2007).

\bibitem{telep-nG} Y. Yang and F.-L. Li, ``Entanglement properties of non-Gaussian resources generated via photon subtraction and addition and continuous-variable quantum-teleportation improvement,'' \emph{Physical Review A} 80, 022315 (2009).

\bibitem{Zhang-Loock} S. L. Zhang, and P. van Loock, ``Distillation of mixed-state continuous-variable entanglement by photon subtraction,'' \emph{Physical Review A} 82, 062316 (2010).

\bibitem{Navarrete}	C. Navarrete-Benlloch, R. Garcia-Patron, J. H. Shapiro, and N. J. Cerf, ``Enhancing quantum entanglement by photon addition and subtraction,'' \emph{Physical Review A} 86, 012328 (2012).

\bibitem{Oxford2013} T. J. Bartley, P. J. D. Crowley, A. Datta, J. Nunn, L. Zhang, and I. Walmsley, ``Strategies for enhancing quantum entanglement by local photon subtraction,'' \emph{Physical Review A} 87, 022313 (2013).

\bibitem{added} S. L. Zhang, Y. Dong, X. Zou, B. Shi, and G. C. Guo, ``Continuous-variable-entanglement distillation with photon addition,'' \emph{Physical Review A} 88, 032324 (2013).

\bibitem{ME} J. Lee and H. Nha, ``Entanglement distillation for continuous variables in a thermal environment: Effectiveness of a non-Gaussian operation,'' \emph{Physical Review A} 87, 032307 (2013).

\bibitem{Seshadreesan}	K. P. Seshadreesan, J. P. Dowling, and G. S. Agarwal, ``Non-Gaussian entangled states and quantum teleportation of Schrodinger-cat states,'' \emph{Physica Scripta} 90, 074029 (2015).

\bibitem{Oxford} T. J. Bartley and I. A. Walmsley, ``Directly comparing entanglement-enhancing non-Gaussian operations, \emph{New Journal of Physics} 17, 023038 (2015).

\bibitem{Allegra}	M. Allegra, P. Giorda, and M. G. A. Paris, ``Role of initial entanglement and non-Gaussianity in the decoherence of photon-number entangled states evolving in a noisy channel,'' \emph{Physical Review Letters} 105, 100503 (2010).

\bibitem{Adesso-proof}	G. Adesso, ``Simple proof of the robustness of Gaussian entanglement in bosonic noisy channels,'' \emph{Physical Review A} 83, 024301 (2011).

\bibitem{Sabapathy}	K. K. Sabapathy, J. Solomon Ivan, and R. Simon, ``Robustness of non-Gaussian entanglement against noisy amplifier and attenuator environments,'' \emph{Physical Review Letters} 107, 130501 (2011).

\bibitem{Filippov}	S. N. Filippov, and M. Ziman, ``Entanglement sensitivity to signal attenuation and amplification,'' \emph{Physical Review A} 90, 010301(R) (2014).

\bibitem{nG1} P. Huang, G. He, J. Fang, and G. Zeng, ``Performance improvement of continuous-variable quantum key distribution via photon subtraction,'' \emph{Physical Review A} 87, 012317 (2013).

\bibitem{nG2} Z. Li, Y. Zhang, X. Wang, B. Xu, X. Peng, and H. Guo, ``Non-Gaussian postselection and virtual photon subtraction in continuous-variable quantum key distribution,'' \emph{Physical Review A} 93, 012310 (2016).

\bibitem{nG-coherent} L. F. M. Borelli, L. S. Aguiar, J. A. Roversi, and A. Vidiella-Barranco, ``Quantum key distribution using continuous-variable non-Gaussian states,'' \emph{Quantum Information Processing} 15, pp. 893-904 (2016).

\bibitem{Amp1} G. S. Agarwal, S. Chaturvedi, and A. Rai, ``Amplification of maximally-path-entangled number states,'' \emph{Physical Review A} 81, 043843 (2010).

\bibitem{Amp2} H. Nha, G. J. Milburn, and H. J. Carmichael, ``Linear amplification and quantum cloning for non-Gaussian continuous variables,'' \emph{New Journal of Physics} 12, 103010 (2010).

\bibitem{nG-modulation} A. Leverrier, and P. Grangier, ``Continuous-variable quantum-key-distribution protocols with a non-Gaussian modulation,'' \emph{Physical Review A} 83, 042312 (2011).

%\bibitem{NOON0} J. Fiurasek, ``Conditional generation of N-photon entangled states of light,'' \emph{Physical Review A} 65, 053818 (2002).

%\bibitem{NOON1} P. Kok, H. Lee, and J. P. Dowling, ``Creation of large-photon-number path entanglement conditioned on photodetection,'' \emph{Physical Review A} 65, 052104 (2002).

%\bibitem{NOON2}  H. Cable and J. P. Dowling, ``Efficient generation of large number-path entanglement using only linear optics and feed-forward,'' \emph{Physical Review Letters} 99, 163604 (2007).

\bibitem{exp0} K. Wakui, H. Takahashi, A. Furusawa, and M. Sasaki, ``Photon subtracted squeezed states generated with periodically poled $KTiOP{O_4}$,'' \emph{Optics Express} 15, pp. 3568-3574 (2007).

\bibitem{exp1} H. Takahashi, J. S. Neergaard-Nielsen, M. Takeuchi, M. Takeoka, K. Hayasaka, A. Furusawa, and M. Sasaki, ``Entanglement distillation from Gaussian input states,'' \emph{Nature Photonics} 4, pp. 178-181 (2010).

\bibitem{exp2} Y. Kurochkin, A. S. Prasad, and A. I. Lvovsky, ``Distillation of the two-mode squeezed state,'' \emph{Physical Review Letters} 112, 070402 (2014).

\bibitem{added_exp1} A. Zavatta, S. Viciani, and M. Bellini, ``Quantum-to-classical transition with single-photon-added coherent states of light,'' \emph{Science} 306, pp. 660-662 (2004).

\bibitem{added_exp2} A. Zavatta, V. Parigi, and M. Bellini, ``Experimental nonclassicality of single-photon-added thermal light states,'' \emph{Physical Review A} 75, 052106 (2007).

%\bibitem{NOON2015} M. Bohmann, J. Sperling, and W. Vogel, ``Entanglement and phase properties of noisy NOON states,'' \emph{Physical Review A} 91, 042332 (2015).

%\bibitem{NOON_exp2} I. Afek, O. Ambar, and Y. Silberberg, ``High-NOON states by mixing quantum and classical light,'' \emph{Science} 328, pp. 879-881 (2010).

%\bibitem{NOON_exp1} Y. Israel, I. Afek, S. Rosen, O. Ambar, and Y. Silberberg, ``Experimental tomography of NOON states with large photon numbers,'' \emph{Physical Review A} 85, 022115 (2012).

\bibitem{Kraus} J. Solomon Ivan, K. K. Sabapathy, and R. Simon, ``Operator-sum representation for bosonic Gaussian channels,'' \emph{Physical Review A} 84, 042311 (2011).

%\bibitem{noiseref} G. Nocerino, D. Buono, A. Porzio, and S. Solimeno, ``Survival of continuous variable entanglement over long distances,'' \emph{Physica Scripta} 2013, 014049 (2013).

%\bibitem{Neda6} N. Hosseinidehaj and R. Malaney,  ``Quantum entanglement distribution in next-generation wireless communication systems,'' \emph{IEEE Vehicular Technology Conference (VTC)}, accepted for publication, arXiv:1608.05188 (2017).

%\bibitem{Distil2} H. M. Chrzanowski, N. Walk, S. M. Assad, J. Janousek, S. Hosseini, T. C. Ralph, T. Symul, and P. K. Lam, ``Measurement-based noiseless linear amplification for quantum communication,'' \emph{Nature Photonics} 8, pp. 333-338 (2014).
%
%\bibitem{After2} A. E. Ulanov, I. A. Fedorov, A. A. Pushkina, Y. V. Kurochkin, T. C. Ralph, and A. I. Lvovsky, ``Undoing the effect of loss on quantum entanglement,'' \emph{Nature Photonics} 9, pp. 764-768 (2015).
%
%\bibitem{Distil1} G. Y. Xiang, T. C. Ralph, A. P. Lund, N. Walk, and G. J. Pryde, ``Heralded noiseless linear amplification and distillation of entanglement,'' \emph{Nature Photonics} 4, pp. 316-319 (2010).

%\bibitem{adaptive} A. J. Hashmi, A. A. Eftekhar, A. Adibi, and F. Amoozegar, ``Analysis of adaptive optics-based telescope arrays in a deep-space inter-planetary optical communications link between Earth and Mars,'' \emph{Optics Communications} 333, pp. 120-128 (2014).

\bibitem{WCP1} H. Weier, T. Schmitt-Manderbach, N. Regner, C. Kurtsiefer, and H. Weinfurter, ``Free space quantum key distribution: towards a real life application,'' \emph{Fortschr. Phys.} 54, pp. 840-845 (2006).

\bibitem{WCP2} M. Jofre, A. Gardelein, G. Anzolin, W. Amaya, J. Capmany, R. Ursin, L. Penate, D. Lopez, J. L. San Juan, J. A. Carrasco, F. Garcia, F. J. Torcal-Milla, L. M. Sanchez-Brea, E. Bernabeu, J. M. Perdigues, T. Jennewein, J. P. Torres, M. W. Mitchell, and V. Pruneri, ``Fast optical source for quantum key distribution based on semiconductor optical amplifiers,'' \emph{Optics Express} 19, pp. 3825-3834 (2011).

\bibitem{hybrid} U. L. Andersen, J. S. Neergaard-Nielsen, P. van Loock, and A. Furusawa, ``Hybrid discrete- and continuous-variable quantum information,'' \emph{Nature Physics} 11, pp. 713-719 (2015).

%%% future directions

\bibitem{Semonov2016-DV} M. O. Gumberidze, A. A. Semenov, D. Vasylyev, and W. Vogel, ``Bell nonlocality in the turbulent atmosphere,'' \emph{Physical Review A} 94, 053801 (2016).

\bibitem{Semonov2010-DV} A. A. Semenov, and W. Vogel, ``Entanglement transfer through the turbulent atmosphere,'' \emph{Physical Review A} 81, 023835 (2010).

\bibitem{scintillation2016} O. O. Chumak and R. A. Baskov, ``Strong enhancing effect of correlations of photon trajectories on laser beam scintillations,'' \emph{Physical Review A} 93, 033821 (2016).

\bibitem{Bohmann3} M. Bohmann, R. Kruse, J. Sperling, C. Silberhorn, and W. Vogel, ``Probing free-space quantum channels with laboratory-based experiments,'' \emph{Physical Review A} 95, 063801 (2017).

\bibitem{LDPC-DV1} D. Elkouss, A. Leverrier, R. Alleaume, and J. J. Boutros, ``Efficient reconciliation protocol for discrete-variable quantum key distribution,'' \emph{IEEE International Symposium on Information Theory (ISIT)}, Seoul, South Korea, pp. 1879-1883 (2009).

\bibitem{LDPC-DV2}  M. Milicevic, C. Feng, L. M. Zhang, and P. Glenn Gulak, ``Key reconciliation with low-density parity-check codes for long-distance quantum cryptography,'' arXiv:1702.07740 (2017).

\bibitem{LDPC-CV1}  X. Wang, Y. Zhang, S. Yu, and H. Guo, ``High speed information reconciliation for long distance continuous-variable quantum key distribution system,'' \emph{Frontiers in Optics 2017, OSA Technical Digest (online) (Optical Society of America, 2017)} paper JW4A.36.

\bibitem{LDPC-CV2} M. Bloch, A. Thangaraj, and S. W. McLaughlin, ``Efficient reconciliation of correlated continuous random variables using LDPC codes,'' arXiv:cs/0509041 (2005).

%\bibitem{LDPC} J. Lodewyck, M. Bloch, R. Garcıa-Patron, S. Fossier, E. Karpov, E. Diamanti, T. Debuisschert, N. J. Cerf, R. Tualle-Brouri, S. W. McLaughlin, and P. Grangier, ``Quantum key distribution over 25 km with an all-fiber continuous-variable system},'' \emph{Physical Review A} 76, 042305 (2007).

\bibitem{LDPC-CV3} A. Leverrier, R. Alleaume, J. Boutros, G. Zemor, and P. Grangier, ``Multidimensional reconciliation for a continuous-variable quantum key distribution,'' \emph{Physical Review A} 77, 042325 (2008).

\bibitem{LDPC-CV4} P. Jouguet, S. Kunz-Jacques, A. Leverrier, P. Grangier, and E. Diamanti, ``Experimental demonstration of long-distance continuous-variable quantum key distribution,'' \emph{Nature Photonics} 7, pp. 378-381 (2013).

\bibitem{Finite0} M. Tomamichel, C. C. W. Lim, N. Gisin, and  R. Renner, ``Tight finite-key analysis for quantum cryptography,'' \emph{Nature Communications} 3, 634 (2012).

\bibitem{Finite1} A. Leverrier, F. Grosshans, and P. Grangier, ``Finite-size analysis of a continuous-variable quantum key distribution,'' \emph{Physical Review A} 81, 062343 (2010).

\bibitem{Finite2} P. Jouguet, D. Elkouss, and S. Kunz-Jacques, ``High-bit-rate continuous-variable quantum key distribution,'' \emph{Physical Review A} 90, 042329 (2014).

\bibitem{Finite3} F. Furrer, T. Franz, M. Berta, A. Leverrier, V. B. Scholz, M. Tomamichel, and R. F. Werner, ``Continuous variable quantum key distribution: Finite-key analysis of composable security against coherent attacks,'' \emph{Physical Review Letters} 109, 100502 (2012).

\bibitem{Finite4} F. Furrer, ``Reverse-reconciliation continuous-variable quantum key distribution based on the uncertainty principle,'' \emph{Physical Review A} 90, 042325 (2014).

%\bibitem{Finite5} T. Gehring, V. Handchen, Jorg Duhme, F. Furrer, T. Franz, C. Pacher, R. F. Werner, and R. Schnabel, ``Implementation of continuous-variable quantum key distribution with composable and one-sided-device-independent security against coherent attacks,'' \emph{Nature Communications} 6 (2015).

\bibitem{Finite-coherent} A. Leverrier, ``Composable security proof for continuous-variable quantum key distribution with coherent states,'' \emph{Physical Review Letters} 114, 070501 (2015).

\bibitem{Finite-MDI1} P. Papanastasiou, C. Ottaviani, and S. Pirandola, ``Finite-size analysis of measurement-device-independent quantum cryptography with continuous variables,'' \emph{Physical Review A} 96, 042332 (2017).

\bibitem{Finite-MDI2} X. Zhang, Y.-C. Zhang, Y. Zhao, X. Wang, S. Yu, and H. Guo, ``Finite-size analysis of continuous-variable measurement-device-independent quantum key distribution,'' Phys. Rev. A 96, 042334  (2017).

\bibitem{Finite-DI} R. Arnon-Friedman, R. Renner, and T. Vidick, ``Simple and tight device-independent security proofs,'' arXiv:1607.01797 (2016).

\bibitem{adaptive-codes1} D. Elkouss, J. Martinez-Mateo, and V. Martin, ``Secure rate-adaptive reconciliation,'' \emph{In Proceedings of International Symposium on Information Theory and its Applications (ISITA)}, Taichung, Taiwan, pp. 179-184 (2010).

\bibitem{adaptive-codes2} X. Wang, Y.-C. Zhang, Z. Li, B. Xu, S. Yu, and H. Guo, ``Efficient rate-adaptive reconciliation for continuous-variable quantum key distribution,'' arXiv:1703.04916 (2017).

\bibitem{adaptive-codes3} X.-Q. Jiang, P. Huang, D. Huang, D. Lin, and G. Zeng, ``Secret information reconciliation based on punctured low-density parity-check codes for continuous-variable quantum key distribution,'' \emph{Physical Review A} 95, 022318 (2017).


\bibitem{tc_teq_st_2:book} L. Hanzo, T. H. Liew, B. L. Yeap, R. Y. S. Tee and S. X. Ng, ``Turbo coding, Turbo equalisation and space-time coding: EXIT-chart-aided near-capacity designs for wireless channels,'' 2nd Edition, (John Wiley IEEE Press,
New York, USA, 2011).


 \bibitem{DBLP:journals/corr/cs-IT-0406001} C. Nguyen,
               G. Van Assche and
               N. J. Cerf, ``Side-information coding with Turbo Codes and its application to quantum
               key distribution,'' \emph{International Symposium on Information Theory and its Applications, ISITA2004},
Parma, Italy  (2004).


 \bibitem{turbo-dv-qkd-2014} N. Benletaief, H.  Rezig,  and A. Bouallegue,  ``Toward efficient quantum key distribution reconciliation,'' \emph{Journal of Quantum Information Science}, 4, pp. 117-128 (2014).

  \bibitem{turbo-dv-qkd-2017} W. Y. Liu, X. F. Zhong, T. Wu, F. Z. Li, B. Jin, Y. Tang, H. M. Hu, Z. P. Li, L. Zhang, W. Q. Cai, S. K. Liao, Y. Cao, and C. Z. Peng, ``Experimental free-space quantum key distribution with efficient error correction'' \emph{Opt. Express}, 25, pp. 10716-10723 (2017).

 \bibitem{5075875} E. Arikan, ``Channel polarization: A method for constructing
capacity-achieving codes for symmetric binary-input memoryless
channels,''
\emph{IEEE Transactions on Information Theory}, 55, 7,  pp. 3051-3073 (2009).



 \bibitem{Jouguet:2014:HPE:2600508.2600516} P. Jouguet and S. Kunz-Jacques, ``High performance error correction for quantum key distribution using polar codes,'' \emph{Quantum Information Computatation}, 14, 3-4,  pp. 329-338 (2014).




\bibitem{no-go-error-correction} J. Niset, J. Fiurasek, and N. J. Cerf, ``No-Go theorem for Gaussian quantum error correction,'' \emph{Physical Review Letters} 102, 120501 (2009).


\bibitem{error-correction1} S. L. Braunstein, ``Quantum error correction for communication with linear optics,'' \emph{Nature} 394, 47–49 (1998).

\bibitem{error-correction2} S. Lloyd, and J. E. Slotine, ``Analog quantum error correction,'' \emph{Physical Review Letters} 80, pp. 4088-4091 (1998).

\bibitem{error-correction3} S. L. Braunstein, ``Error correction for continuous variables,'' \emph{Physical Review Letters} 80, pp. 4084-4087 (1998).

\bibitem{error-correction4} T. A. Walker, and S. L. Braunstein, ``Five-wave-packet linear optics quantum-error-correcting code,'' \emph{Physical Review A} 81, 062305 (2010).

\bibitem{error-correction5} M. M. Wilde, H. Krovi, and T. A. Brun, ``Entanglement-assisted quantum error correction with linear optics,'' \emph{Physical Review A} 76, 052308 (2007).

\bibitem{error-correction6} J. Niset, U. L. Andersen, and N. J. Cerf, ``Experimentally feasible quantum erasure-correcting code for continuous variables,'' \emph{Physical Review Letters} 101, 130503 (2008).

%\bibitem{error-correction7} T. C. Ralph, ``Quantum error correction of continuous-variable states against Gaussian noise,'' \emph{Physical Review A} 84, 022339 (2011).

\bibitem{error-correction8} T. Aoki, G. Takahashi, T. Kajiya, J.-i. Yoshikawa, S. L. Braunstein, P. van Loock, and A. Furusawa, ``Quantum error correction beyond qubits,'' \emph{Nature Physics} 5, pp. 541-546 (2009).

\bibitem{error-correction9} M. Lassen, M. Sabuncu, A. Huck, J. Niset, G. Leuchs, N. J. Cerf, and U. L. Andersen, ``Quantum optical coherence can survive photon loss using a continuous-variable quantum erasure-correcting code,'' \emph{Nature Photonics} 4, pp. 700-705 (2010).

\bibitem{error-correction10} M. Lassen, A. Berni, L. S. Madsen, R. Filip, and U. L. Andersen, ``Gaussian error correction of quantum states in a correlated noisy channel,'' \emph{Physical Review Letters} 111, 180502 (2013).

\bibitem{error-correction11} S. Hao, X. Su, C. Tian, C. Xie, and K. Peng, ``Five-wave-packet quantum error correction based on continuous-variable cluster entanglement,'' \emph{Scientific Reports} 5, 15462, (2015).


\bibitem{MWE1} C. Eichler, D. Bozyigit, C. Lang, M. Baur, L. Steffen, J. M. Fink, S. Filipp, and A. Wallraff, ``Observation of two-mode squeezing in the microwave frequency domain,'' \emph{Physical Review Letters} 107, 113601 (2011).

\bibitem{MWE2} E. P. Menzel, R. Di Candia, F. Deppe, P. Eder, L. Zhong, M. Ihmig, M. Haeberlein, A. Baust, E. Hoffmann, D. Ballester, K. Inomata, T. Yamamoto, Y. Nakamura, E. Solano, A. Marx, and R. Gross, ``Path entanglement of continuous-variable quantum microwaves,'' \emph{Physical Review Letters} 109, 250502 (2012).

\bibitem{MWE3} E. Flurin, N. Roch, F. Mallet, M. H. Devoret, and B. Huard, ``Generating entangled microwave radiation over two transmission lines,'' \emph{Physical Review Letters} 109, 183901 (2012).

\bibitem{MW-T1} S. Barzanjeh, M. Abdi, G. J. Milburn, P. Tombesi, and D. Vitali, ``Reversible optical-to-microwave quantum interface,'' \emph{Physical Review Letters} 109, 130503 (2012).

\bibitem{MW-T2} R. Di Candia, K. G. Fedorov, L. Zhong, S. Felicetti, E. P. Menzel, M. Sanz, F. Deppe, A. Marx, R. Gross and E. Solano, ``Quantum teleportation of propagating quantum microwaves,'' \emph{EPJ Quantum Technology} 2, 25 (2015).

\bibitem{MW-S} M. Abdi, P. Tombesi, and D. Vitali, ``Entangling two distant non-interacting microwave modes,'' \emph{Annalen der Physik (Berlin)} 527, pp. 139-146 (2015).

\bibitem{MW-R} S. Barzanjeh, S. Guha, C. Weedbrook, D. Vitali, J. H. Shapiro, and S. Pirandola, ``Microwave quantum illumination,'' \emph{Physical Review Letters} 114, 080503 (2015).







\end{thebibliography}
\end{document}